\documentclass[twocolumn, twocolappendix, tighten]{aastex631}
\usepackage{float}
\usepackage{tablefootnote}

\defcitealias{Inoueetal2014}{I14}
\defcitealias{Bassettetal2021}{B21}

\DeclareRobustCommand{\ion}[2]{%
\relax\ifmmode
\ifx\testbx\f
{{#1\,\mathsc{#2}}}\else
{\mathrm{#1\,\mathsc{#2}}}\fi
\else\textup{#1\,{\mdseries\textsc{#2}}}%
\fi}

\begin{document}

\title{LyC Leakers in the \textsl{AstroSat} UV Deep Field: Extreme UV emitters at the Cosmic Noon}

\correspondingauthor{Suraj Dhiwar}
\email{suraj@iucaa.in}

\author[0000-0001-8650-205X]{Suraj Dhiwar}
\affiliation{Inter University Center for Astronomy and Astrophysics, Ganeshkhind, Pune 411007, India}
\affiliation{Department of Physics, Savitribai Phule Pune University, Pune 411007, India}
\author[0000-0002-8768-9298]{Kanak Saha}
\affiliation{Inter University Center for Astronomy and Astrophysics, Ganeshkhind, Pune 411007, India}
\author[0009-0003-8568-4850]{Soumil Maulick}
\affiliation{Inter University Center for Astronomy and Astrophysics, Ganeshkhind, Pune 411007, India}
\author[0000-0002-0648-1699]{Brent M. Smith} 
\affiliation{School of Earth and Space Exploration, Arizona State University, Tempe, AZ 85287-1404, USA}
\author[0000-0003-4531-0945]{Chayan Mondal}
\affiliation{Inter University Center for Astronomy and Astrophysics, Ganeshkhind, Pune 411007, India}
\author[0000-0002-7064-5424]{Harry I. Teplitz} 
\affiliation{IPAC, Mail Code 314-6, Caltech, 1200 E. California Blvd., Pasadena, CA 91125, USA}
\author[0000-0002-9946-4731]{Marc Rafelski} 
\affiliation{Space Telescope Science Institute, 3700 San Martin Drive, Baltimore, MD 21218, USA}
\affiliation{Department of Physics and Astronomy, Johns Hopkins University, Baltimore, MD 21218, USA}
\author[0000-0001-8156-6281]{Rogier A. Windhorst} 
\affiliation{School of Earth and Space Exploration, Arizona State University,
Tempe, AZ 85287-1404, USA}

\shorttitle{L\lowercase{y}C Leakers at Cosmic Noon}
\shortauthors{Suraj Dhiwar et al.}


\begin{abstract}

We report the direct detection of Lyman Continuum (LyC) emission from 9 galaxies and 1 Active Galactic Nuclei (AGN)  at $z$ $\sim$ 1.1-1.6 in the GOODS-North field using deep observations from the Ultraviolet Imaging Telescope (UVIT) onboard AstroSat. The absolute escape fraction of the sources estimated from the far-ultraviolet (FUV) and H$\alpha$ line luminosities using Monte Carlo (MC) analysis of two Inter-Galactic Medium (IGM) models span a range $\sim$ 10 - 55 $\%$. The restframe UV wavelength of the sources falls in the extreme-ultraviolet (EUV) regime $\sim$ 550-700 \AA, the shortest LyC wavelength range probed so far.
This redshift range remains devoid of direct detections of LyC emission due to the instrumental limitations of previously available facilities. 
With UVIT having a very low detector noise, each of these sources are detected with an individual signal-to-noise ratio (SNR) $>$ 3 while for the stack of six sources, we achieve an SNR $\sim$ 7.4. 
The LyC emission is seen to be offset from the optical centroids and extended beyond the UVIT PSF of 1.$^{\prime\prime}6$ in most of the sources. This sample fills an important niche between GALEX and Cosmic Origins Spectrograph (COS) at low-$z$, and HST WFC3 at high-$z$ and is crucial in understanding the evolution of LyC leakers.



\end{abstract}



\section{Introduction}
\label{sec:intro}
After the Big Bang, the Universe expanded and cooled until the electrons and protons could combine to form atoms (at $z\sim$1090) and remained in the neutral form until the Cosmic Dawn ($z\sim$20-15) when the first galaxies started forming. During the period from $z\sim$12 to $z\sim$6, when the LyC radiation from the galaxies started ionizing the surrounding Inter-Galactic Medium (IGM), the Universe underwent a transformation from a completely neutral state  to an almost completely ionised state; see \citet{Robertsonetal2010} for a review. Detecting the ionizing radiation from this early epoch of galaxies is extremely important to understand the nature and formation of these galaxies and the process of reionization. 

However, direct detection of LyC photons from the era of reionization is challenging because of the increasing IGM opacity \citep{Madau95, Inoueetal2014} and plausible faintness of the ionizing sources. Nevertheless, there have been detection of LyC emission from local analogues of the high redshift galaxies at $z$ $<$ 0.4 using the Cosmic Origins Spectrograph on the Hubble Space Telescope \citep{Leitetetal2013, Borthakuretal2014, Izotovetal2016, Leithereretal2016, Izotovetal2018,Flury22}, as well at higher redshift $z$ $>$ 3 using HST and ground-based imaging \citep{Shapleyetal2016, Vanzellaetal2016, Bianetal2017, Steideletal2018, Vanzellaetal2018, Fletcheretal2019, Prichardetal2022,Liuetal2023}. 

There was no direct detection of LyC emission in the intermediate redshift range 0.5 $\lesssim z \lesssim$ 3 until the UVIT onboard the Indian multiwavelength satellite AstroSat detected EUV radiation from a clumpy galaxy (AUDFs01) at $z$ $\simeq$ 1.42 \citep{Sahaetal2020}. Previously, extensive stacking analyses have been carried out using HST/WFC3 deep field data and provided upper limits on the LyC escape fraction in this redshift range, however, with no significant detection of LyC emission \citep{Sianaetal2007, Iwataetal2009, Smithetal2018, Smithetal2020, Alavietal2020}. A larger sample of EUV emitters like AUDFs01 is required to investigate if such galaxies are representative of the high-redshift galaxies that contributed to the reionization of the Universe.

In this letter, we report the direct detection of LyC emission in the EUV domain ($550-700$ \AA) from nine star-forming (SF) galaxies and one hosting an Active Galactic Nucleus (AGN) at the cosmic noon (z $\sim$ 1.1-1.6) in the GOODS-North field using deep UV imaging from UVIT \citep{Mondaletal2023}. 

Throughout this work, all distance dependent quantities use $H_{0}$ = 70 km s$^{-1}$ Mpc$^{-1}$, $\Omega_{m}$ = 0.3, and $\Omega_{\Lambda}$ = 0.7 following $\Lambda$CDM cosmology. 

\section{Data}
\label{sec:data} 

\subsection{UVIT Imaging data}
\label{sec:uvit}
The far-ultraviolet (FUV) and near-ultraviolet (NUV) imaging data come from AstroSat observations of the GOODS-North field (GT08-77; principal investigator, Kanak Saha). These observations were carried out by UVIT (with a field of view (FOV) of 28$^{\prime}$ diameter) onboard AstroSat, which performed simultaneous observation in the F154W ($\lambda_{mean}$=1541 \AA, $\Delta \lambda$=380 \AA) and N242W ($\lambda_{mean}$=2418 \AA, $\Delta \lambda$=785 \AA) bands during 10 - 12 March, 2018. The observations were carried out for 40 kilosec, which corresponds to $\sim 32$ orbits. During each orbit, FUV and NUV observations in photon counting mode were taken every 33 ms, resulting in about 45,000–50,000 frames (in each band) accumulation in a good orbit. 
The final science-ready images of the AstroSat Ultra-violet Deep Field North (AUDFn) had a total exposure time of t$_{exp}$ = 34022 s in the FUV and 19228~s in the NUV with 5$\sigma$ limiting magnitude 26.79 and 26.73, respectively. We ask the reader to refer to \citet{Mondaletal2023} for more details on the data reduction.

\subsection{3DHST grism and Imaging data}
\label{sec:3dhst}

We make use of the archival grism spectral data from the 3D-HST survey \citep[PI
: Weiner,][]{Brammeretal2012, Momchevaetal2016}. 3D-HST provides rest-frame optical spectra for a sample of $\sim$ 7000 galaxies at 1 $< z <$ 3.5. It covers three-quarters (625 arcmin$^2$) of the CANDELS Treasury survey area with two orbits of primary WFC3/G141 grism coverage and two to four orbits with the ACS/G800L grism in parallel. The WFC3/G141 spectra provide continuous wavelength coverage from 1.1 to 1.65 $\mu$m at a spatial resolution of $\sim$ 0.13$^{\prime\prime}$. We also  make use of the publicly available deep imaging data in optical HST ACS \citep{Giavaliscoetal2004}, IR from WFC3 \citep{Skeltonetal2014, Koekemoeretal2011, Groginetal2011} and U-band from the Mosaic camera on the Kitt Peak 4-m telescope \citep{Capaketal2004} of the GOODS-North field.


\begin{figure*}[hbt!]
    \centering
    
    \includegraphics[width=0.31\columnwidth]{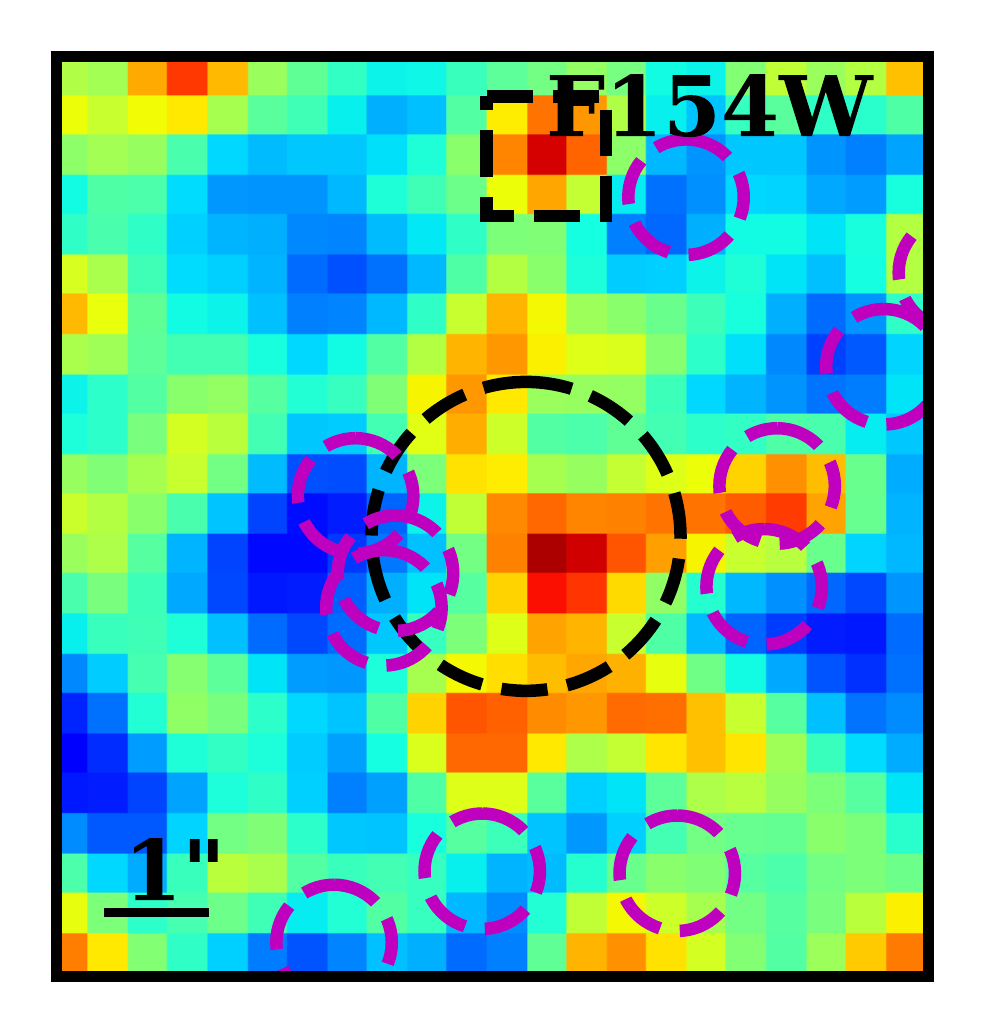}
    \includegraphics[width=0.312\columnwidth]{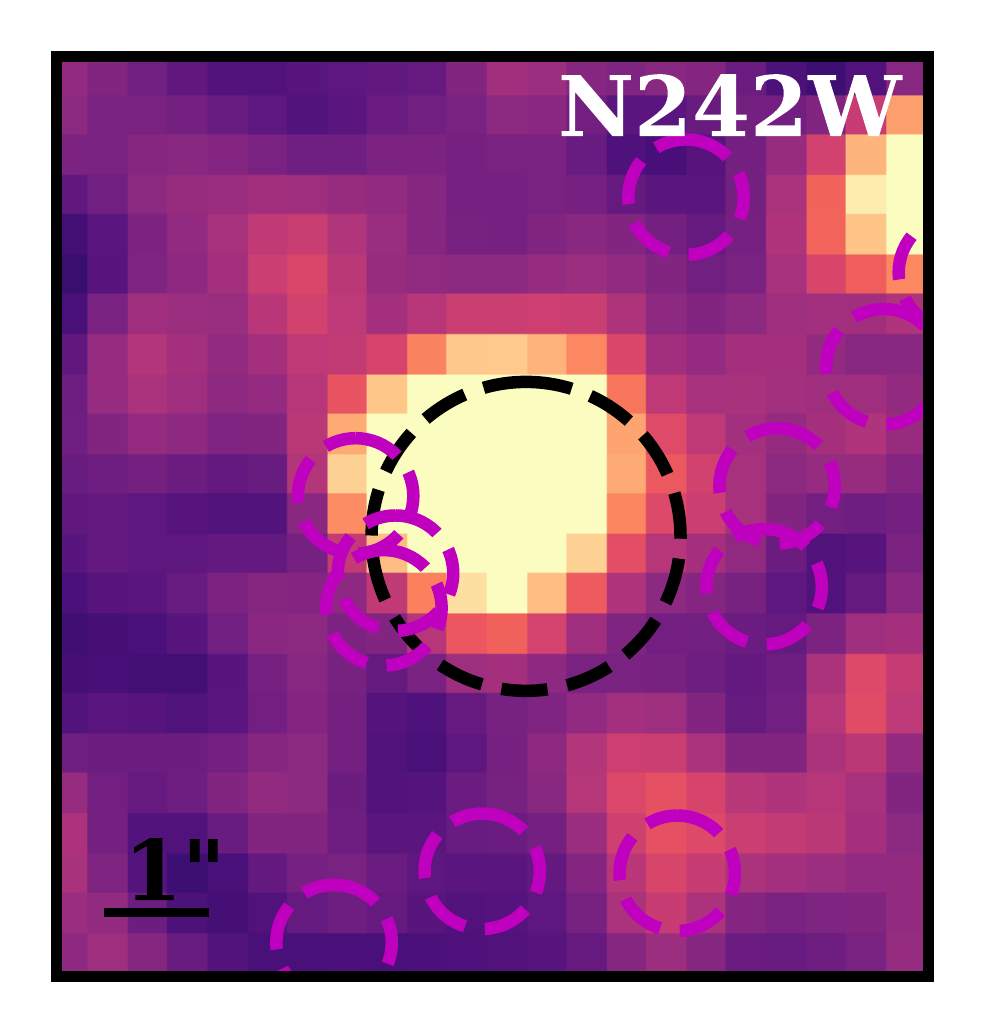}
    \includegraphics[width=0.32\columnwidth]{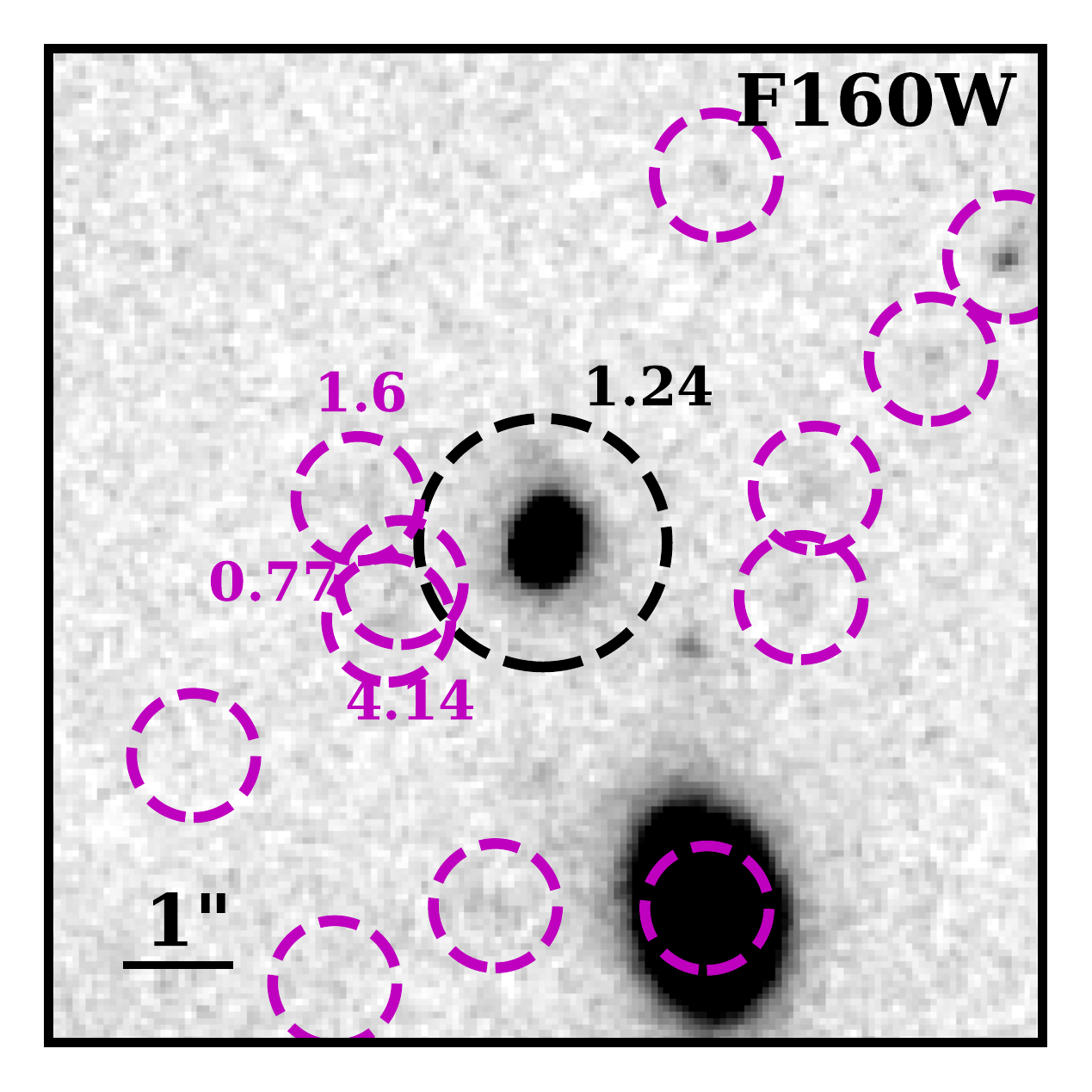}
    \includegraphics[width=0.32\columnwidth]{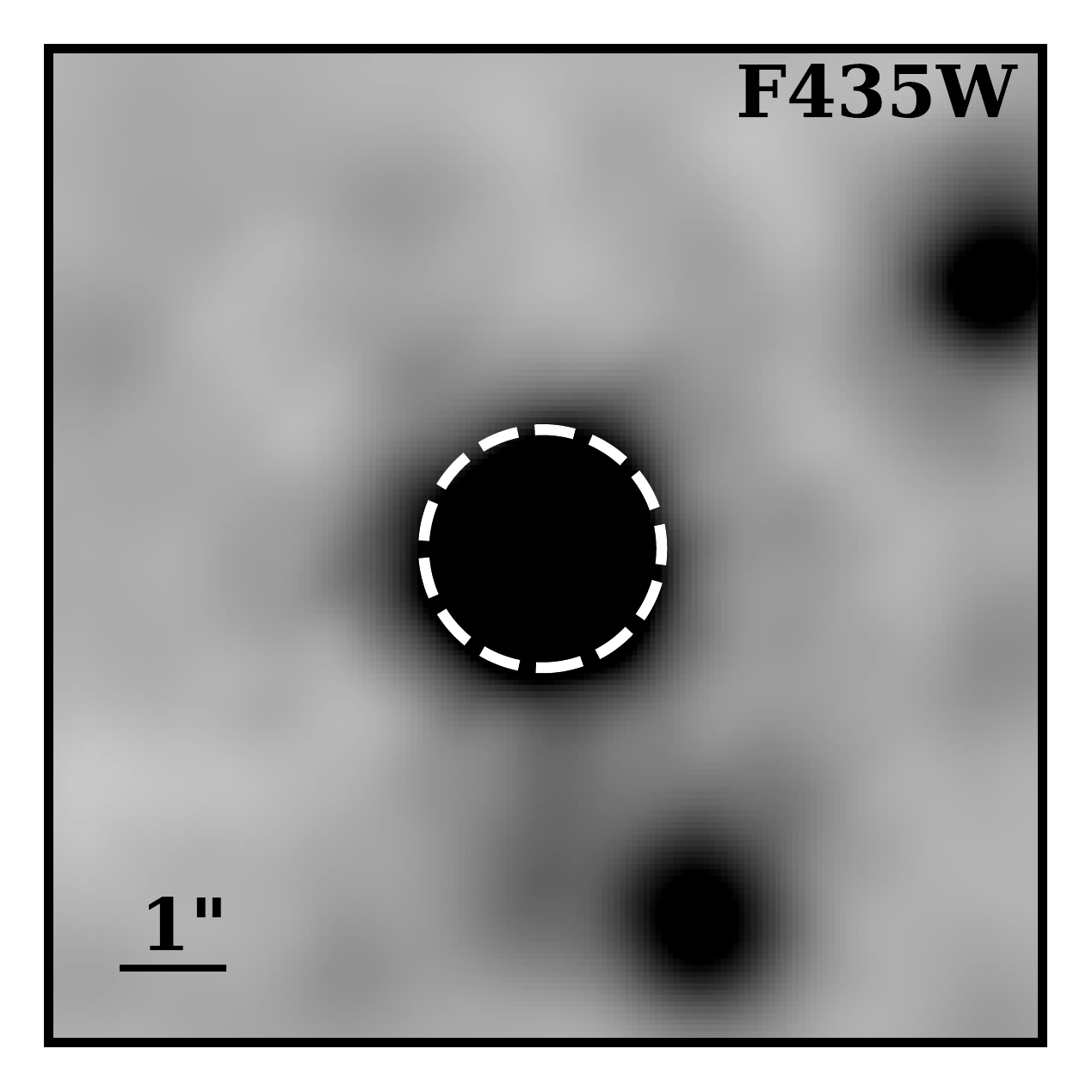}
    \includegraphics[width=0.80\columnwidth]{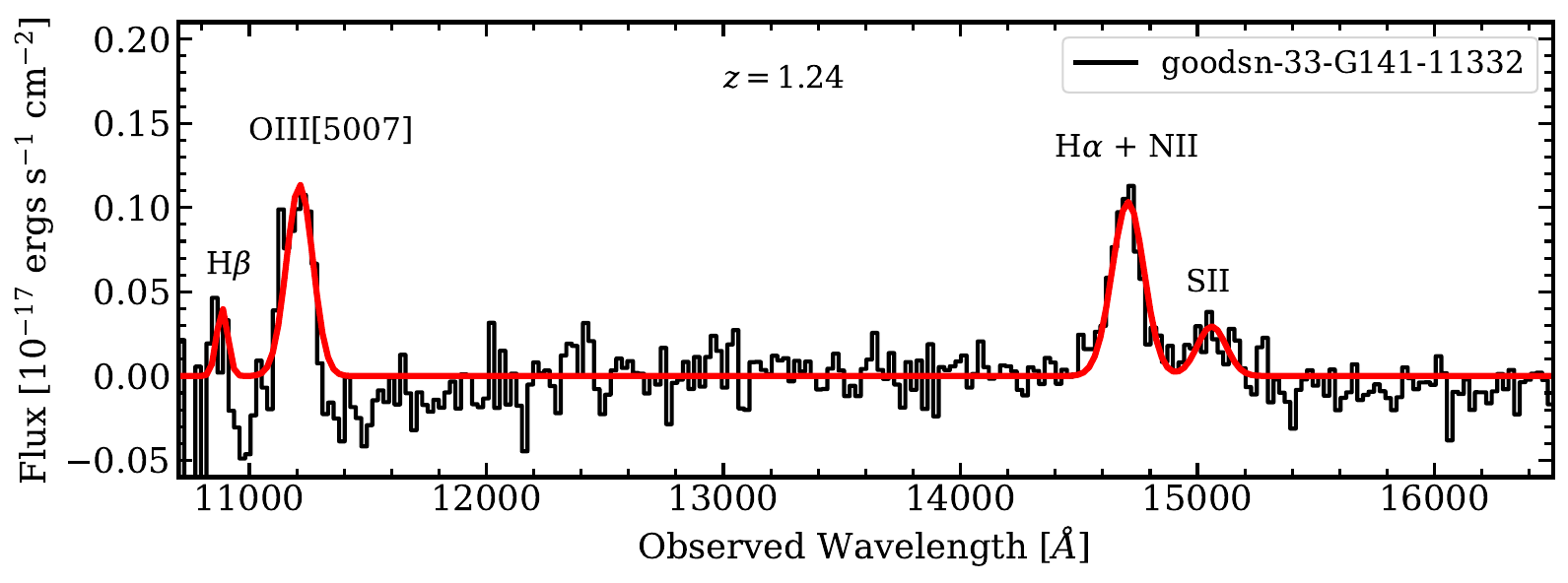}

    \includegraphics[width=0.31\columnwidth]{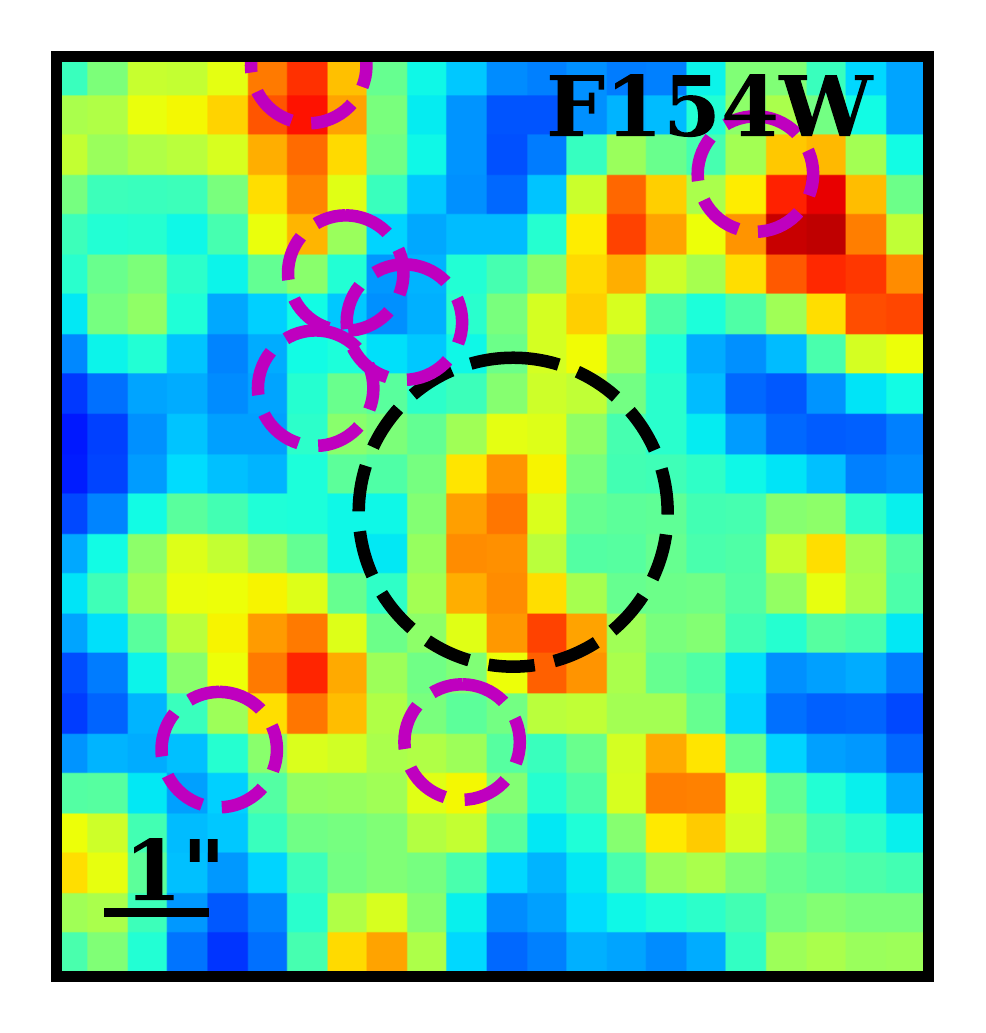}
    \includegraphics[width=0.32\columnwidth]{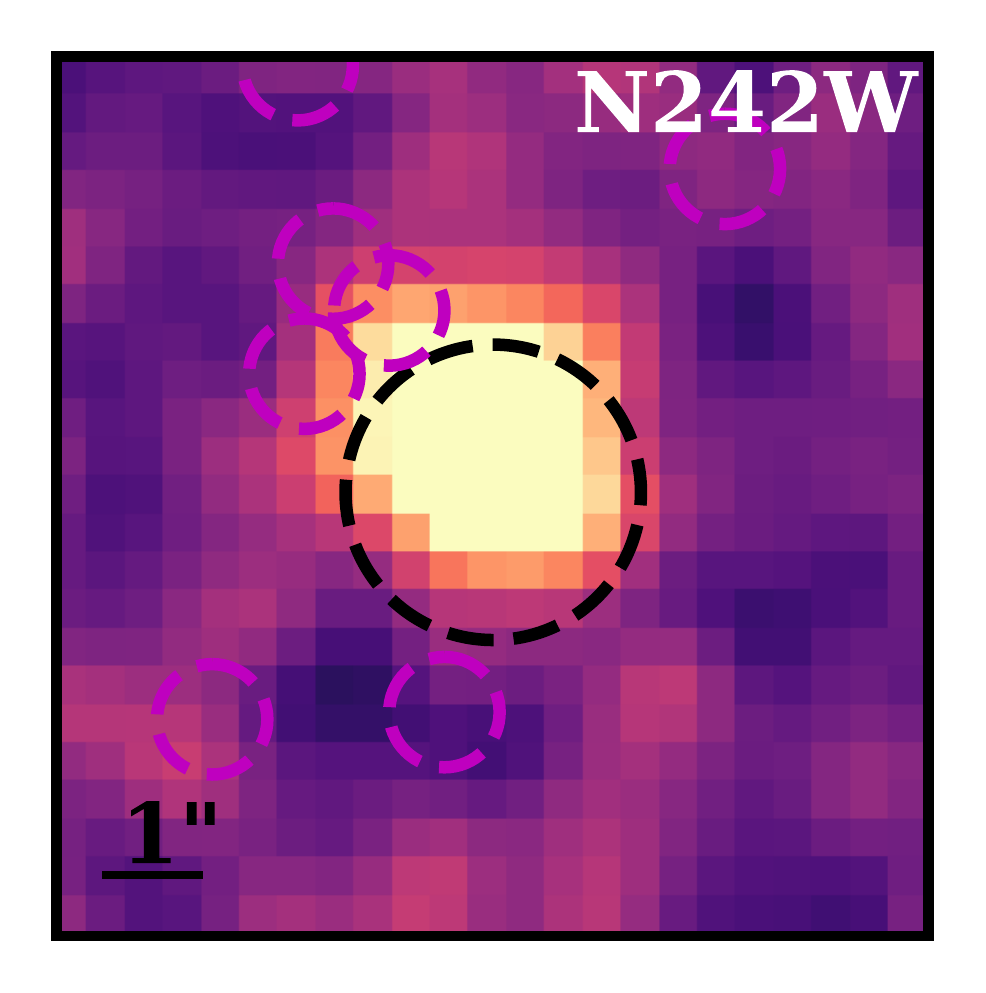}
    \includegraphics[width=0.32\columnwidth]{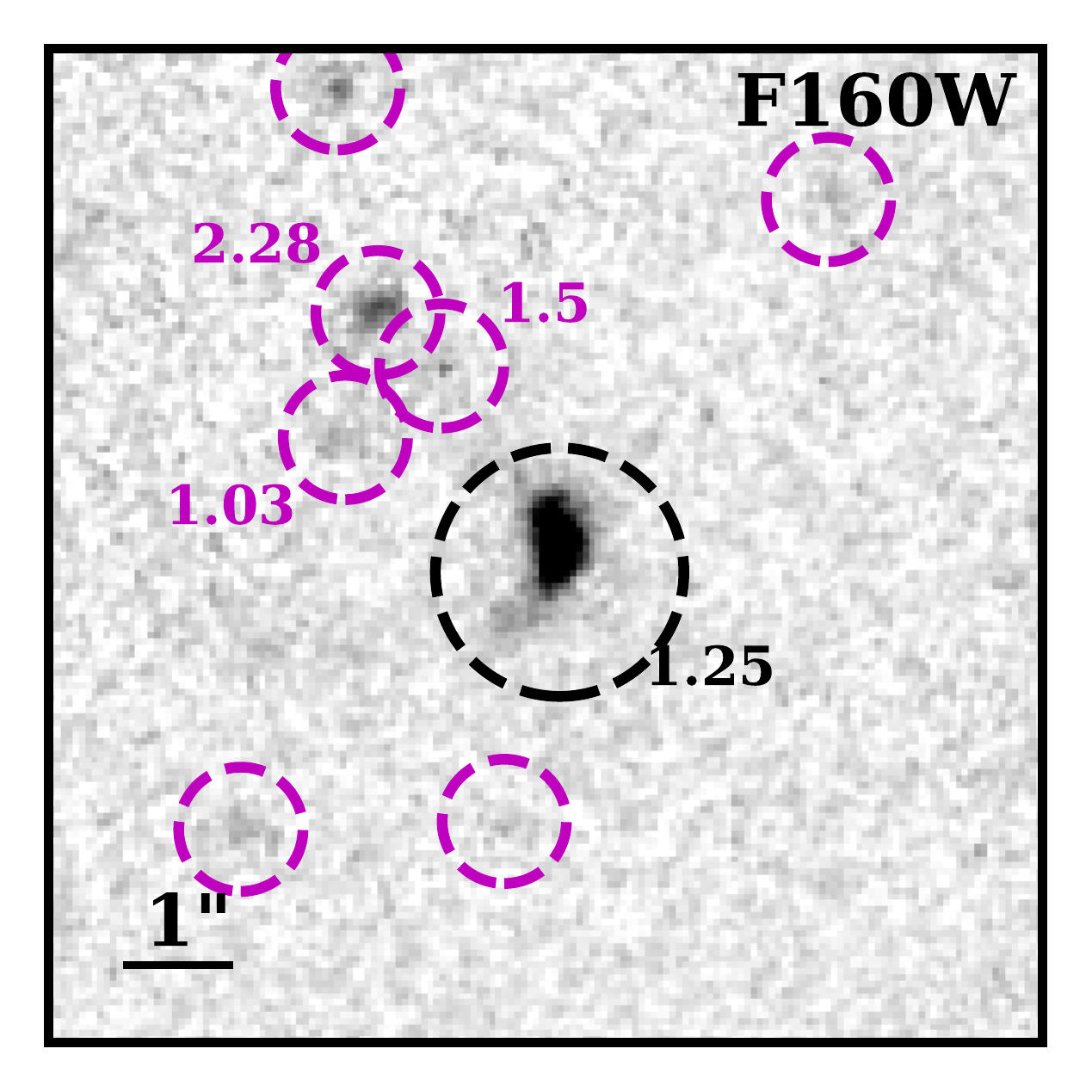}
    \includegraphics[width=0.32\columnwidth]{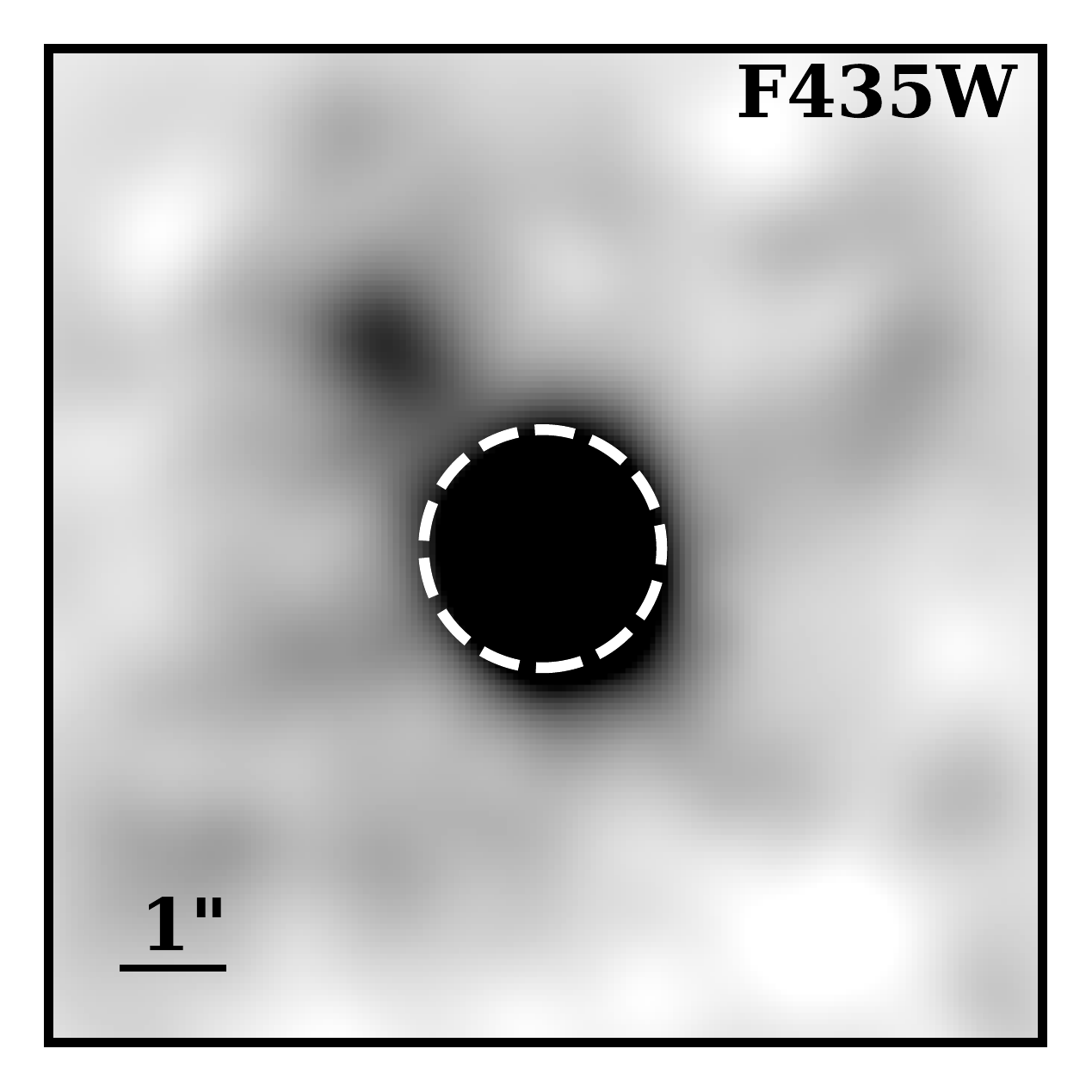}
    \includegraphics[width=0.80\columnwidth]{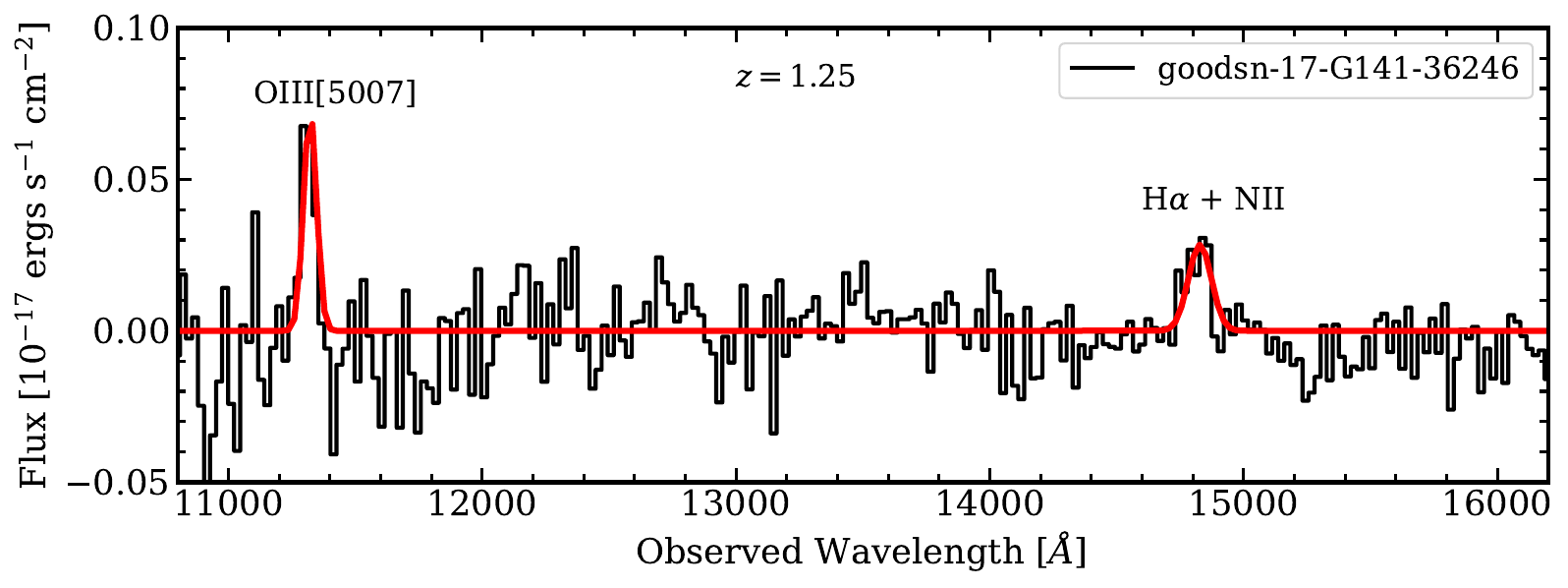}

    \includegraphics[width=0.31\columnwidth]{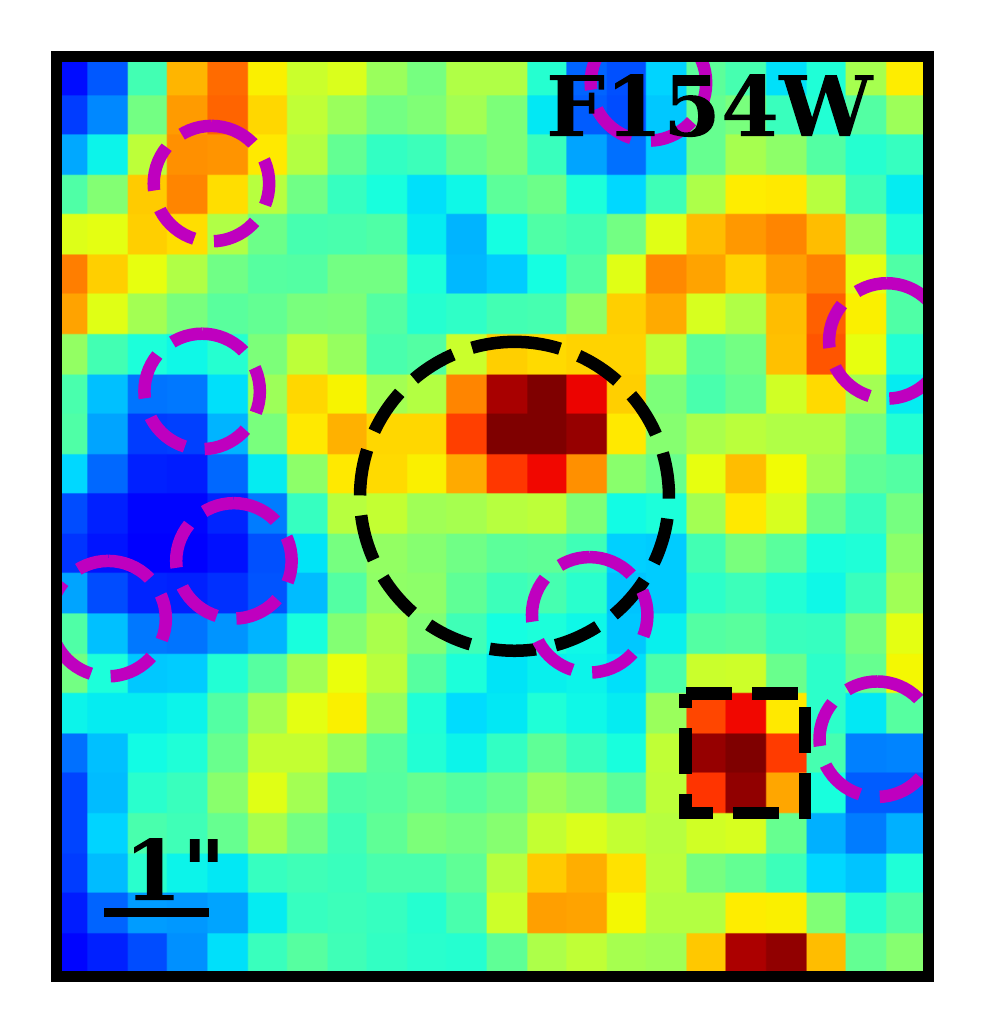}
    \includegraphics[width=0.312\columnwidth]{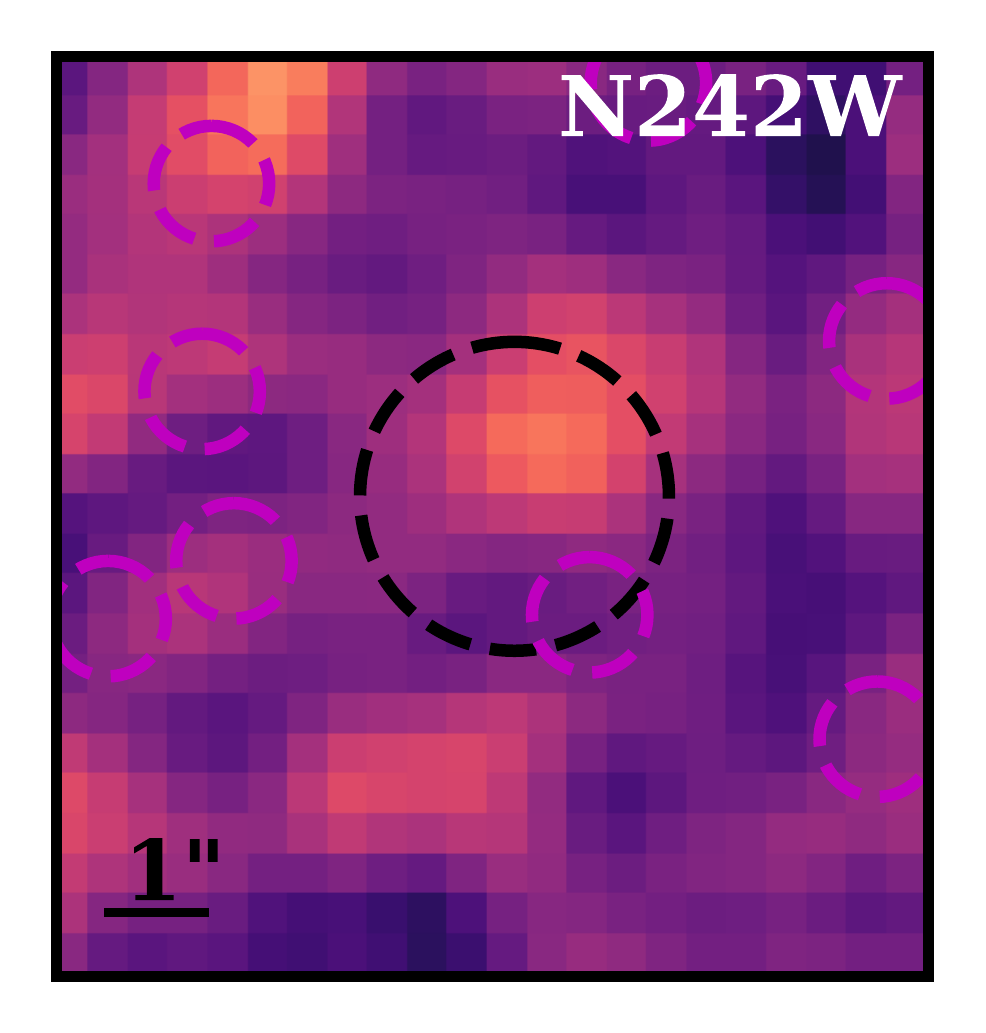}
    \includegraphics[width=0.32\columnwidth]{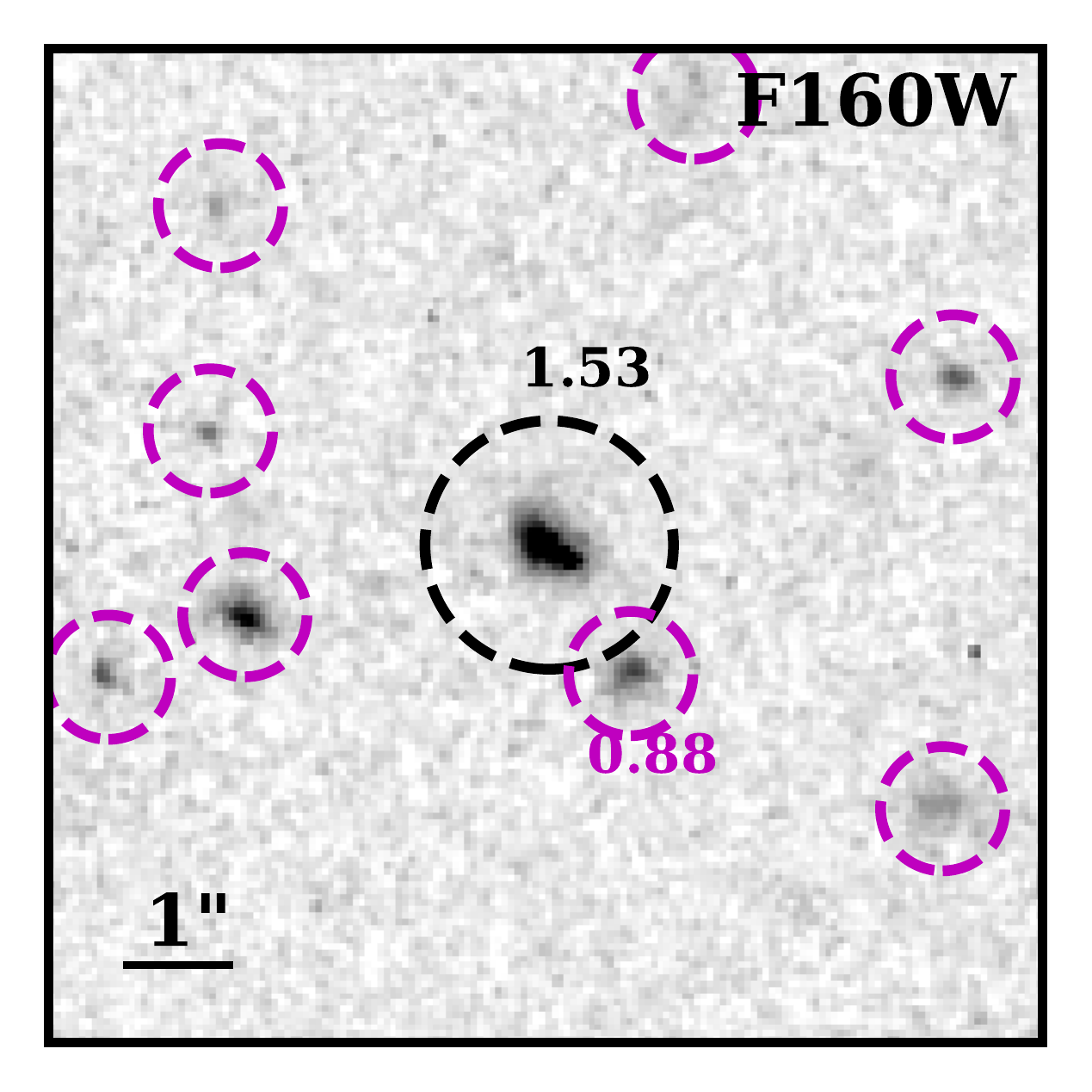}
    \includegraphics[width=0.32\columnwidth]{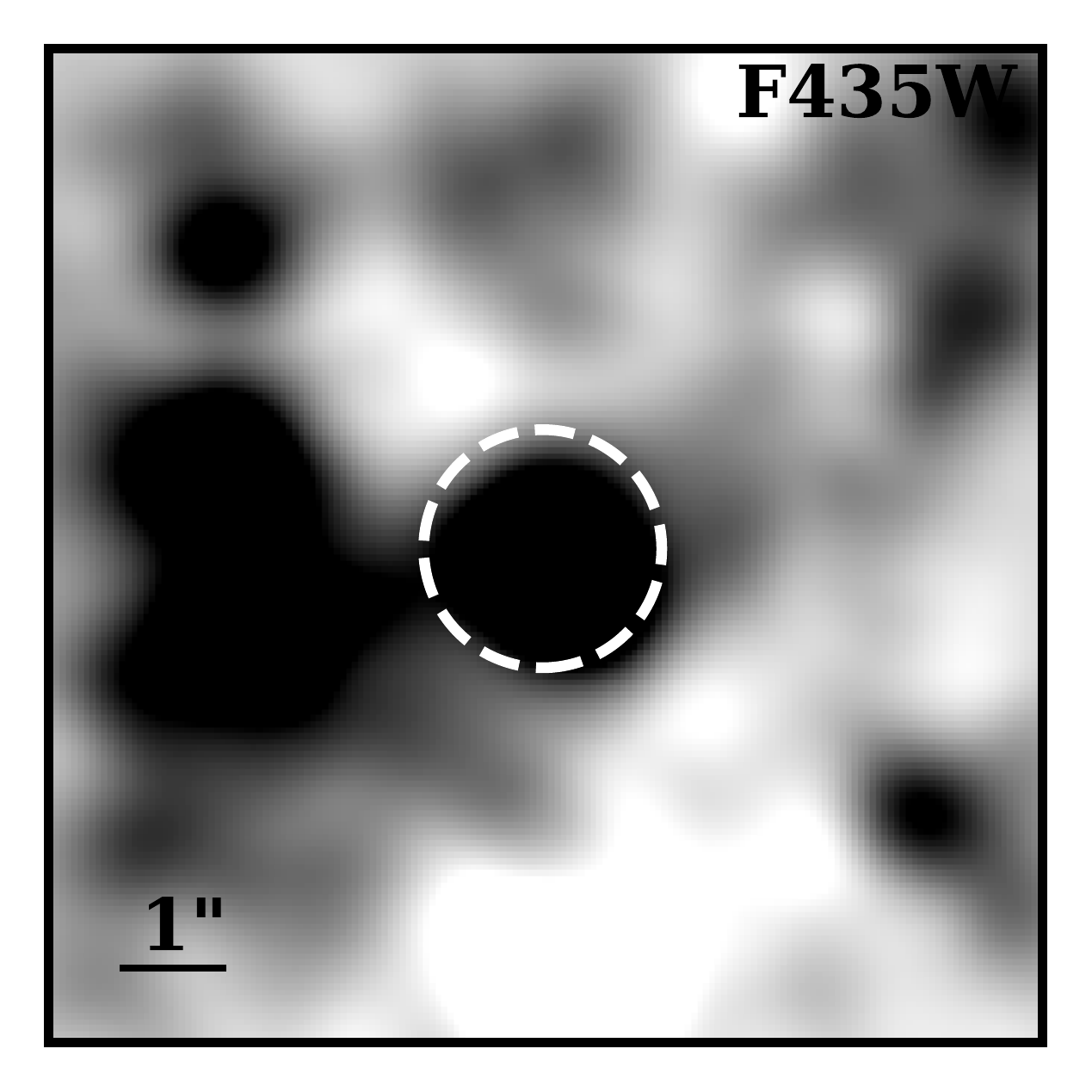}
    \includegraphics[width=0.80\columnwidth]{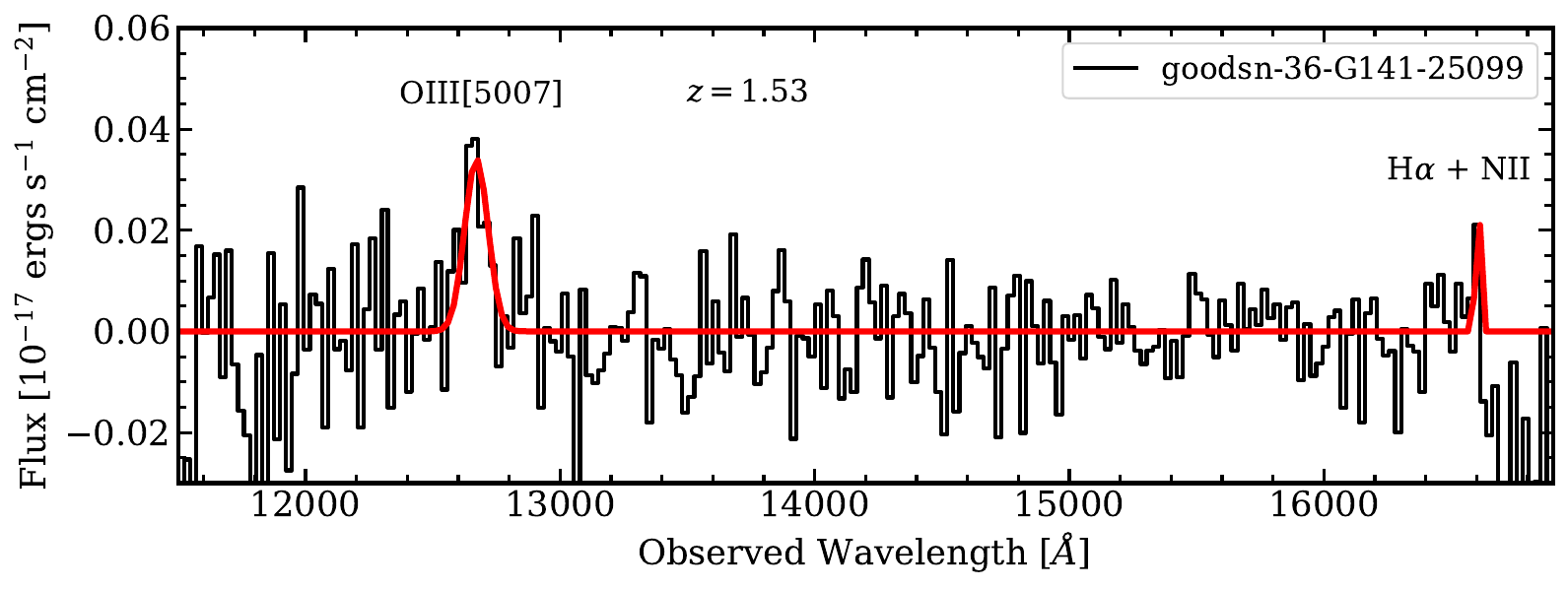}

    \includegraphics[width=0.31\columnwidth]{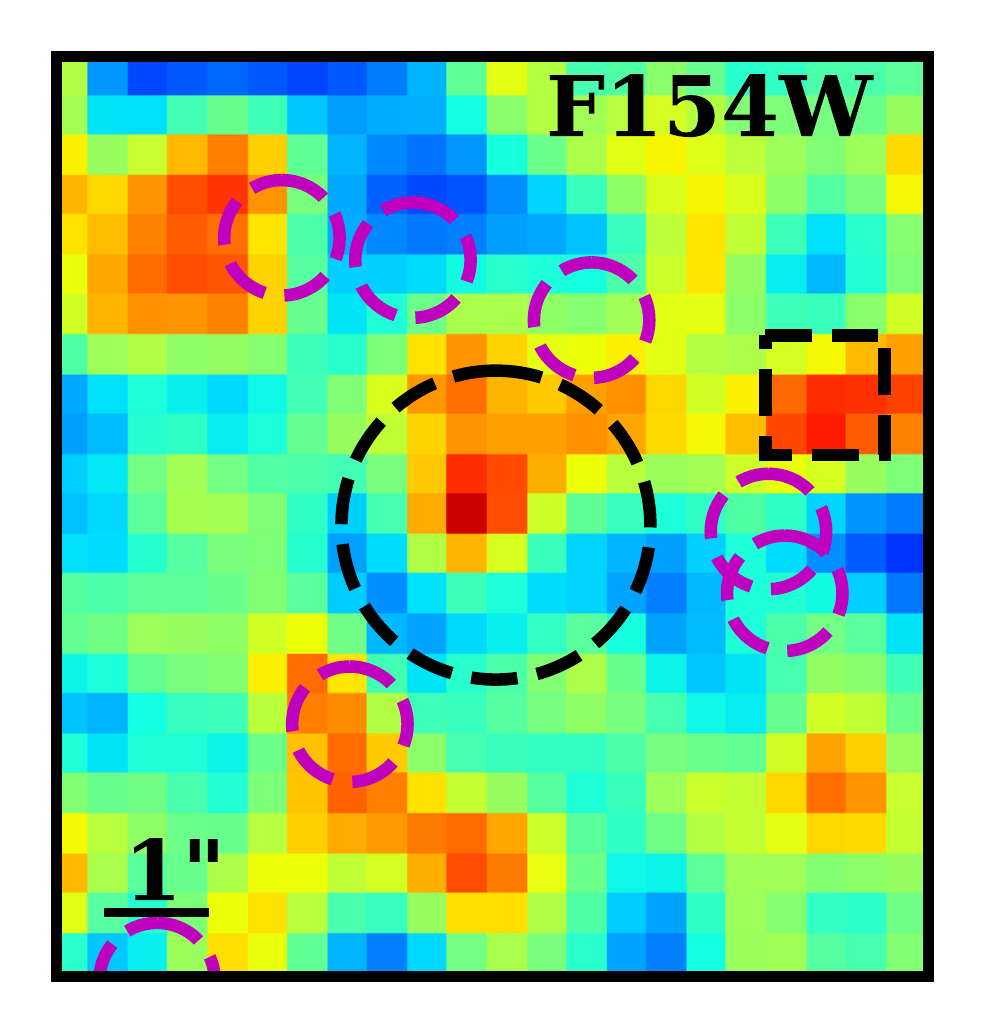}
    \includegraphics[width=0.312\columnwidth]{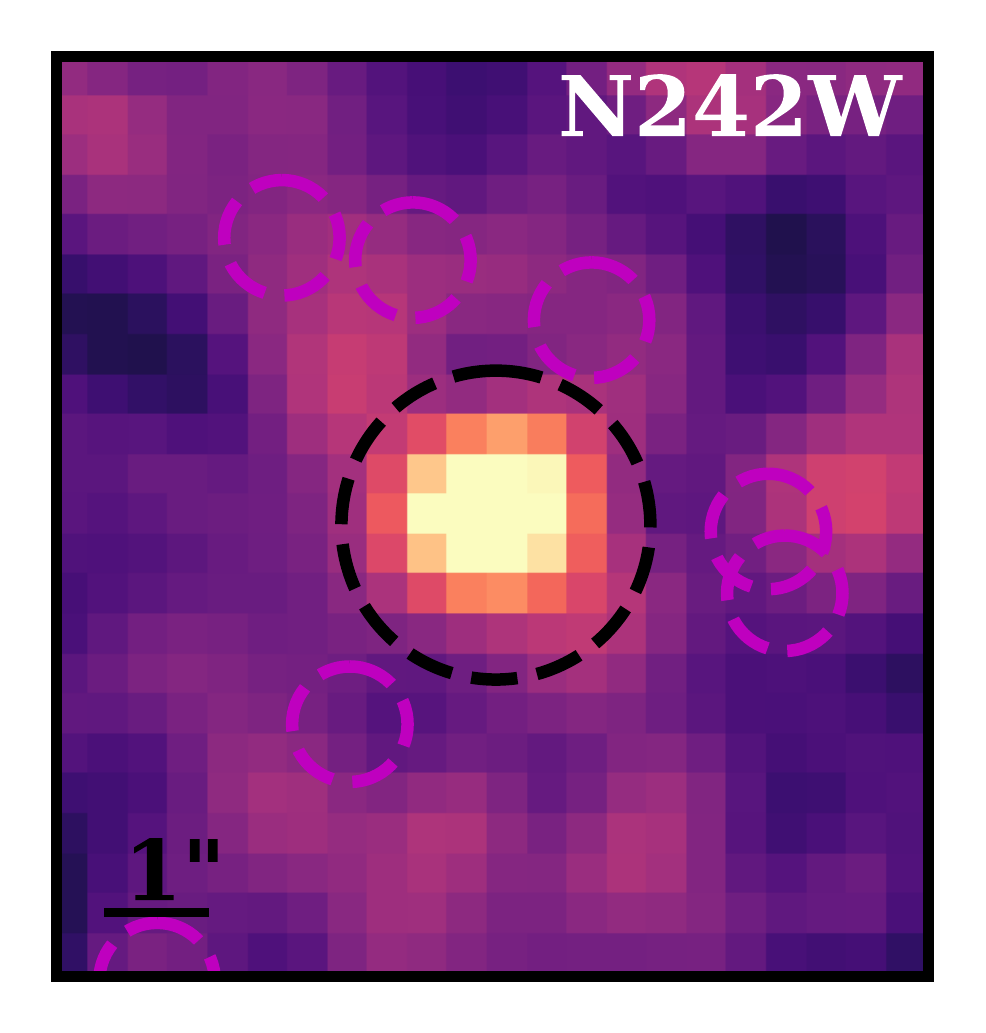}
    \includegraphics[width=0.32\columnwidth]{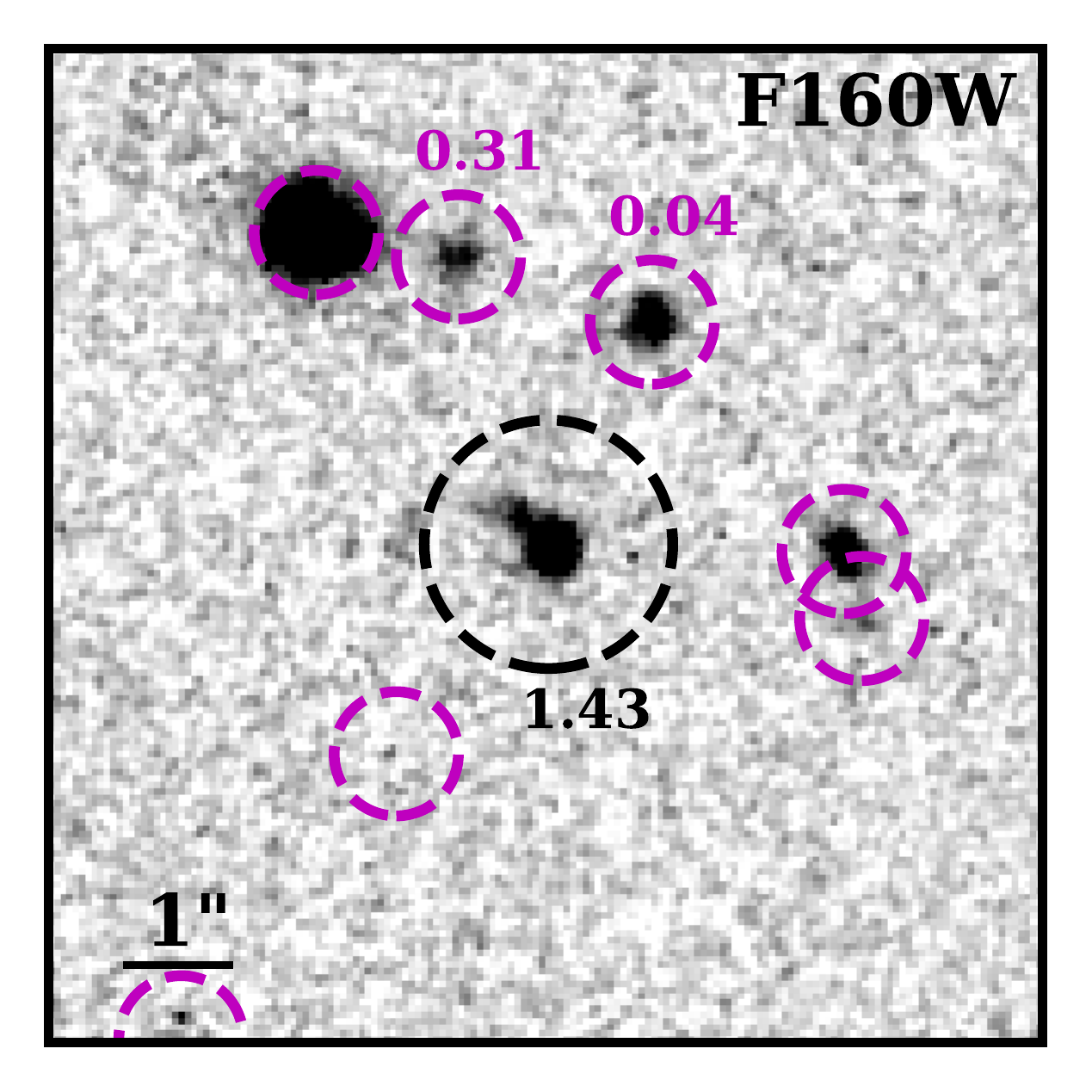}
    \includegraphics[width=0.32\columnwidth]{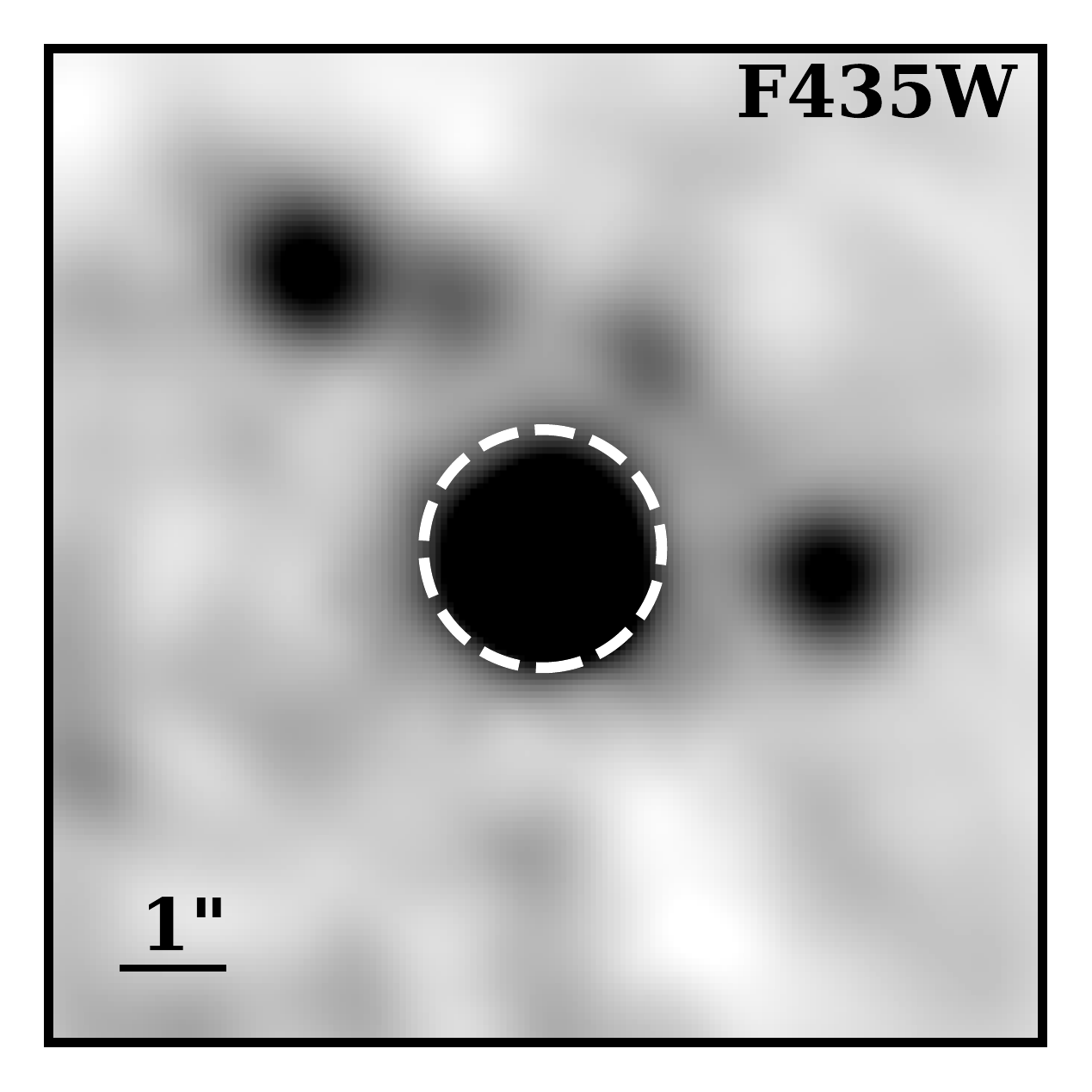}
    \includegraphics[width=0.80\columnwidth]{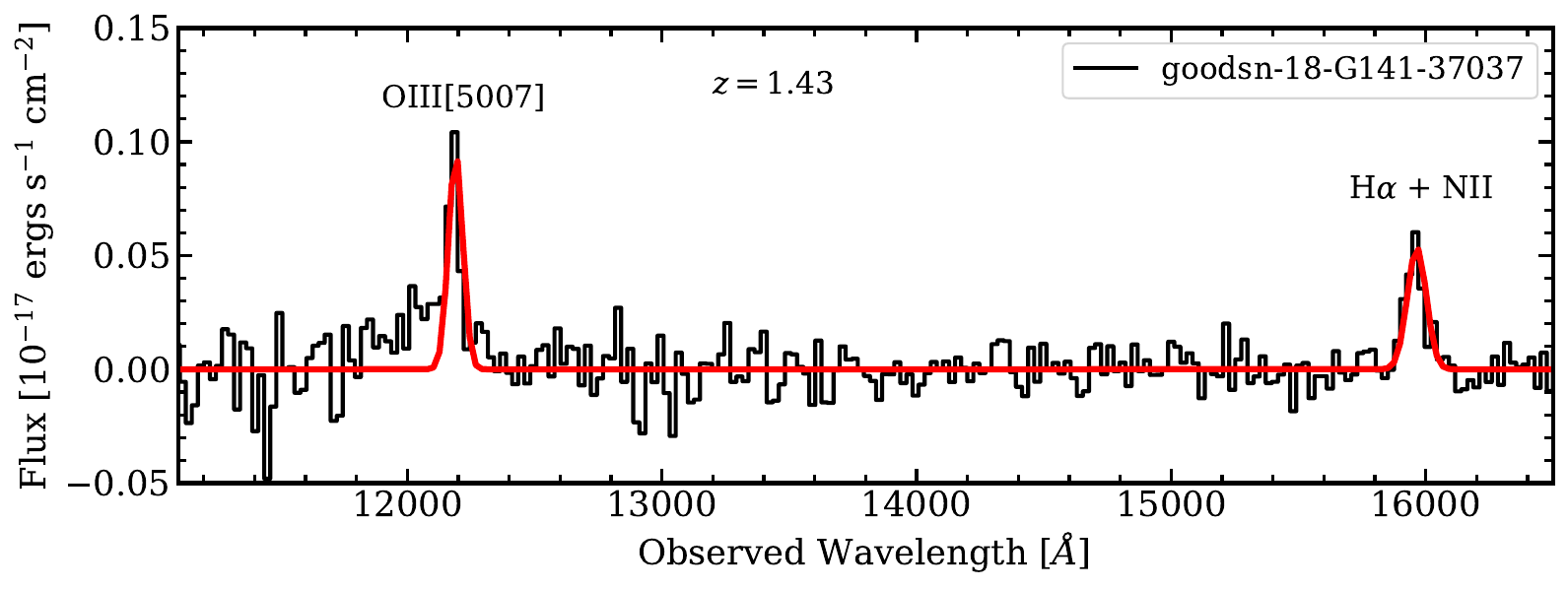}

    \includegraphics[width=0.31\columnwidth]{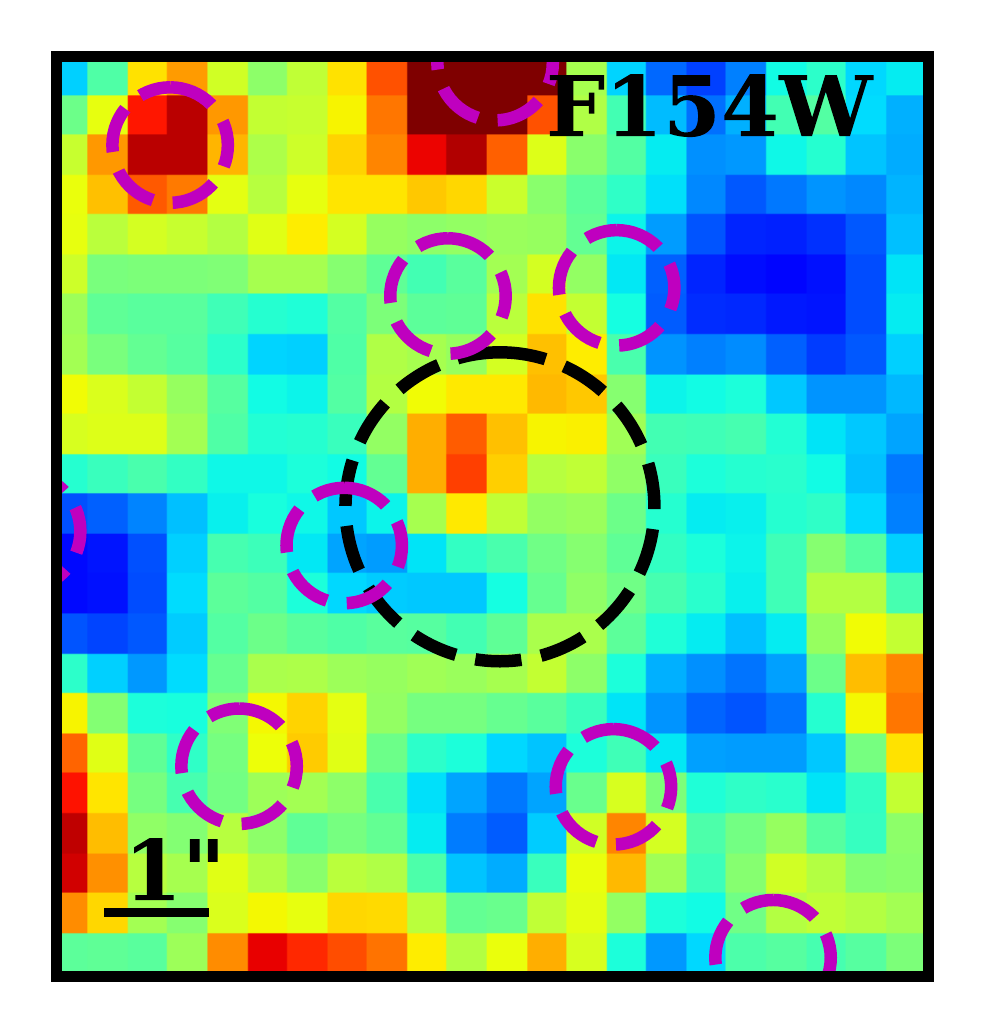}
    \includegraphics[width=0.312\columnwidth]{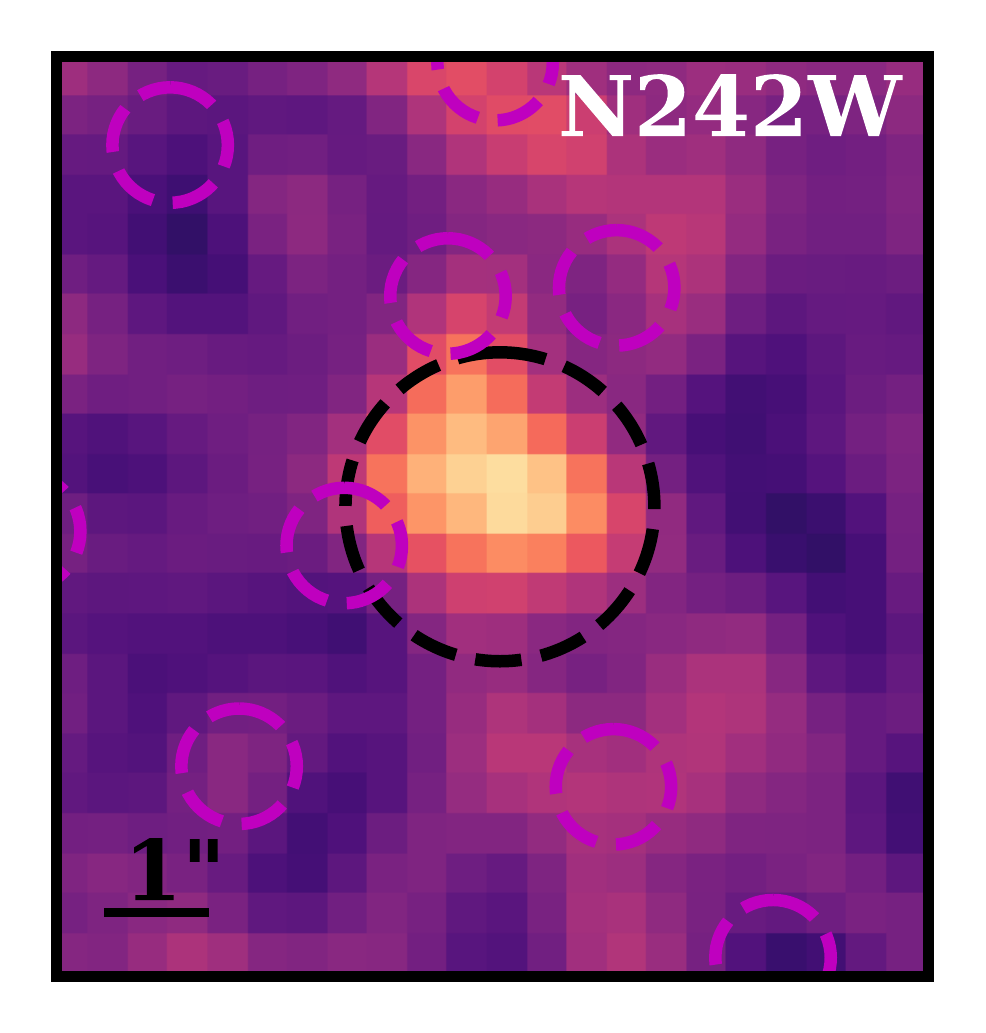}
    \includegraphics[width=0.32\columnwidth]{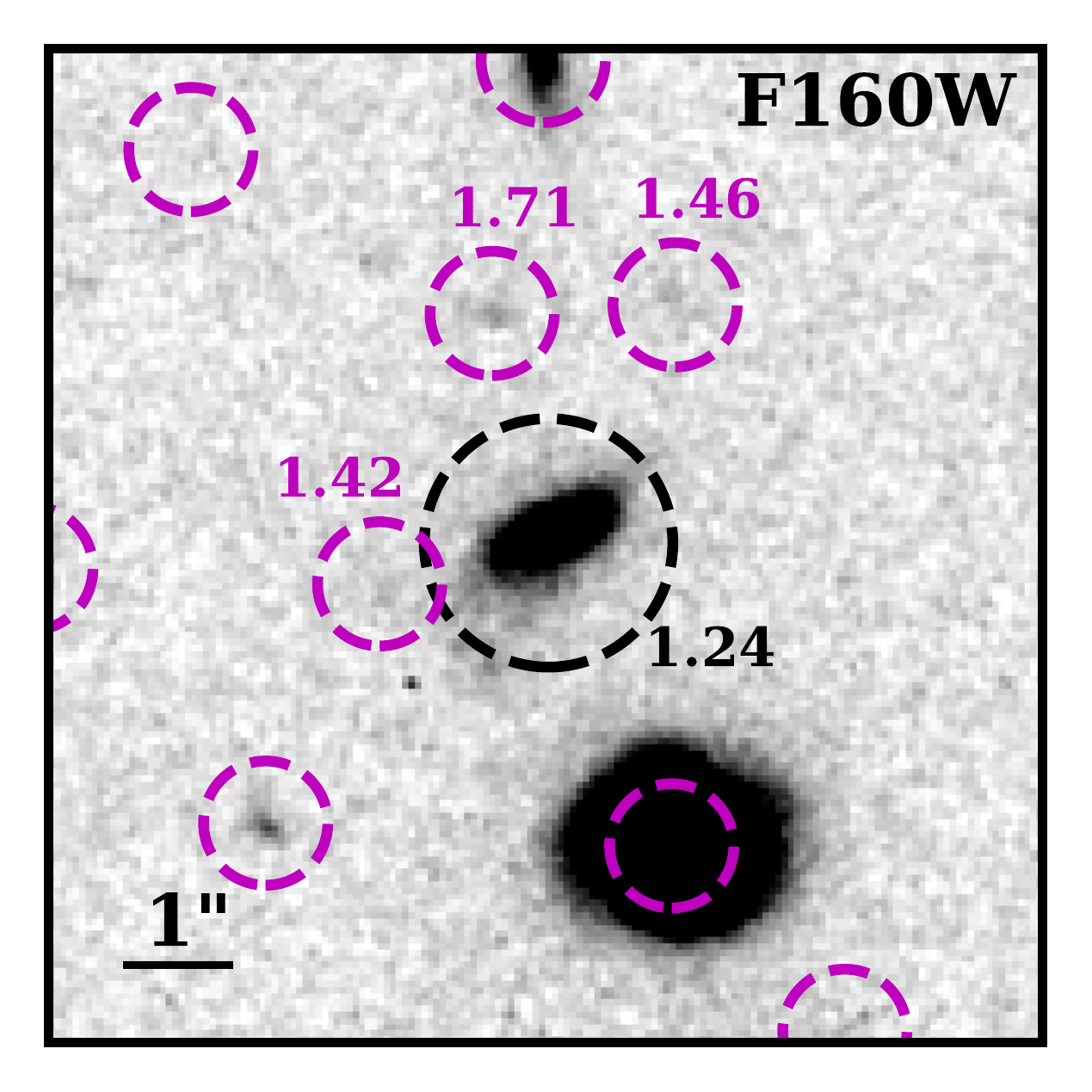}
    \includegraphics[width=0.32\columnwidth]{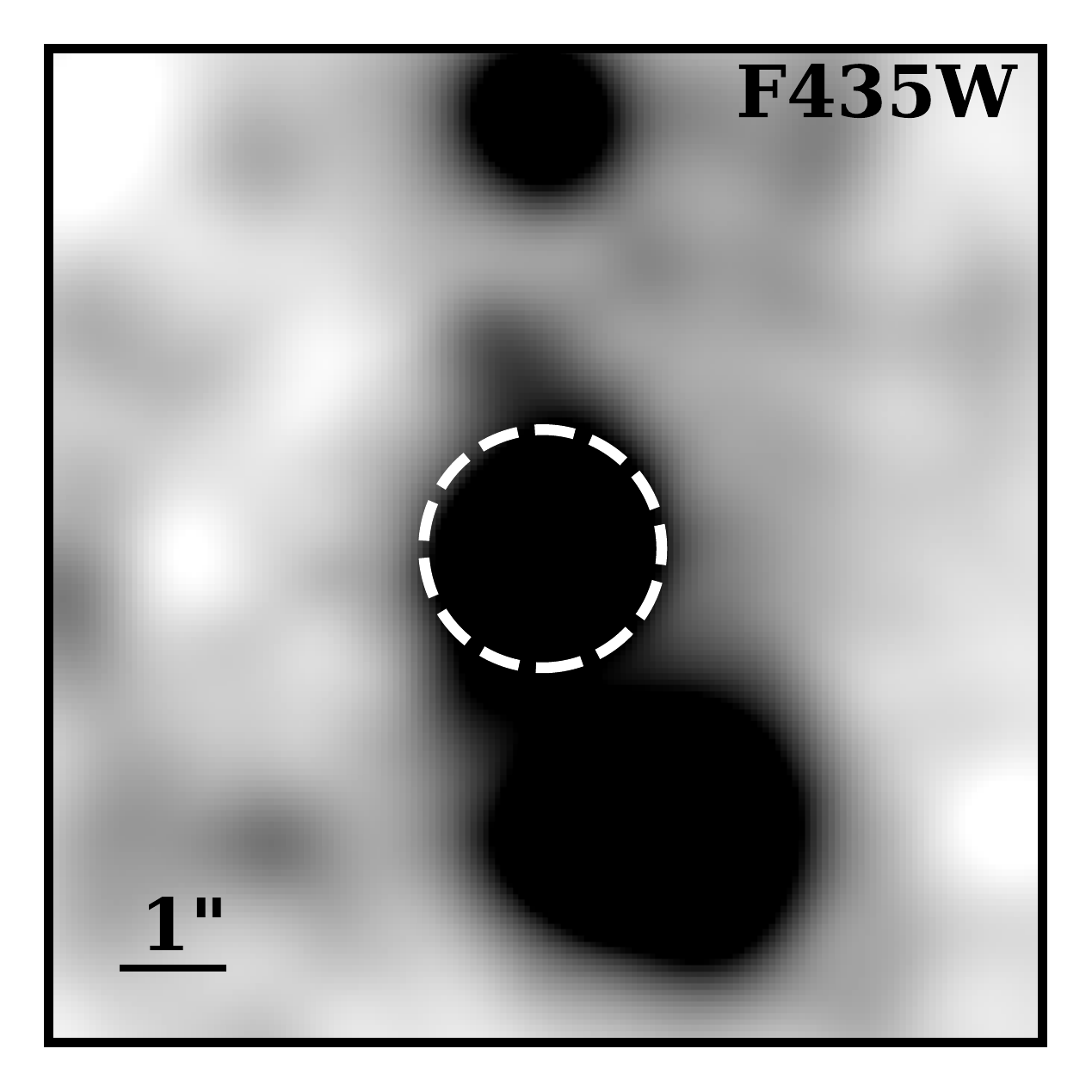}
    \includegraphics[width=0.80\columnwidth]{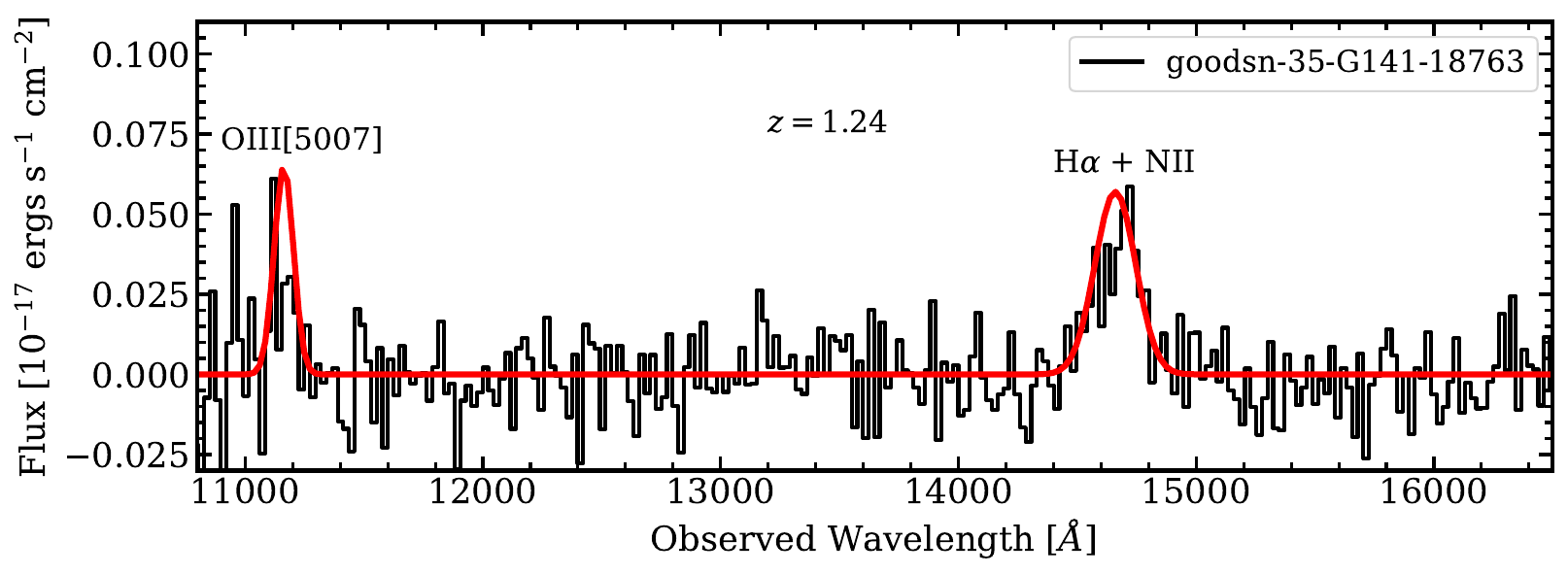}

    \includegraphics[width=0.31\columnwidth]{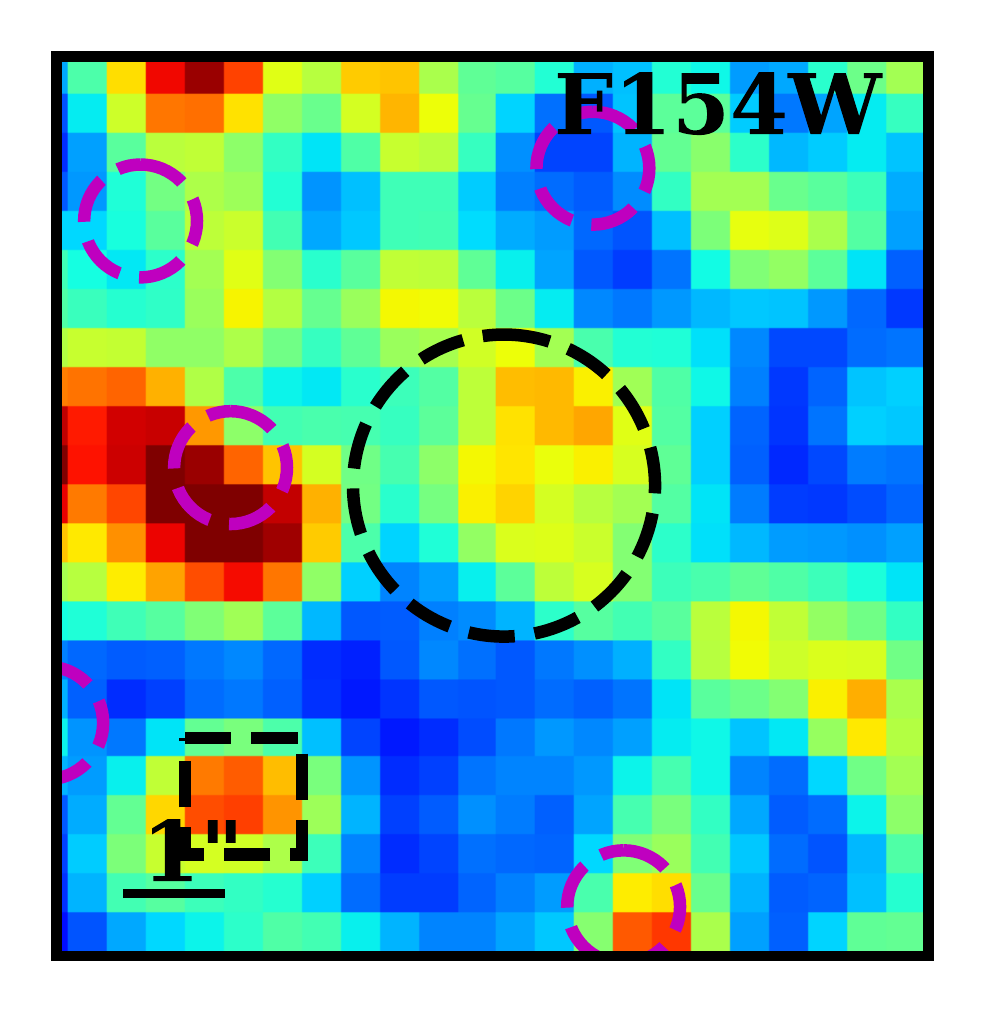}
    \includegraphics[width=0.312\columnwidth]{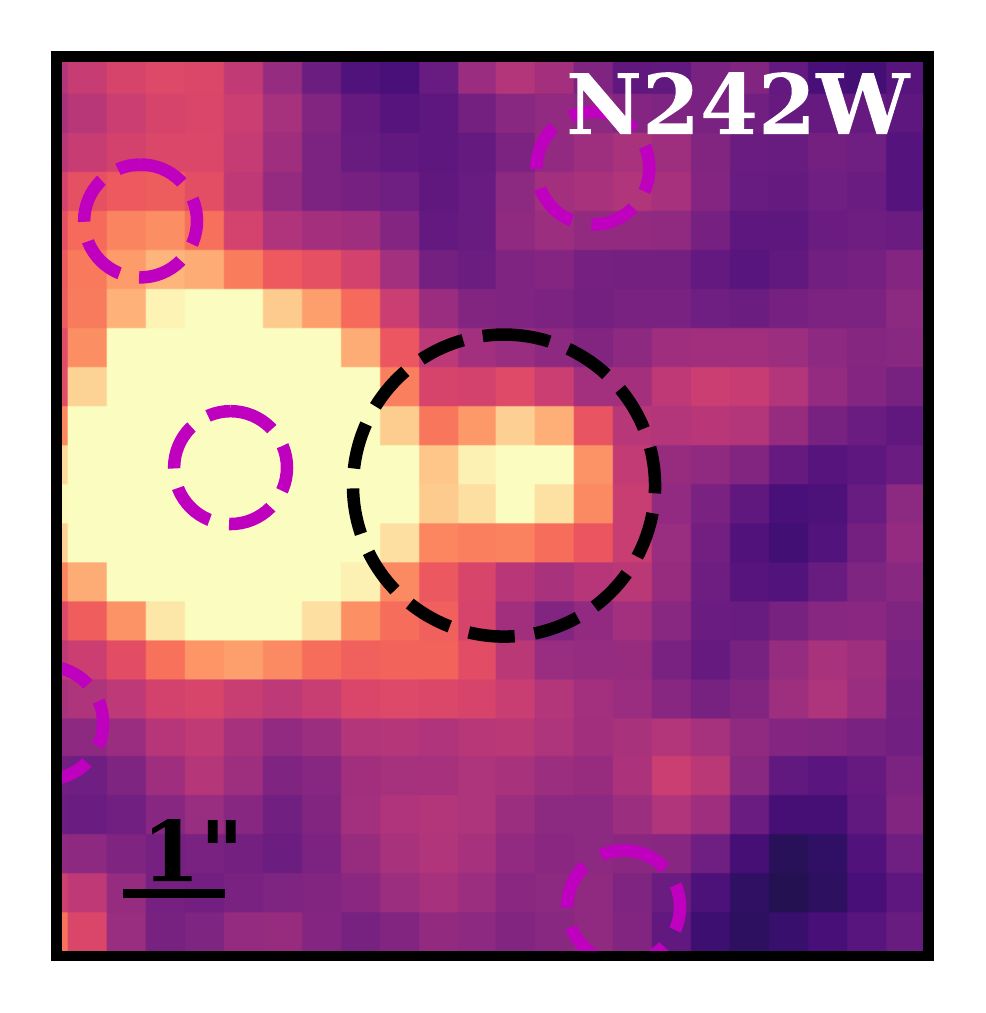}
    \includegraphics[width=0.32\columnwidth]{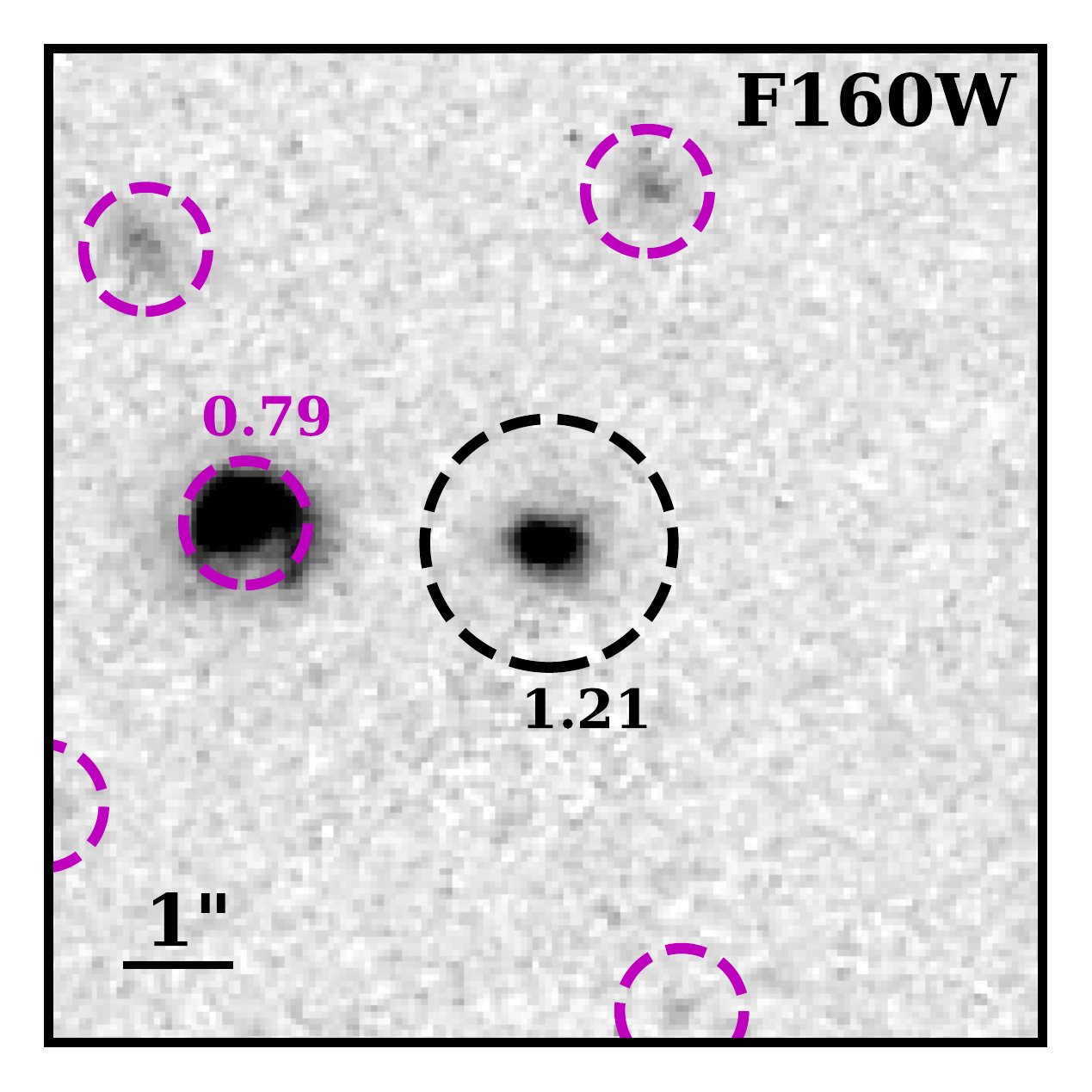}
    \includegraphics[width=0.32\columnwidth]{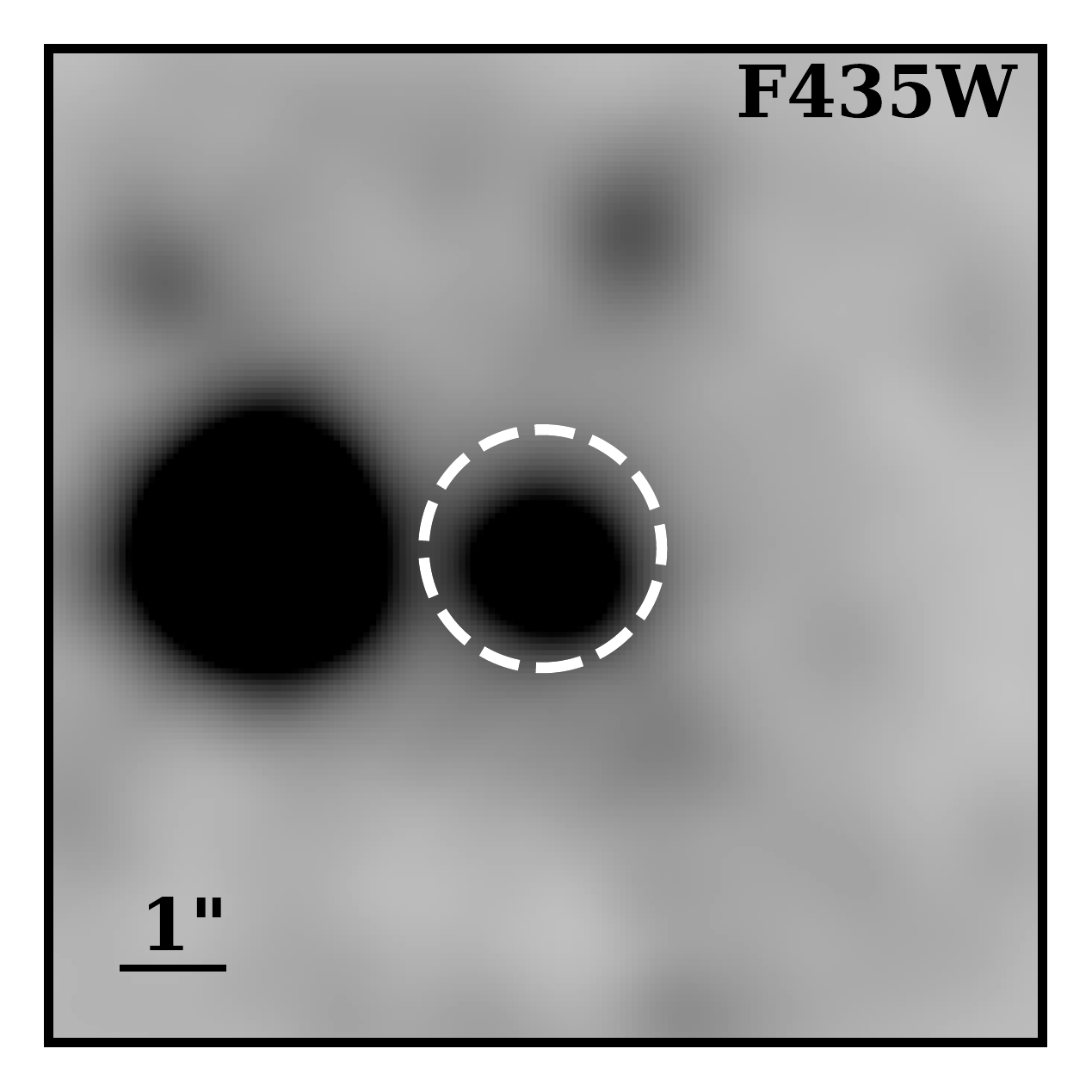}
    \includegraphics[width=0.80\columnwidth]{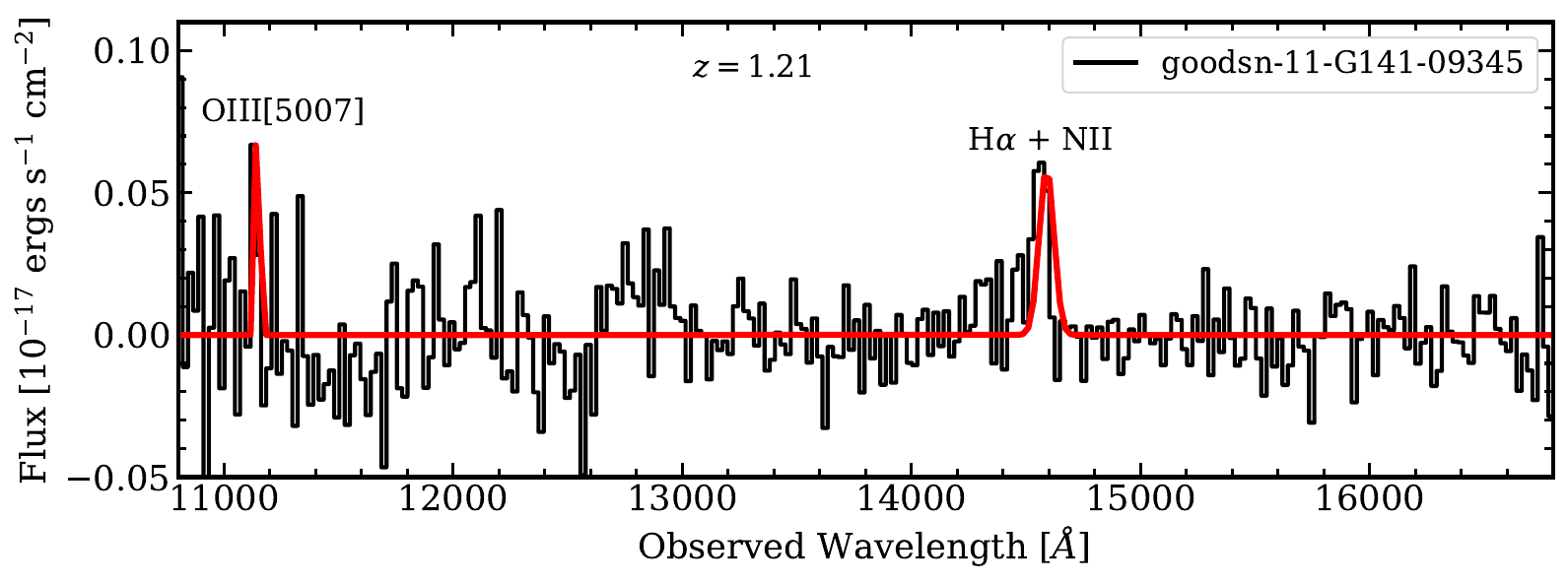}

     \includegraphics[width=0.315\columnwidth]{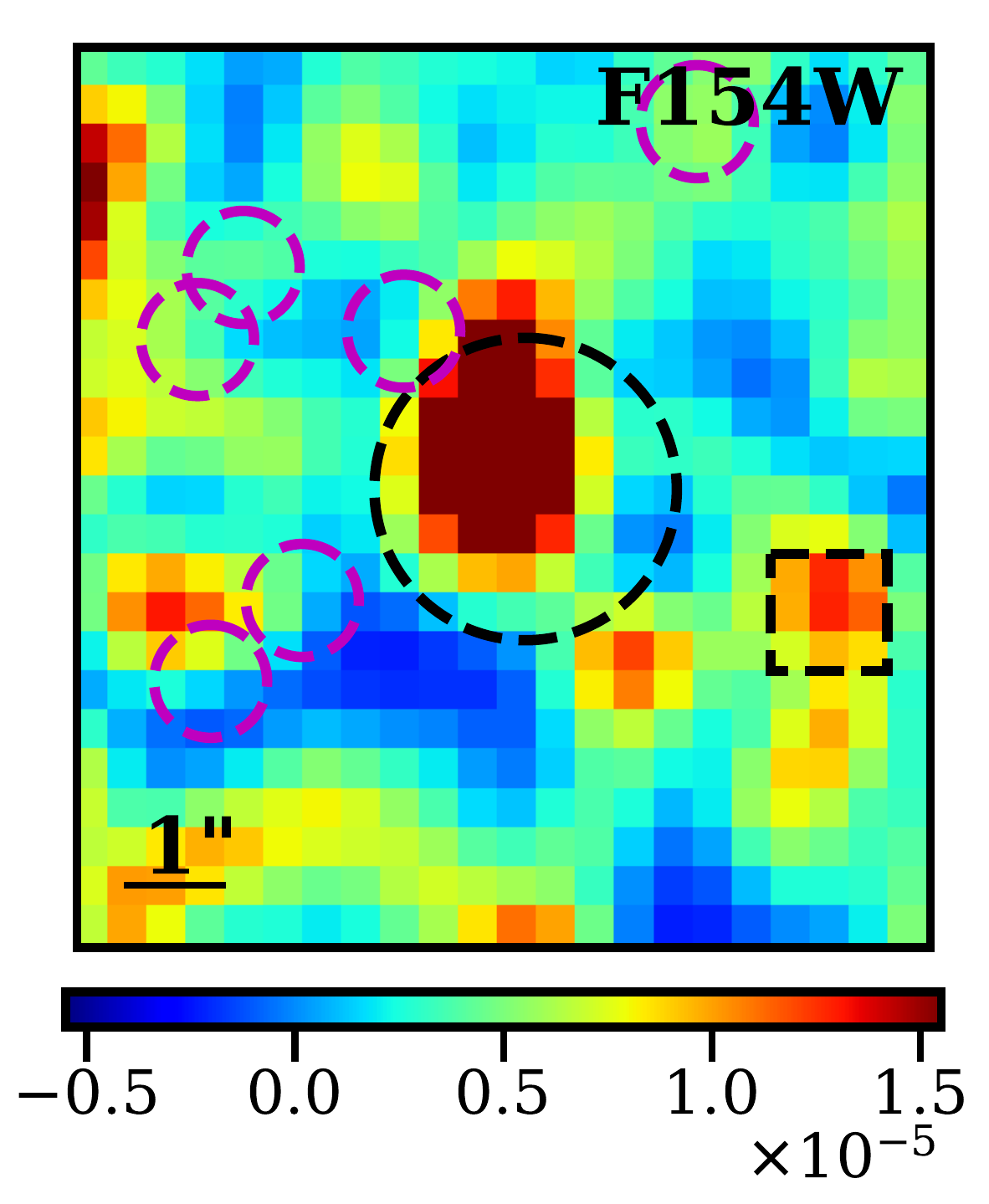}
     \includegraphics[width=0.315\columnwidth]{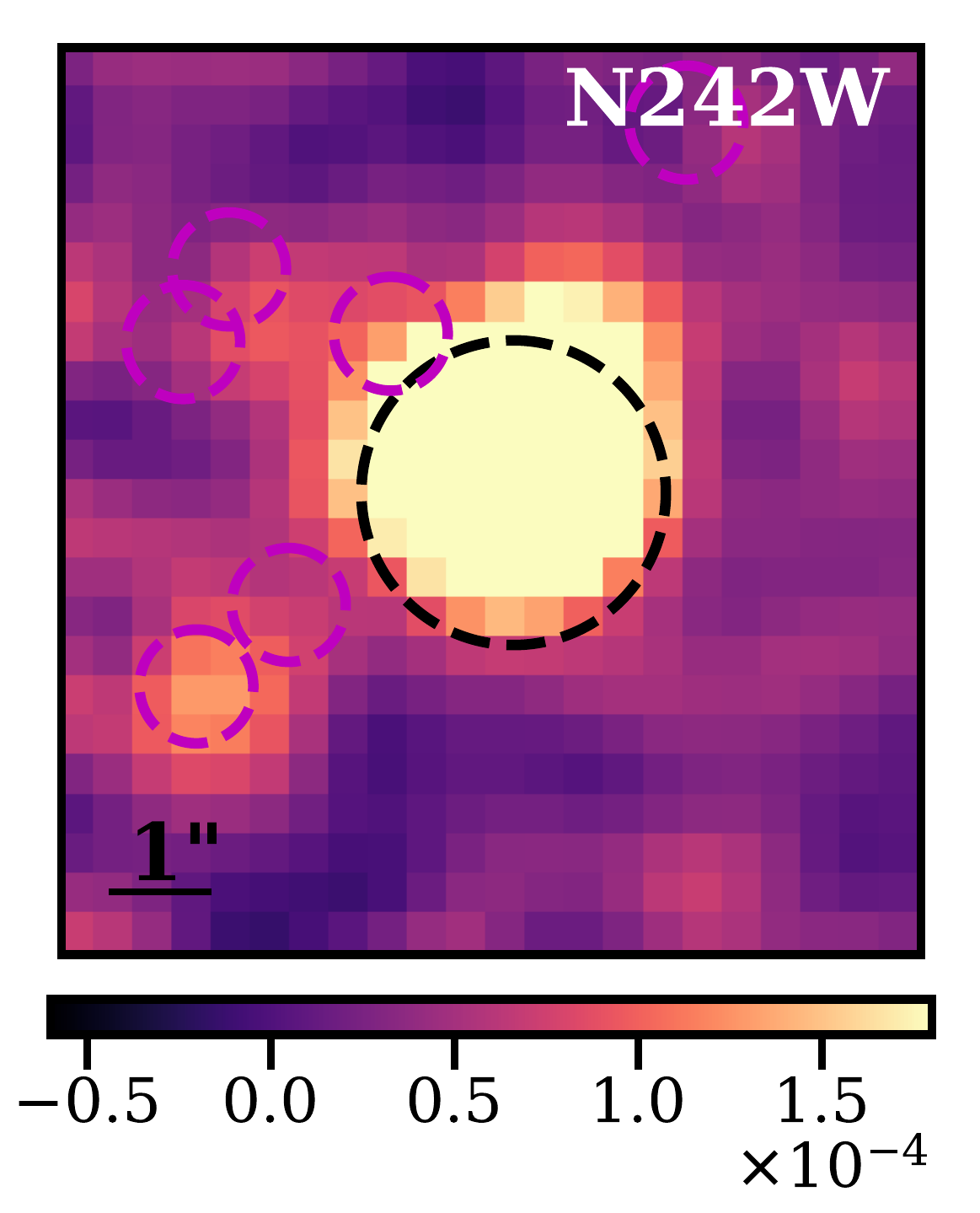}
     \includegraphics[width=0.32\columnwidth]{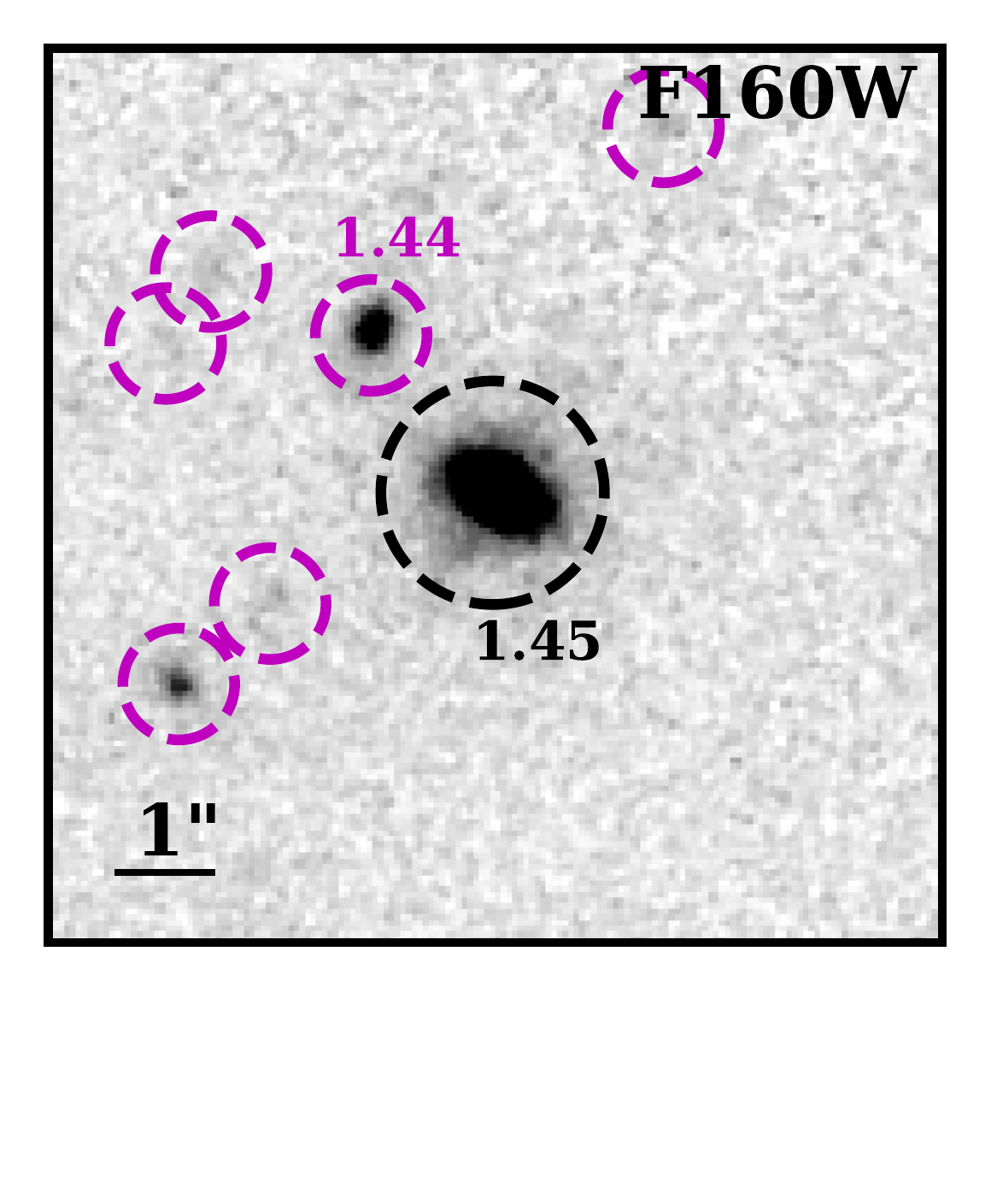}
     \includegraphics[width=0.32\columnwidth]{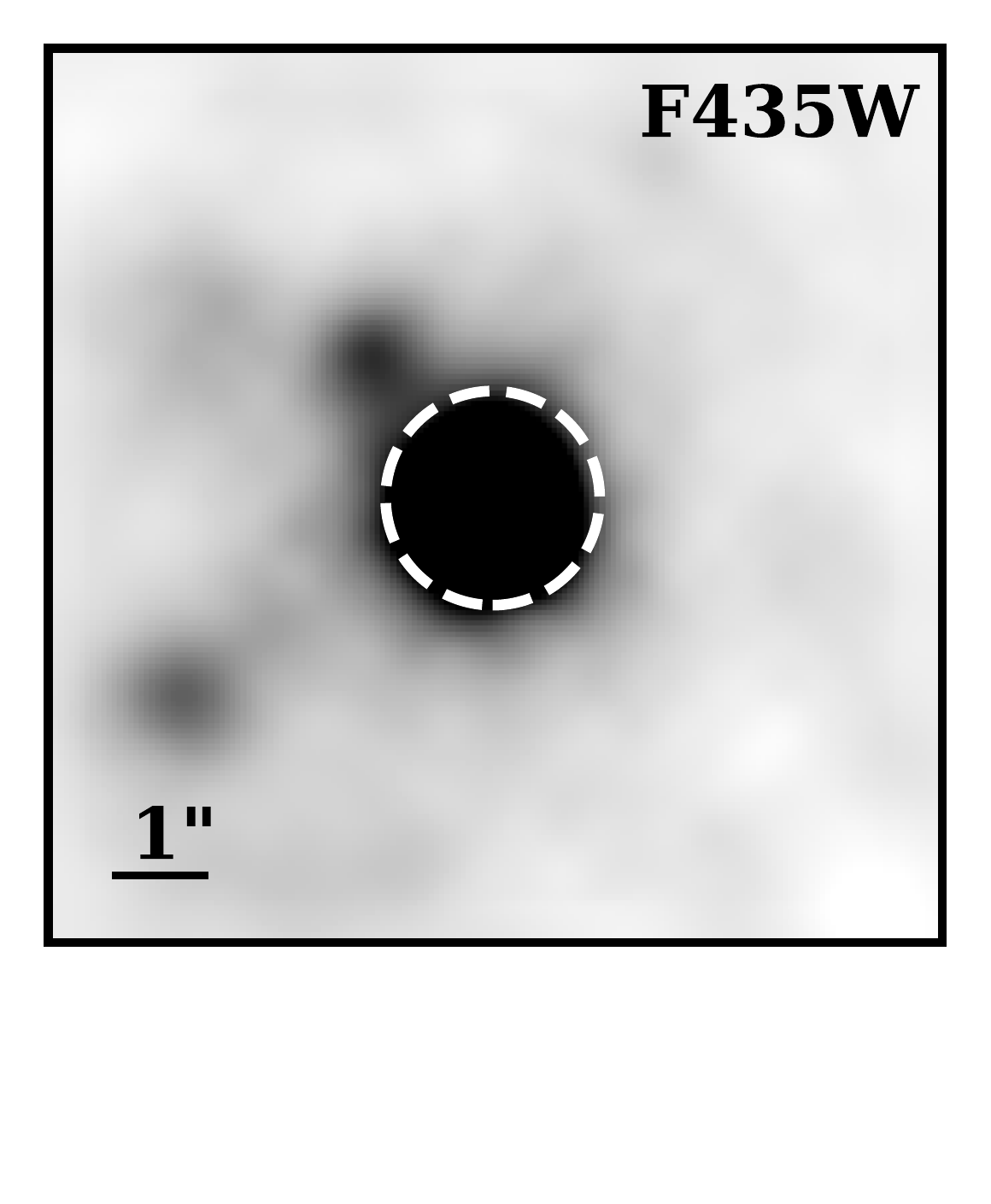}
     \includegraphics[width=0.80\columnwidth]{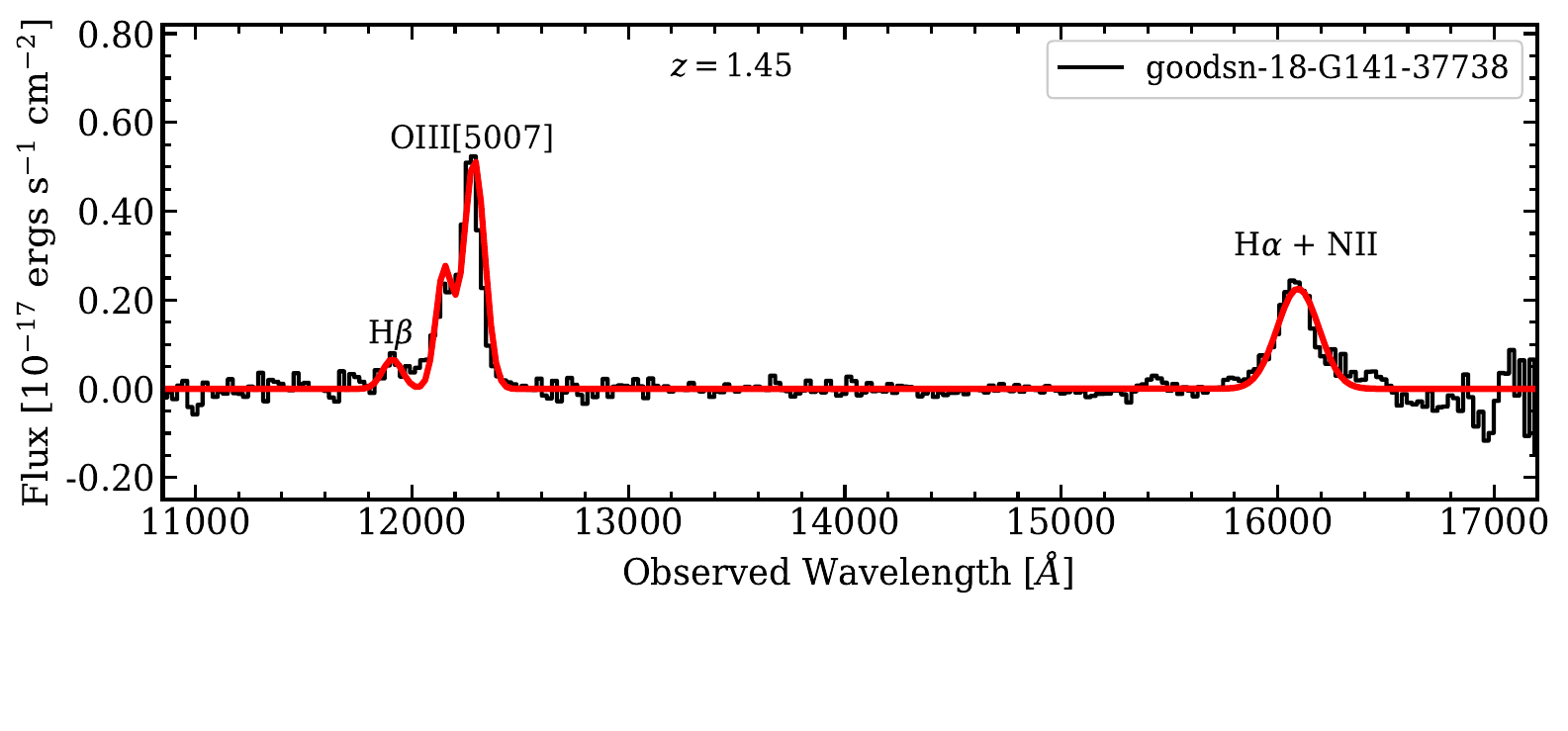}

    \caption{The first two panels from the left show the UVIT F154W and N242W band images of the LyC leakers. In the images, North is up and East is towards the left. The black dashed circles on the UVIT images represent apertures of size 1$^{\prime\prime}$.6 with the same centres as in the HST/F160W images on the right (third) panel. In the HST/F160W band images, these apertures are of size 1$^{\prime\prime}$ and are centred about the LyC emitters. The magenta circles indicate neighbouring objects. 
    The numbers outside the apertures indicate the redshift of the respective objects close to the LyC leakers. The fourth panel shows the HST/F435W images convolved with the F154W PSF for a size comparison. Spurious sources (see Section \ref{sec:det}) are highlighted by dashed black squares in the F154W band. We quote the magnitude (SNR) of such sources from top to bottom. On the first row in the north: 27.07 (1.88), third row in the south-west: 26.50 (2.77); fourth row in the west 26.54 (2.69); sixth row in the south-east: 27.05 (1.93) and seventh row in the south-west: 26.76 (2.38).
    On the extreme right are the 3D-HST spectra of the LyC leakers. The black solid curves are the original spectra after contamination and continuum subtraction. The red curves indicate the fitted models. }
    \label{fig:spec}
\end{figure*}

\subsection{Sample selection}
\label{sec:sample}
The spectral data from 3DHST G141 grism is used to identify sources at redshift 1.0 $<z<$ 2.0. We use the H$\alpha$ and [OIII]5007 (hereafter [OIII]) lines having line fluxes with SNR $>$ 3 to identify emission line sources. In the G141 grism with a wavelength coverage of 1.10 - 1.65 $\mu$m, the H$\alpha$ and [OIII] lines are simultaneously observed only between the redshift range $z \sim$1.1-1.6, restricting the redshift range of the sources to 1.1 $<z<$ 1.6 (the redshifts are confirmed with the H$\alpha$ and OIII[5007] lines in the grism spectra, whose analysis is discussed in detail in Section~\ref{sec:spec}. 
Of the 178 sources found from 3DHST grism data, 56 sources are identified in UVIT F154W band images. 
We perform photometry and the appropriate background estimation of these 56 potential LyC leaker candidates (described in Section~\ref{sec:phot}). To avoid contamination, we consider only the sources having no neighboring companions within 1.2$^{\prime\prime}$ radius in the \citet{Skeltonetal2014} catalog. In addition to checking the catalog, we also visually inspect the HST F435W and F160W bands for any contaminations within 1.2$^{\prime\prime}$ radius. With these criteria and requiring the FUV SNR $>3$, we are left with 11 sources. One source in these 11 is observed to have an H$\alpha$ line at the edge of the spectral coverage of G141. To avoid the uncertainty in redshift and line flux measurement, we exclude this source from our sample. We consider the remaining sample of 10 sources as our final sample of LyC leakers. The locations of the 10 detected LyC leakers in GOODS-North field are shown in Figure~\ref{fig:audf-lyc}. We further segregate the sample of 10 LyC leakers into the ones having LyC emission within UVIT PSF FWHM $\sim1^{\prime\prime}$.6 (7 sources) and sources having extended emission beyond 1$^{\prime\prime}$.6 (3 sources). The UVIT F154W, N242W, HST F160W and HST F435W (convolved with the UVIT FUV PSF) images of the 7 sources are shown in Figure~\ref{fig:spec}, while the extended sources are shown in Figure~\ref{fig:spec1}.

We note that one of the LyC leakers in the sample (goodsn-36-G141-25099) has a very narrow H$\alpha$ line and rather looks like a noise peak, which could affect the LyC escape fraction and SFR calculations, however the redshift estimated (1.52) from the grism spectra ([OIII] and H$\alpha$ lines) is very close to the photometric redshift (1.56) provided in the \citet{Skeltonetal2014} catalog. This implies that the line identification is correct and redshift of the source is likely to be $z\sim1.52$.

\begin{figure*}[hbt!]
    \centering
    \includegraphics[width=0.32\textwidth]{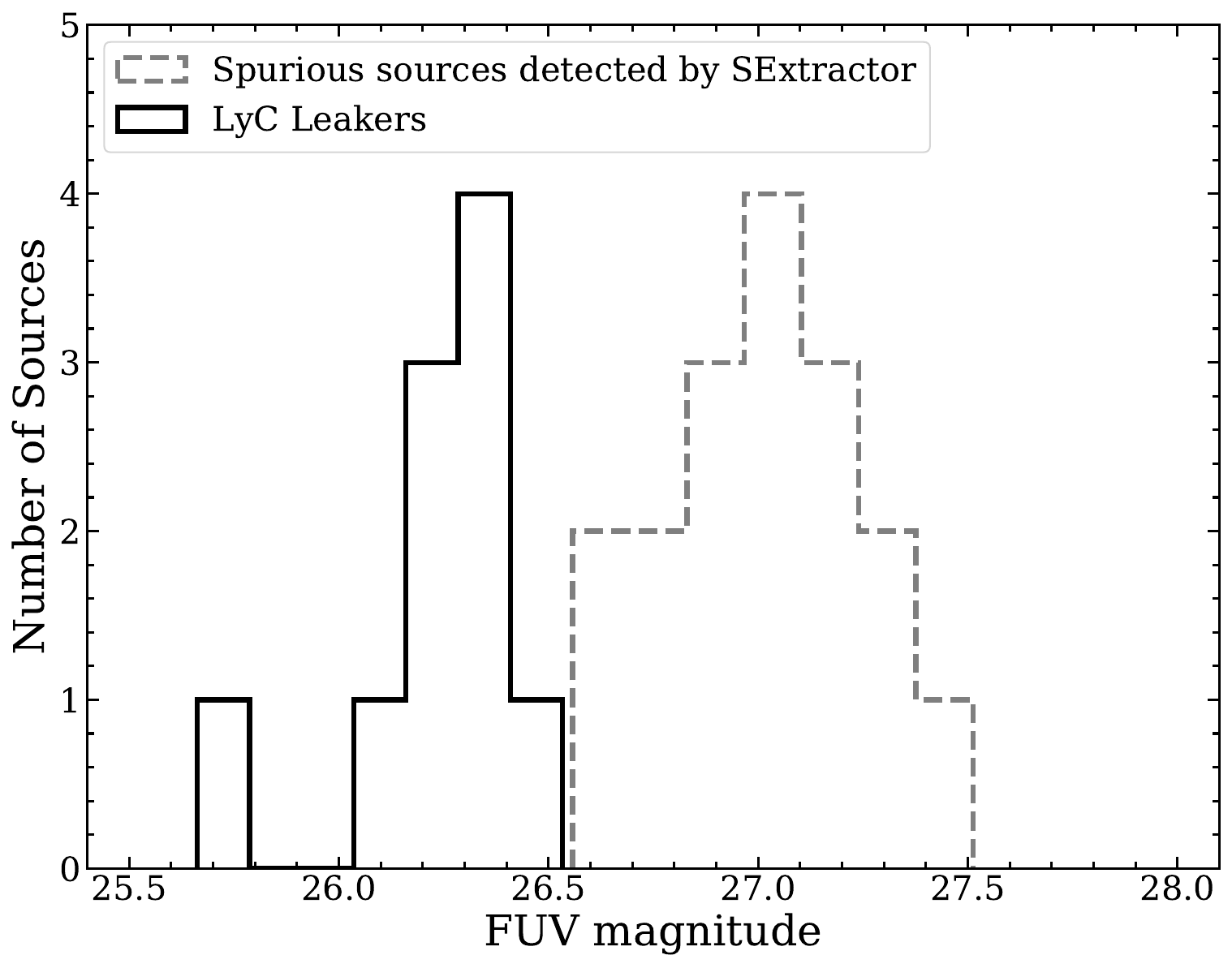}
    \includegraphics[width=0.32\textwidth]{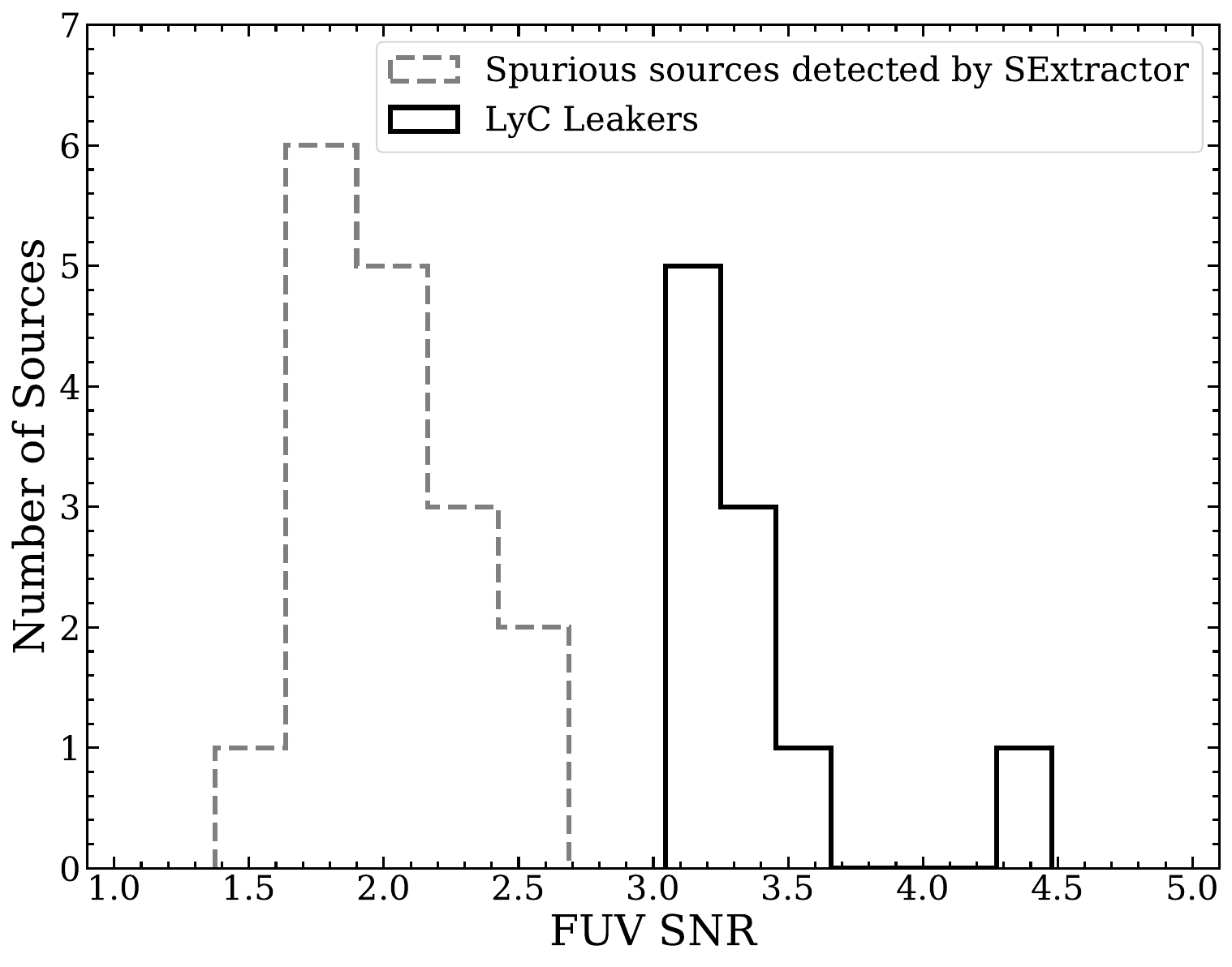}
    \includegraphics[width=0.33\textwidth]{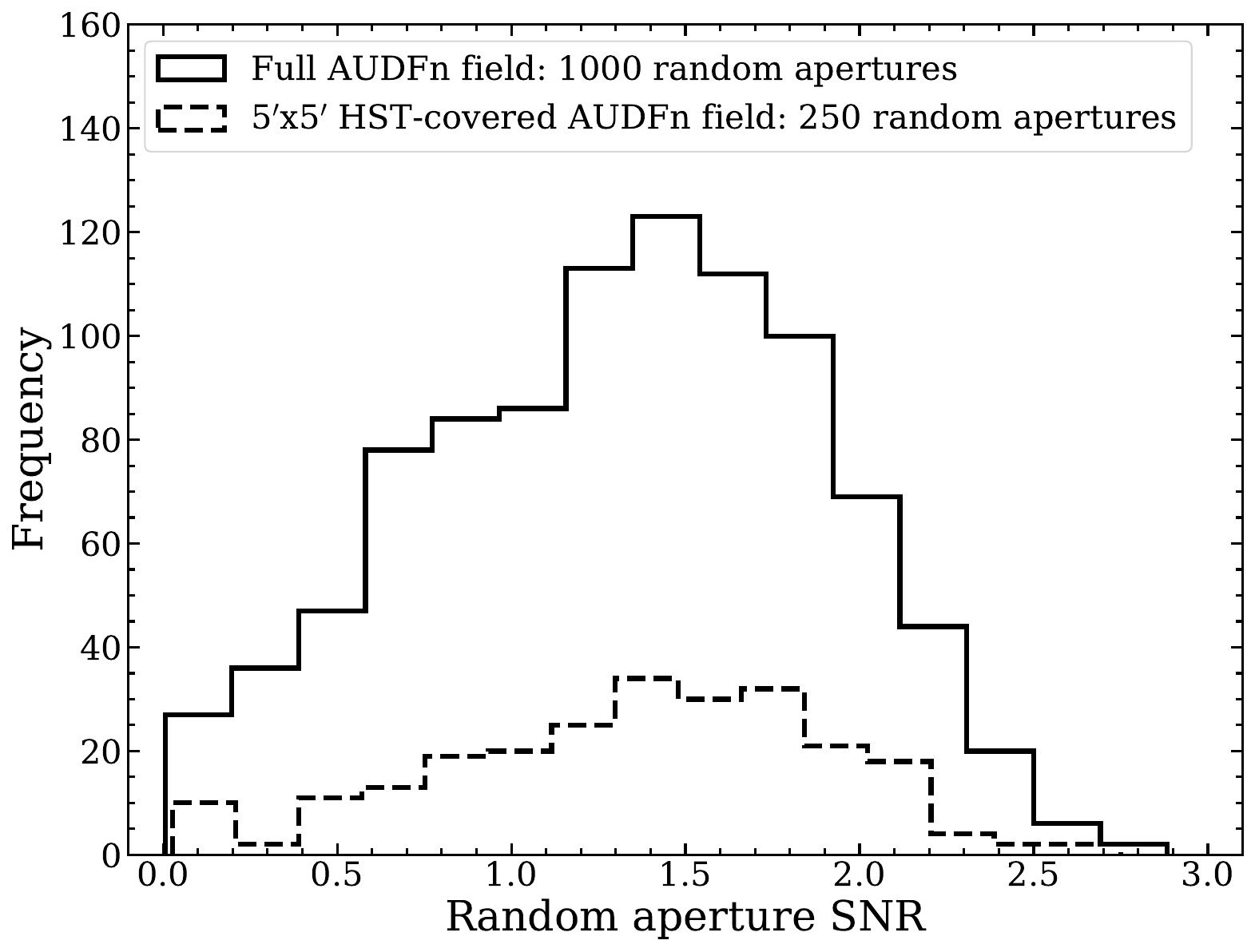}
    \caption{\textit{Left panel}:  The F154W magnitude of sources detected within $10^{\prime\prime}\times10^{\prime\prime}$ field around the 10 LyC leakers by SExtractor. The spurious sources are the ones detected by SExtractor in F154W band around the LyC leakers but not having any counterpart either in HST F435W or F160W band. \textit{Middle panel}: The F154W SNR histogram of the same LyC leakers and the spurious sources as in the left panel. \textit{Right panel}: The SNR distribution of 1000 and 250 circular apertures placed randomly on the entire AUDFn field of 28$^{\prime}$ diameter (solid line) and  5$^{\prime}\times5^{\prime}$ HST-covered patch (dashed line) respectively, after segmenting the detected sources. Both the histograms show apertures only with positive flux values. All magnitudes and SNRs are calculated based on aperture of radius = 1.6$^{\prime\prime}$ unless mentioned otherwise.}
    \label{fig:significance}
\end{figure*}


\section{Photometric Analysis}
\label{sec:phot}
For the 56 potential LyC candidates chosen using the H$\alpha$ and [OIII] line emission criteria, we estimate the SNR at their respective positions in the F154W filter of the UVIT. The first step involves estimating the local background around each source. For this, 
we run Source Extractor \citep{BertinArnouts1996} on the 30$^{\prime\prime}$ $\times$ 30$^{\prime\prime}$ (~75$\times$75 pixels) cutouts around the source. We then place boxes of size 7$\times$7 pixels at 10 random locations where there are no sources detected. The 7$\times$7 box size is optimally chosen as it is $\sim$twice the PSF in UVIT F154W band. The parameters used for generating the SExtractor segmentation maps are listed in Table~\ref{tab:SExtractor}. The mean of these 10 background values is taken as the local background value of the source (The number of random boxes is chosen to be low to avoid overlap, however, a higher number of sampling boxes does not change the background estimate significantly). Then the flux within 1.6$^{\prime\prime}$ circular aperture placed at the source location is extracted. Using the measured flux, the SNR is then estimated using the relation given in \citet{Sahaetal2020}. For one of the sources (goodsn-11-G141\_09345) the SNR estimated within the 1.6$^{\prime\prime}$ aperture is 2.64, whereas it is 3.10 within an aperture of 1.2$^{\prime\prime}$. This could be because of a larger number of noise pixels in the 1.6$^{\prime\prime}$ aperture. For this particular source we quote the SNR within 1.2$^{\prime\prime}$ aperture in Table~\ref{tab:phot}.
 The detection significance and probability of false detections are discussed in Section~\ref{sec:det}.


Further, AB magnitudes are obtained using the magnitude zero points 17.77 and 19.81 for the F154W and N242W, respectively. The magnitudes are then corrected for foreground dust extinction following \citet{SchlaflyFinkbeiner2011} and internal extinction using the UV beta slope $f_{\lambda}\propto \lambda^{\beta}$ {\citep{Reddy18}}, derived from photometry using the Kitt Peak 4-m telescope U band ($\lambda_{mean}=3828$ \AA) and HST F606W ($\lambda_{mean}=5921$ \AA) falling in the rest-wavelength $\lambda \in$ [1,268 - 2,580] \AA ~using the equation given by \citep{Nordonetal2013}. The magnitudes and SNR in the F154W band for our final sample of ten LyC leakers are provided in Table~\ref{tab:phot}.

\begin{table*}[ht!]
\centering
\caption{UV Photometric properties of LyC leakers- Column(1) 3D-HST ID, Column(2): redshift mentioned in the \cite{Momchevaetal2016} catalog, (the redshift is confirmed by visually inspecting the H$\alpha$ and [OIII] emission lines in the G141 grism spectra),
column(3): FUV magnitude corrected for foreground and internal extinction, column(4): SNR of the LyC signal in F154W band within 1.$^{\prime\prime}$6 aperture, Column(5) \& (6): Mean IGM transmission ($e^{-\tau}$) using \citep{Inoueetal2014} and \citep{Bassettetal2021}, respectively, Column(7): Rest-frame UV wavelength in \AA~ probed by the F154W band, Column(8): LyC (F154W) to UV Continuum (F435W) ratio corresponding to rest frame $\sim$600-700 \AA~ in LyC and $\sim$1500-2300 \AA~ in UVC (Ultraviolet Continuum), Column(9) \& (10): Mean LyC escape fractions for the \citet{Inoueetal2014} and \citep{Bassettetal2021} IGM models respectively. The first 7 sources above the horizontal line are the ones where the LyC emission is within a 1.$^{\prime\prime}$6 radius aperture, while the objects below the line have signs of extended emission.}
\begin{tabular}{cccccccccccccc}
\hline\hline
3D-HST ID & $z$  & {m$_{FUV}$ } & {SNR$_{FUV}$} & Mean T & Mean T & Rest UV $\lambda$ & $\frac{LyC}{UVC}$ & $f_{esc}$ & $f_{esc}$\\
  &  &  mag &  & Inoue & Bassett & (\AA) & & Inoue & Bassett \\
  (1) & (2) & (3) & (4) & (5) & (6) & (7) & (8) & (9) & (10)\\ \hline
goodsn-33-G141\_11332 & 1.24 & 26.20$\pm$0.35 & 3.07 & 0.63 & 0.39 & 691 & 0.62 & 0.15$\pm$0.04 & 0.22$\pm$0.02 \\ 
goodsn-17-G141\_36246 & 1.26 & 26.22$\pm$0.35 & 3.12 & 0.63 & 0.38 & 686 & 0.85 & 0.56$\pm$0.10 & 0.67$\pm$0.06 \\ 
goodsn-36-G141\_25099\footnote{H$\alpha$ line very narrow, $f_{esc}$ estimates highly uncertain} & 1.52 & 26.21$\pm$0.35 & 3.09 & 0.51 & 0.27 & 613 & 4.44 & 0.79$\pm$0.13 & 0.88$\pm$0.08 \\ 
goodsn-18-G141\_37037 & 1.43 & 26.10$\pm$0.33 & 3.29 & 0.54 & 0.29 & 636 & 1.31 & 0.44$\pm$0.08 & 0.60$\pm$0.04 \\ 
goodsn-35-G141\_18763 & 1.24 & 26.23$\pm$0.35 & 3.04 & 0.63 & 0.39 & 690 & 1.62 & 0.19$\pm$0.05 & 0.28$\pm$0.03 \\ 
goodsn-11-G141\_09345 & 1.21 & 26.43$\pm$0.41 & 3.10\footnote{SNR measured within 1$^{\prime\prime}.2$ aperture radius} & 0.64 & 0.40 & 699 & 1.41 & 0.35$\pm$0.09 & 0.46$\pm$0.06 \\ 
goodsn-18-G141\_37738\footnote{AGN} & 1.45 & 25.55$\pm$0.24 & 4.48 & 0.54 & 0.27 & 631 & 0.69 & 0.11$\pm$0.02 & 0.19$\pm$0.01 \\ \hline
goodsn-17-G141\_35906 & 1.30 & 26.11$\pm$0.33 & 3.27 & 0.59 & 0.36 & 672 & 2.70 & 0.36$\pm$0.08 & 0.48$\pm$0.04 \\ 
goodsn-46-G141\_14637 & 1.49 & 26.12$\pm$0.33 & 3.27 & 0.51 & 0.27 & 622 & 1.01 & 0.24$\pm$0.06 & 0.37$\pm$0.04 \\ 
goodsn-24-G141\_18817 & 1.49 & 26.01$\pm$0.31 & 3.46 & 0.51 & 0.27 & 623 & 1.19 & 0.29$\pm$0.06 & 0.44$\pm$0.04 \\ 
 \hline  
\end{tabular}

\label{tab:phot}
\end{table*}

\section{Detection Significance of LyC Leakers}
\label{sec:det}
In order to check the detection significance of the LyC leakers in the F154W band, we carry out an exercise using SExtractor. We produce $10^{\prime\prime}\times10^{\prime\prime}$ cutouts around the 10 detected LyC leakers from the UVIT/F154W deep image. Considering the SExtractor input parameters listed in Table \ref{tab:SExtractor}, we identify sources in each cutout and estimate their magnitude within an aperture of radius 1\farcs6. In Figure~\ref{fig:significance} (left panel), we show the magnitude histogram of the LyC leakers, and the SExtractor detected sources that do not have an HST counterpart (termed spurious sources) from all the cutouts. 
There is a clear segregation in the magnitudes of the spurious sources and the LyC leakers. Further, on the fainter side, we detect sources down to the 3$\sigma$ detection limit (27.35 mag) in F154W filter as reported by \citet{Mondaletal2023}. 
The faintest LyC leaker in our sample is around $\sim0.8$ mag brighter than the 3$\sigma$ detection limit. It is to be noted here, the 3$\sigma$ detection limit (within an aperture of radius 1$^{\prime\prime}$) is estimated using the average rms noise of the background sky sampled from the entire AUDFn field \citep[see][]{Mondaletal2023}. Whereas the SNR in this study is estimated using the source signal, local sky background, and exposure time. Following this, the 3$\sigma$ limit is generally fainter than the magnitude of the source having SNR = 3. In our case, the 3$\sigma$ detection limit in F154W is 27.3 mag, which is fainter than the magnitude of a source having F154W SNR = 3.

This signifies that the leaker candidates are unlikely to be spurious noise sources. In the middle panel of Figure \ref{fig:significance}, we show the SNR distribution of the same sources. The SNR histogram further highlights that the identified LyC leakers have a higher SNR value (i.e., 9 out of 10 with SNR $>$ 3.0 in 1.6$^{\prime\prime}$ aperture) compared to a large percentage of spurious sources detected around them.

We carry out an additional independent exercise to test the possibility of false detections in F154W band image. For this, we identify sources in the entire AUDF-north image (diameter $\sim$ 28$^{\prime}$) using SExtractor with the parameters listed in Table \ref{tab:SExtractor}. Then, 1000 apertures of $1^{\prime \prime}$.6 radii are placed randomly such that they do not enclose any segmented pixels identified in the previous step. The mean local background around our 10 leakers is considered as the background value to construct the SNR distribution of the flux measured within these 1000 random apertures (only positive flux values, Figure~\ref{fig:significance} right panel). We find no apertures out of 1000 containing flux with SNR $>$ 3, which indicates that the random fluctuating background pixels are unlikely to mimic an SNR $>$ 3 source flux; see also \cite{Borgohainetal2022} for a similar outcome from GOODS South F154W deep image.
For quantifying the number of FUV spurious sources with respect to the HST detection, we produced a 5$^{\prime}$ $\times$ 5$^{\prime}$ cutout of the GOODS-N field in UVIT F154W, N242W, HST F435W, and F160W bands. We created a rescaled UVIT PSF convolved HST detection image combining F435W and F160W bands and used it to produce a segmentation map. We placed 250 random apertures in source free regions guided by this segmentation map. We found less than 4\% of the 250 apertures having F154W SNR $>$ 3. These can either be real FUV sources that are faint in the redder bands or noise peaks in the FUV background. The random aperture exercise using segmentation from UVIT detection is further repeated for this 5$^{\prime}$ $\times$ 5$^{\prime}$ cutout with 250 circles and we found a similar result as seen for the entire field (Figure~\ref{fig:significance} right panel). 
Further, in 30$^{\prime\prime} \times$ 30$^{\prime\prime}$ area around the LyC leakers, we found SNR$>$3 F154W sources and identify their counterpart in HST F435W filter image. In most cases, we found one-to-one correspondence except goodsn-11-G141$\_$09345 and goodsn-17-G141$\_$35906 where F154W (F435W) source numbers are 4(3) and 9(8) respectively.

\section{Spectroscopic Analysis}
\label{sec:spec}

We use the WFC3 G141 grism data for the spectral analysis. We first extract the 1D contamination subtracted spectral data for the 10 shortlisted LyC leaking candidates. For continuum subtraction, we fit the 1D contamination subtracted spectra with a polynomial function using the \textit{specutils} package in PYTHON. All 10 shortlisted LyC leaking candidates show H$\alpha$ and [OIII] lines in emission. The {[OIII] is a doublet and is fitted accordingly. The emission line fluxes were estimated by fitting gaussian models to the continuum subtracted observed emission line profiles using the \textit{astropy modelling} package in PYTHON. The uncertainties on the line fluxes were estimated using the error spectrum provided for each object with the 1D G141 grism spectra. Given the error at each wavelength, the error on the line flux was estimated by carrying out standard error propagation of the addition of uncertainties under the line. The region of the spectrum chosen to fall under the line is taken as the same spectral window as that used to fit the Gaussian in the line modeling.
In the G141 grism spectra the H$\alpha$ line is blended with [NII]. We extract H$\alpha$ line flux from the blended [NII] line using a method based on equivalent width following \citep{Sobraletal2012}. Further the [OIII] doublet is also blended and wherever possible we fit a double gaussian to extract the [OIII] line flux (although the estimated [OIII] line fluxes are not used in any of the results in this paper). The obtained line fluxes are further corrected for internal extinction. As most of the galaxies lack H$\beta$ emission, the UV $\beta$ slope technique \citet{Reddy18} is used to correct for the internal extinction. Five of the LyC leakers in the sample show H$\beta$ line emission. For these galaxies we estimated the E(B-V) values using the Balmer decrement method by assuming the theoretical ratio as 2.86 assuming Case-B recombination \citet{OsterbrockBochkarev1989} with an electron
temperature of $\sim$ 10$^{4}$ K and electron density of 100 cm$^{-3}$. Only two of the galaxies (goodsn-33-G141$\_$11332 and goodsn-46-G141$\_$14637) show positive E(B-V) values. Assuming a stellar to nebular extinction conversion factor = 0.44, the E(B-V) for goodsn-33-G141$\_$11332 using Balmer decrement (0.35) is in close agreement with the estimate from UV $\beta$ slope method (0.31), while for goodsn-46-G141$\_$14637 the Balmer decrement gives a lower value. However, it is to be noted that there can be scatter in the relation between stellar and nebular extinction with objects being considerably offset from line with slope $f=0.44$ \citep{Koyam19,Reddy20}. Since the number of sources is not sufficient for a statistical comparison and Balmer decrement is not applicable for most of the sources in the sample we use the E(B-V) estimates from UV $\beta$ slope. The estimated line fluxes and derived quantities like H$\alpha$ SFR and sSFR are provided in Table~\ref{tab:specs} in Appendix \ref{sec:supp_plots}, while the 1D spectra are shown in extreme right panels of Figure~\ref{fig:spec} \&~\ref{fig:spec1}. Further, the H$\alpha$ line is used to confirm the redshift of the sources. 

\begin{figure}[ht!]
    \centering
    \includegraphics[width=\columnwidth]{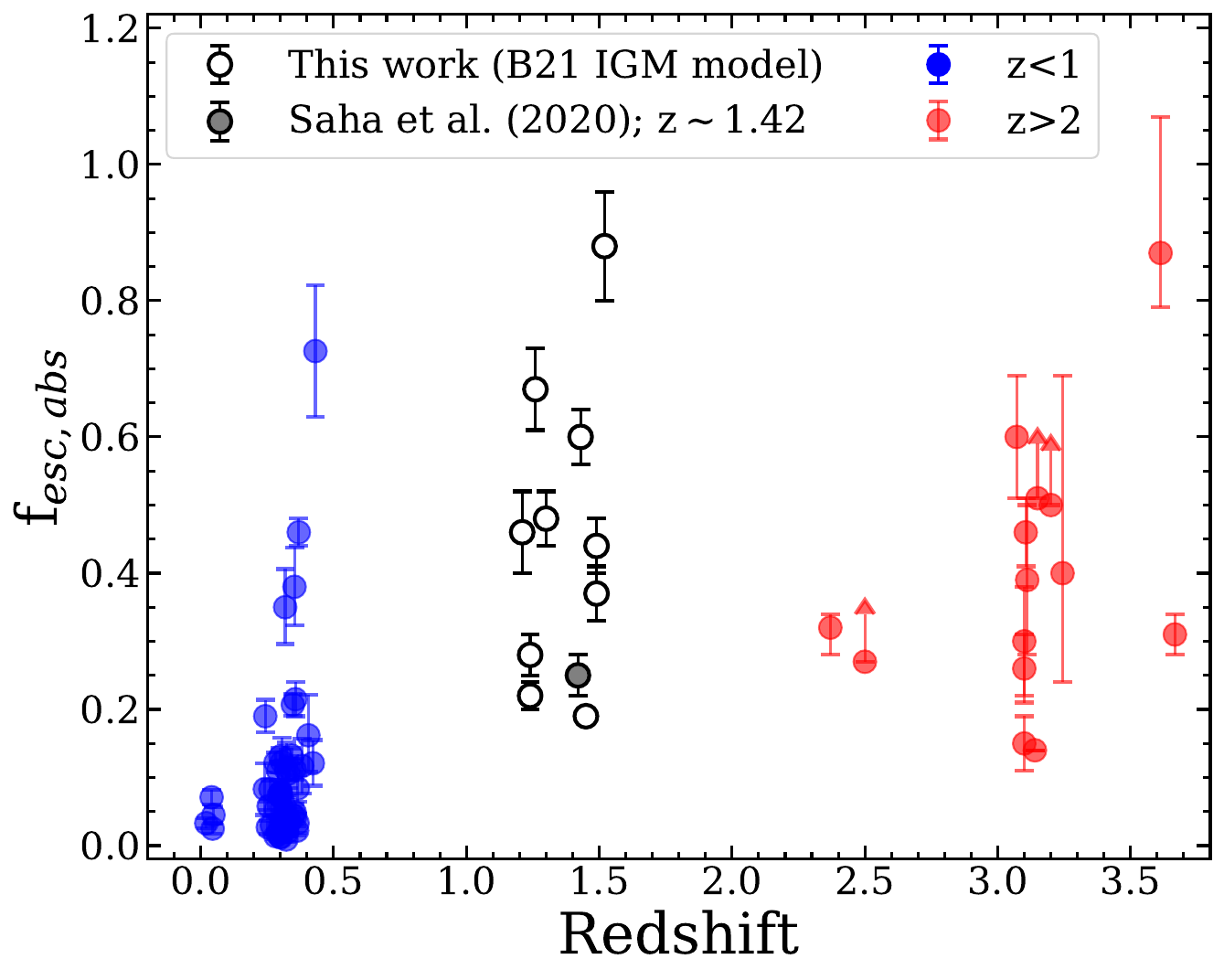}
    
    \caption{The LyC absolute $f_{esc}$ estimated using \citet{Bassettetal2021} IGM model in black open circles for the 10 leakers in our sample compared to individual LyC leakers at lower ($z<1$, blue filled circles) and higher redshifts ($z>2$, red filled circles). The low-z leakers are taken from \citet{Leitetetal2013,Borthakuretal2014,Leithereretal2016,Izotovetal2016,Izotovetal2018, Izotov21,Flury22}. The high-z leakers are taken from \citet{Shapleyetal2016,Vanzellaetal2010,Vanzellaetal2016,Bianetal2017,Fletcheretal2019,RiveraThorsen19,MarquesChaves21,MarquesChaves22}. The individual detection at $z=1.42$ from \citet{Sahaetal2020} is indicated by the grey filled circle.}
    \label{fig:fesc}
\end{figure}

\begin{figure*}[tp!]
    \centering
    \includegraphics[width=\textwidth]{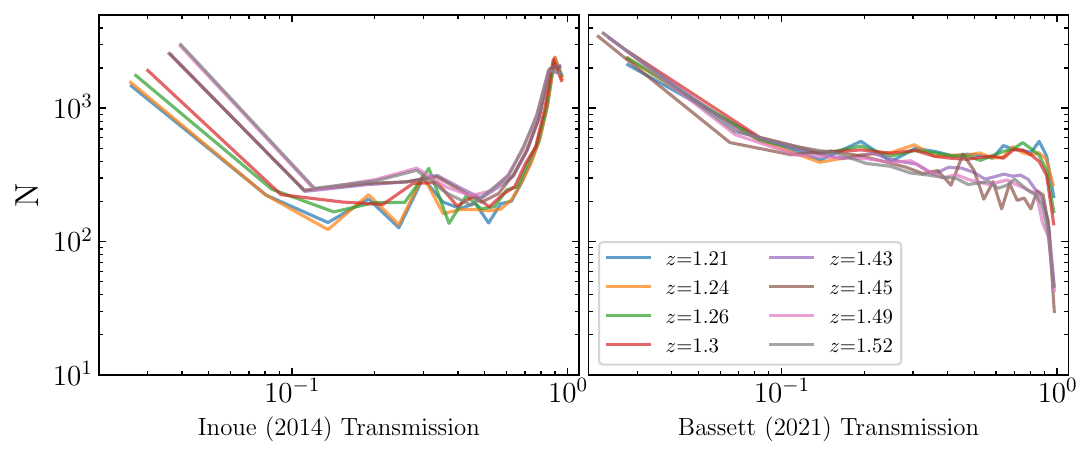}
    \includegraphics[width=0.97\textwidth]{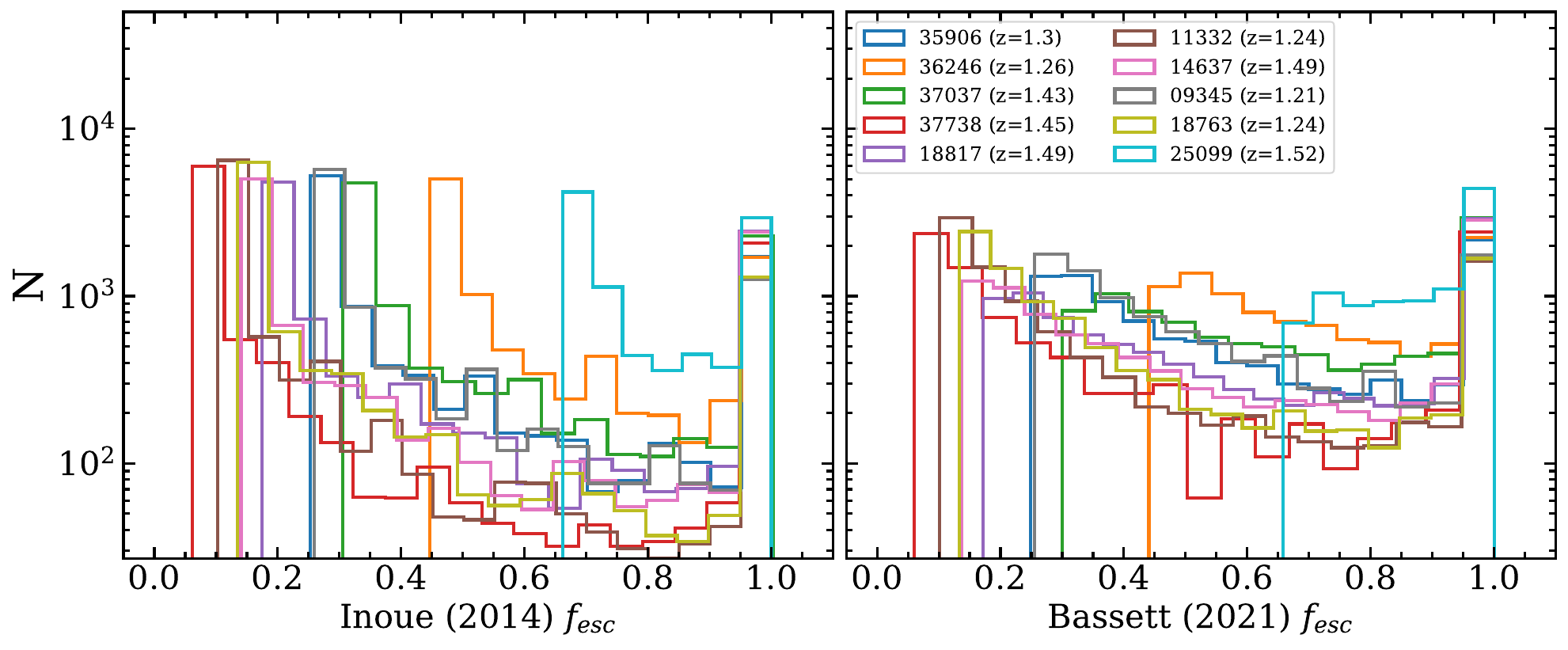}
    \caption{Top: The UVIT F154W-weighted IGM transmission distributions of \citet{Inoueetal2014} and \citet{Bassettetal2021} for the redshifts corresponding to the LyC leakers. Each curve was binned according to Freedman-Diaconis. The \citet{Inoueetal2014} simulations resulted in more clear sightlines compared to \citet{Bassettetal2021} for the same redshifts, which would result in lower $f_{esc}$ (see eqn.~\ref{eq:NescLyC} \& \ref{eq:fesc}). Bottom: Escape fraction distribution of the LyC leakers using IGM attenuation models from \citet{Inoueetal2014} on left and \citet{Bassettetal2021} on the right.\label{fig:igmTran}
}
\end{figure*}

\section{Escape fraction and IGM Attenuation}
\label{sec:fesc}

To derive the absolute LyC escape fraction, we first estimate the number of hydrogen ionizing photons produced that do not escape the galaxy and are in dynamic equilibrium with the ISM using the dust-corrected luminosity of the H$\alpha$ recombination line. Assuming a temperature T = 10$^4$ K and electron density n$_e$ = 100 cm$^{-3}$, we can estimate the number of LyC photons emitted per unit time that do not escape the galaxy \citep{OsterbrockFerland2006,Sahaetal2020} as, 

\begin{equation}
    N^{non-esc}_{LyC} = 7.28 \times 10^{11}L(H\alpha),
\end{equation}

where $L(H\alpha)$ is the dust-corrected H$\alpha$ line luminosity.
We estimate the rate at which LyC photons escape the galaxy, $N_{LyC}^{esc}$ from the FUV flux measured in the F154W filter following \citet{Sahaetal2020}. Note that this gives us a lower limit on the number of escaping LyC photons from the galaxy since the FUV band probes only a part of the rest-frame LyC range for our sample. 

Thus, we have, 

\begin{equation}\label{eq:NescLyC}
    N^{esc}_{LyC} = \frac{L_{F154W}(\lambda^{F154W}_{mean})^2}{hc} e^{\tau_{IGM}}
\end{equation}

where L$_{F154W}$ is the luminosity computed using the observed flux in the F154W filter, $\lambda^{F154W}_{mean}$ = 1541\AA~for the F154W filter and the e$^{\tau_{IGM}}$ term is introduced to correct for the absorption by the IGM. The IGM models used in the analysis are discussed in Section~\ref{sec:igm}. We now use the following expression to denote the absolute escape fraction,

\begin{equation}\label{eq:fesc}
    f_{esc} = \frac{N^{esc}_{LyC}}{N^{esc}_{LyC}+N^{non-esc}_{LyC}}
\end{equation}

The estimated mean escape fractions of the 10 LyC leakers are provided in Table~\ref{tab:phot}.

In Figure~\ref{fig:fesc} we show the LyC $f_{esc}$ of our sample in comparison with LyC leakers at lower and higher redshifts. It is difficult to predict any redshift evolution of $f_{esc}$ from the distribution; however, it is important to note here that different studies use different IGM models and probe LyC at various rest frame wavelengths.

\begin{figure*}[ht!]
    \centering

    \includegraphics[width=0.31\columnwidth]{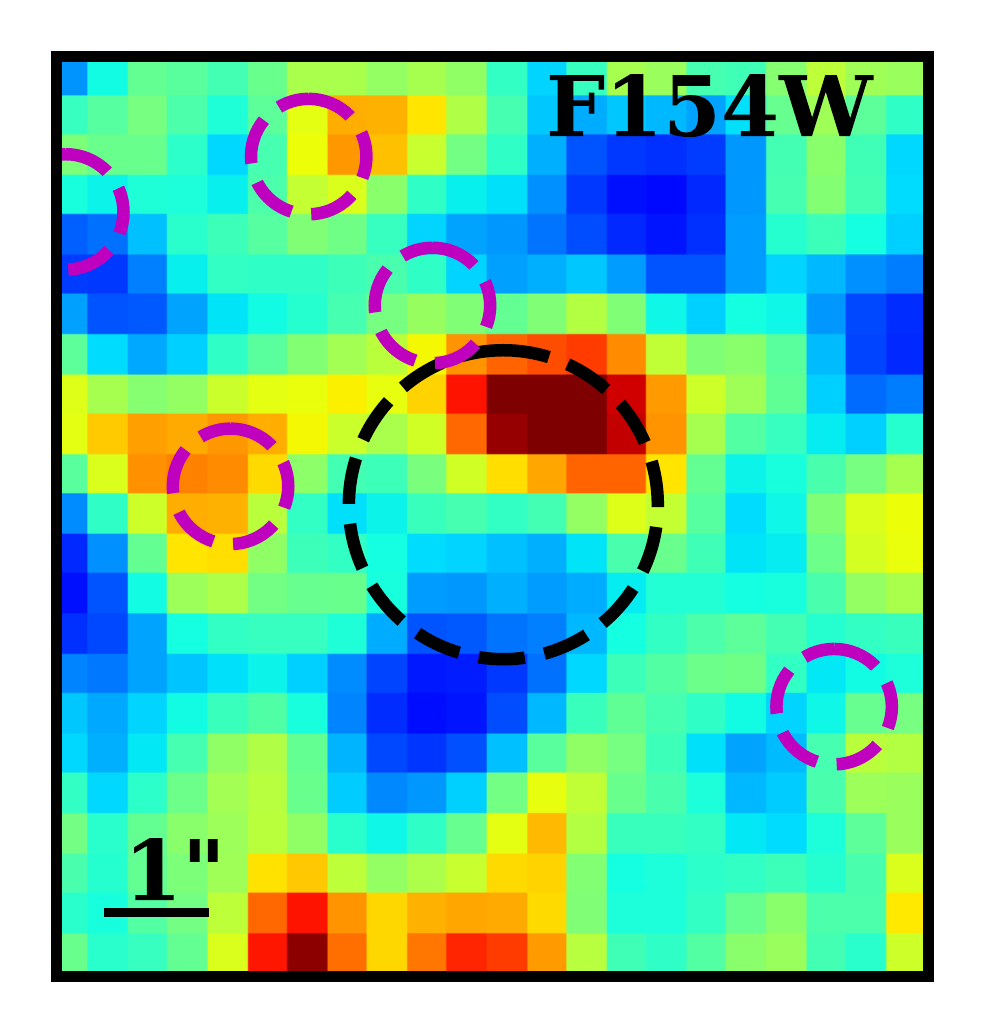}
    \includegraphics[width=0.312\columnwidth]{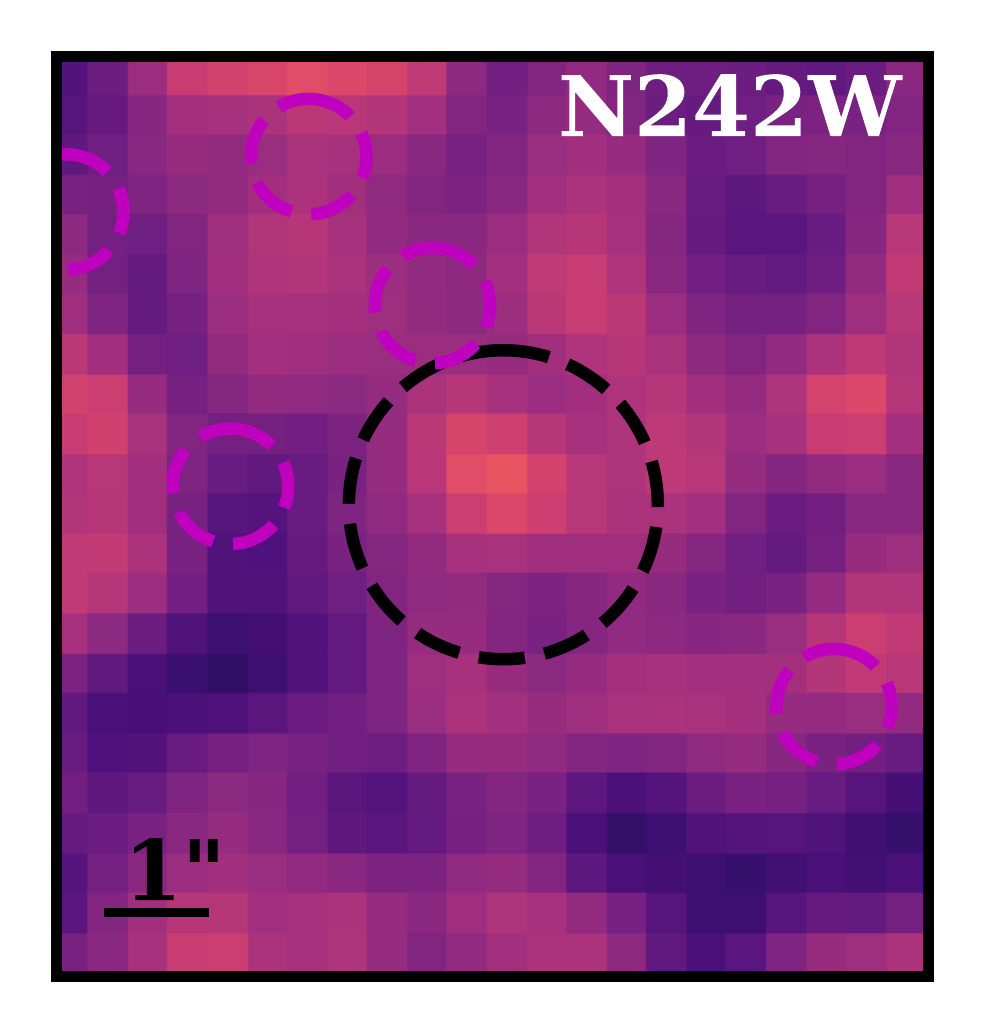}
    \includegraphics[width=0.32\columnwidth]{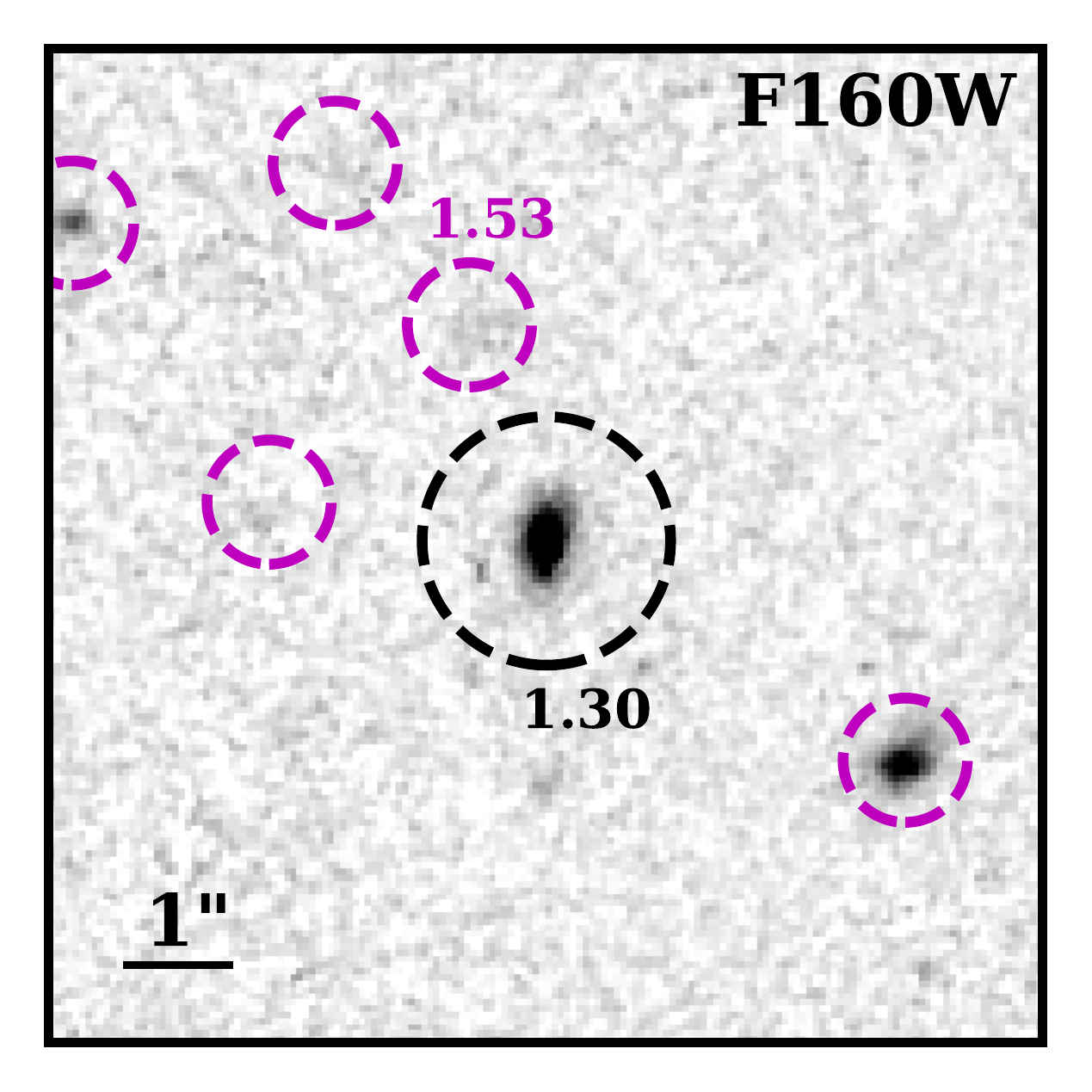}
    \includegraphics[width=0.32\columnwidth]{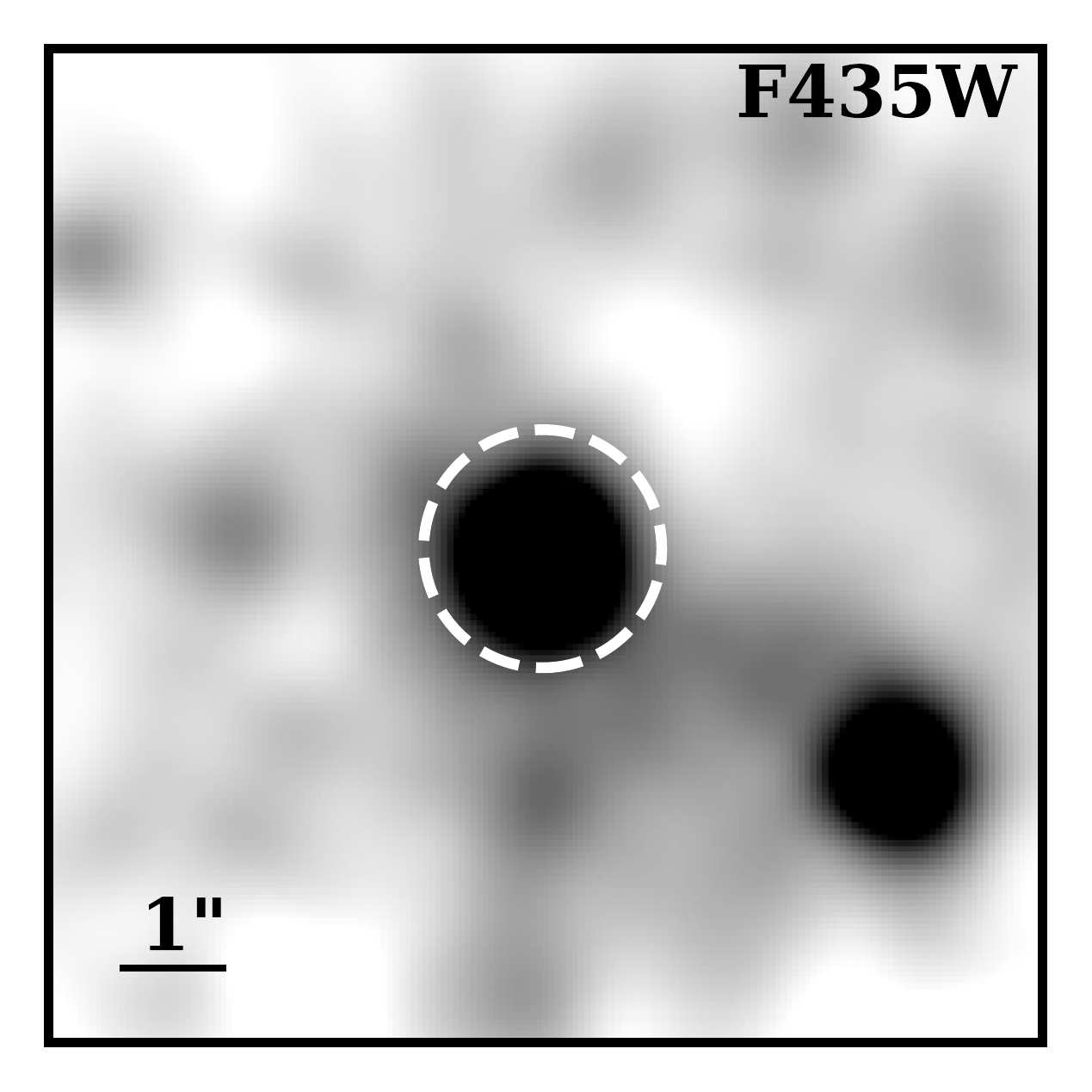}
    \includegraphics[width=0.80\columnwidth]{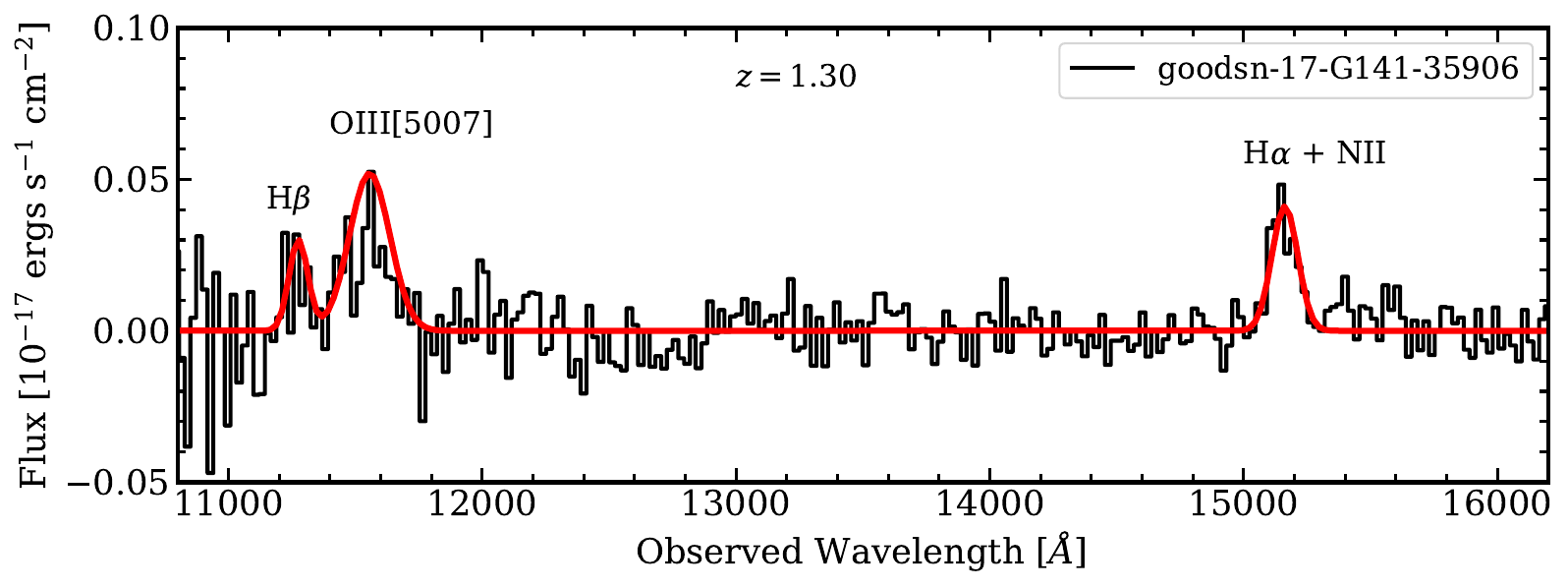}

    \includegraphics[width=0.31\columnwidth]{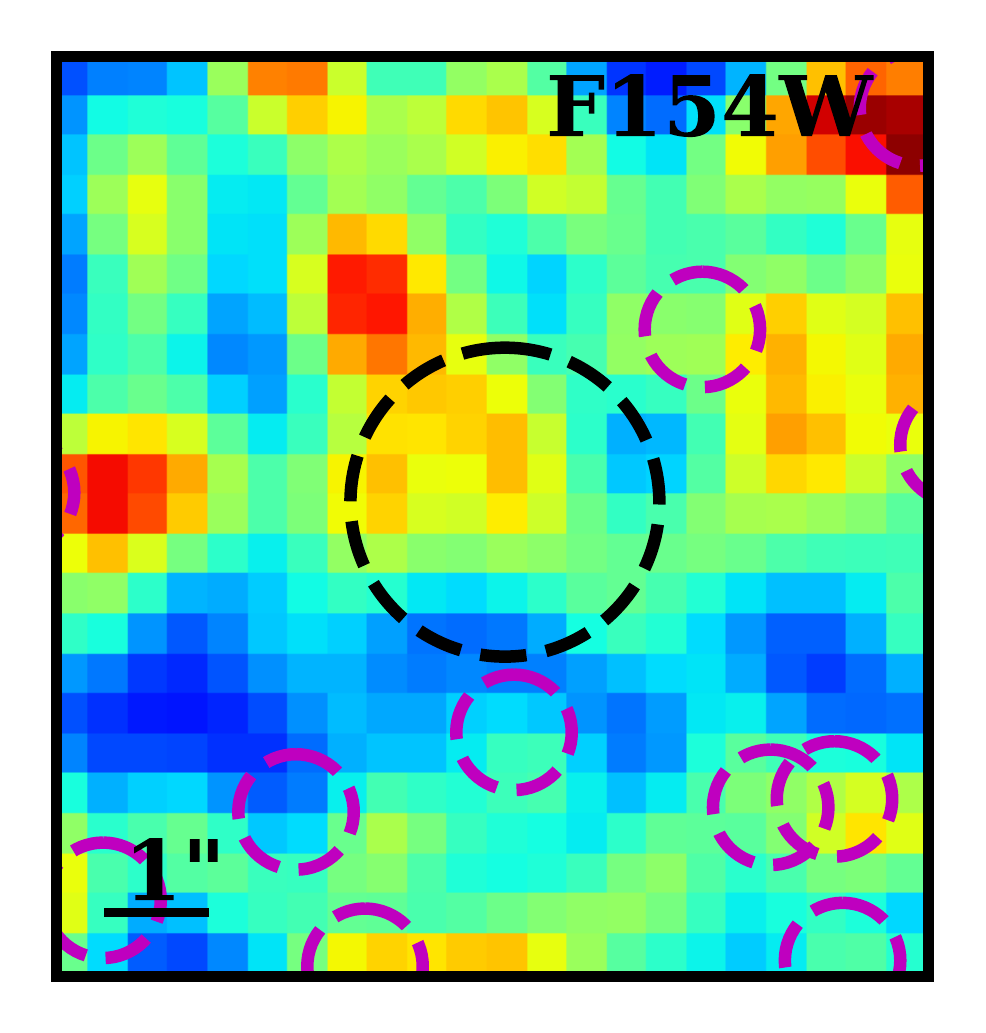}
    \includegraphics[width=0.312\columnwidth]{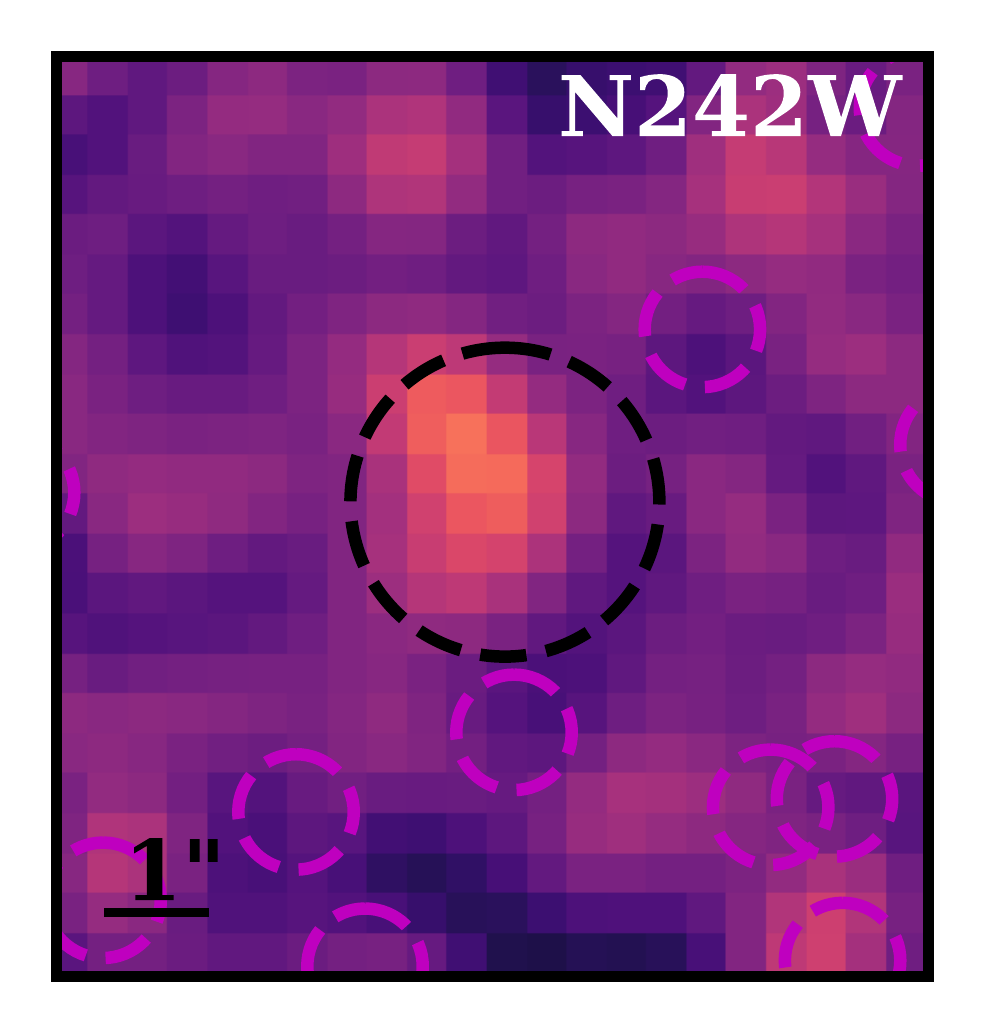}
    \includegraphics[width=0.32\columnwidth]{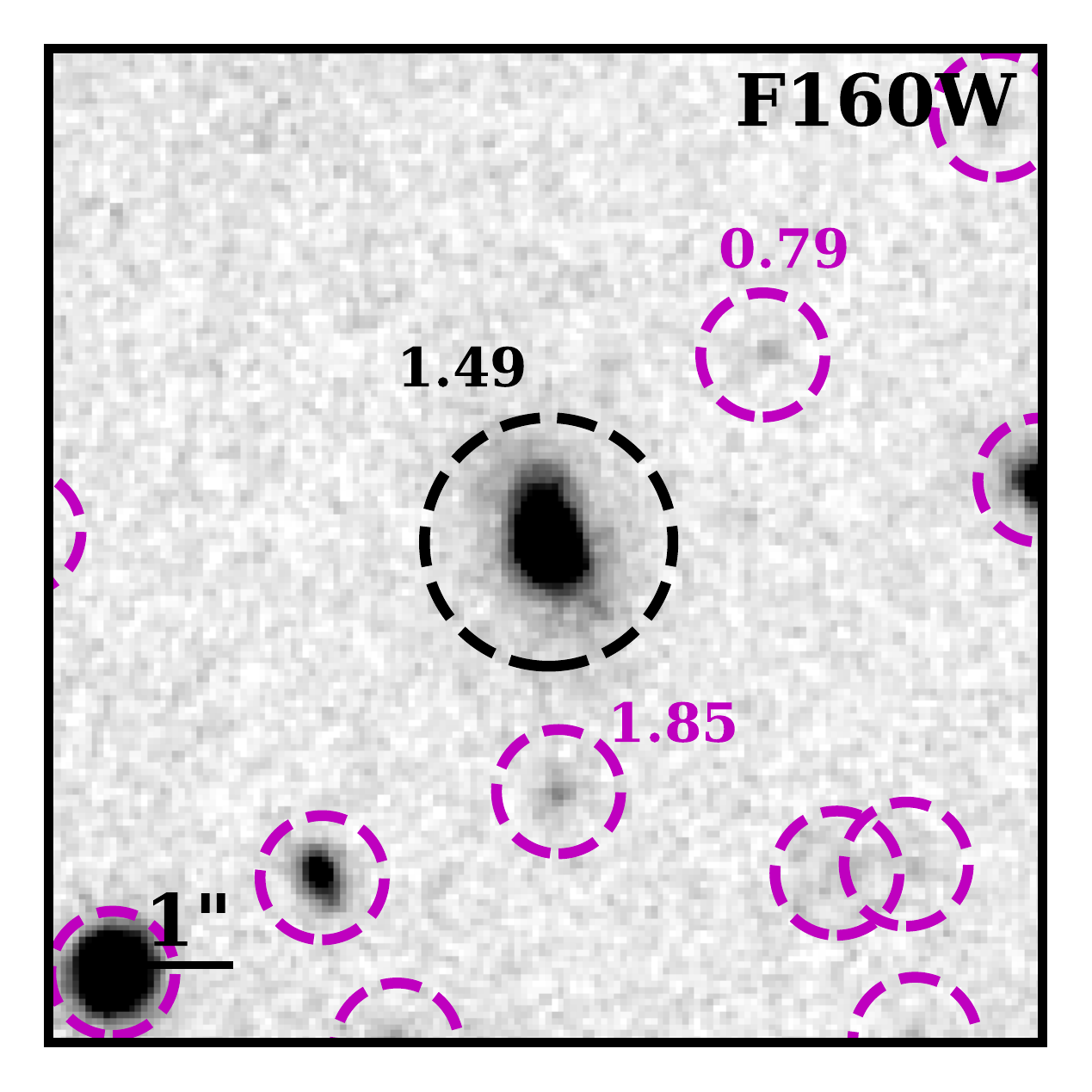}
    \includegraphics[width=0.32\columnwidth]{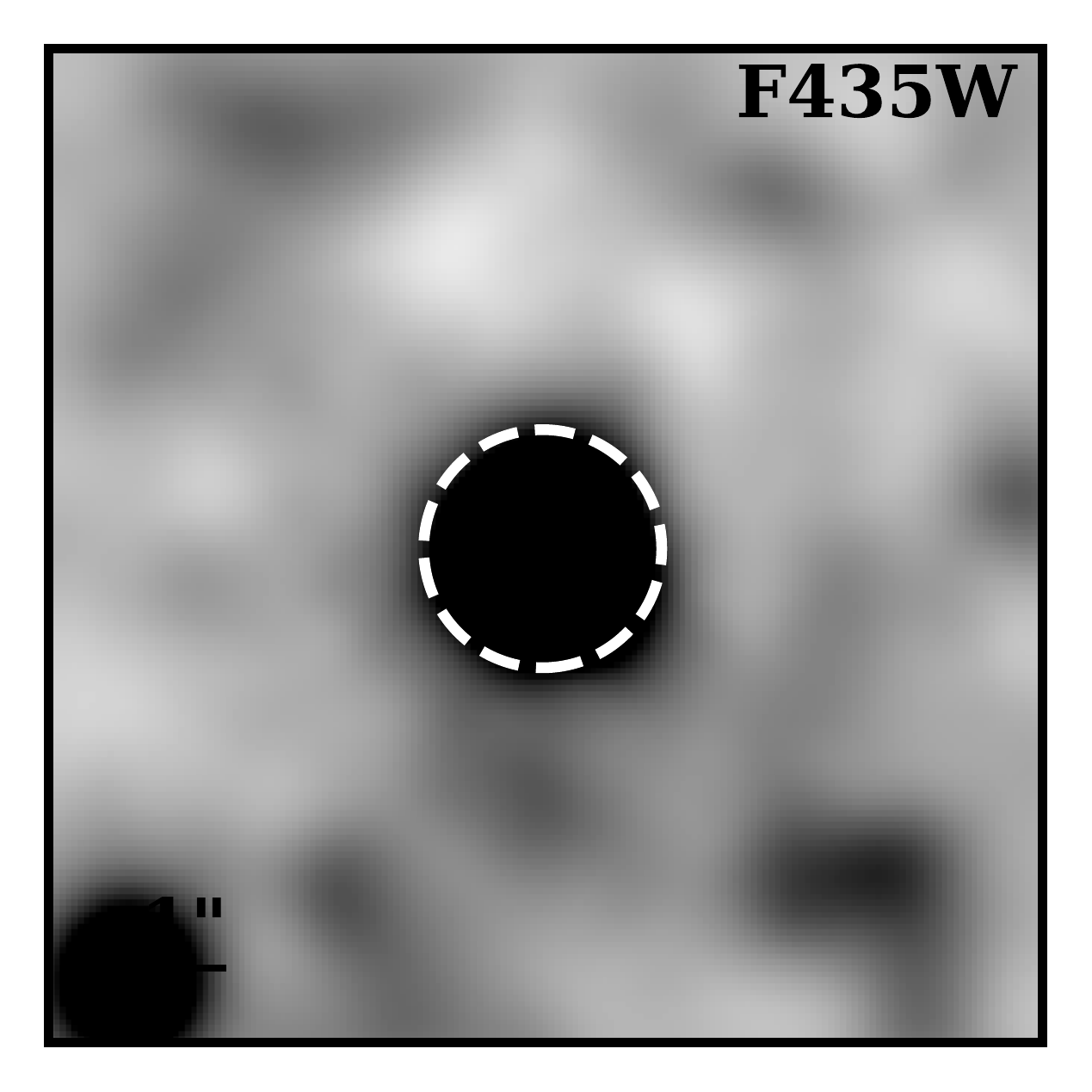}
    \includegraphics[width=0.80\columnwidth]{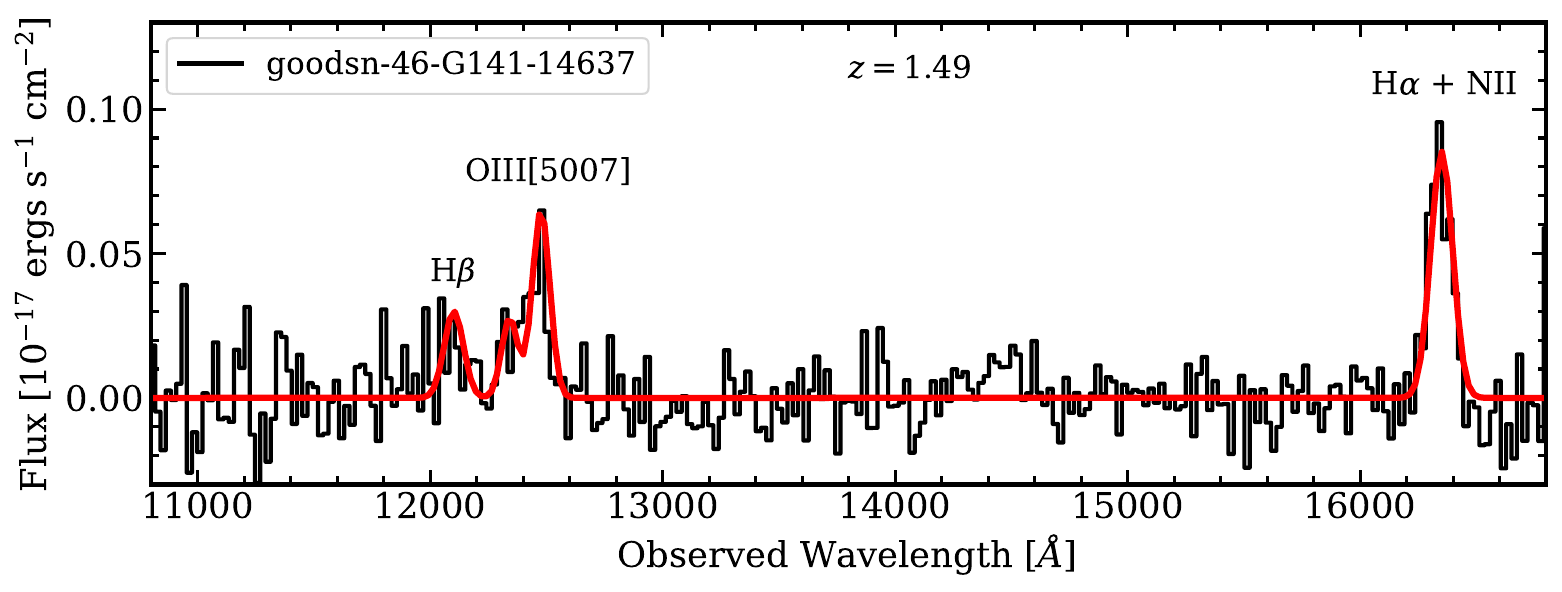}

    \includegraphics[width=0.31\columnwidth]{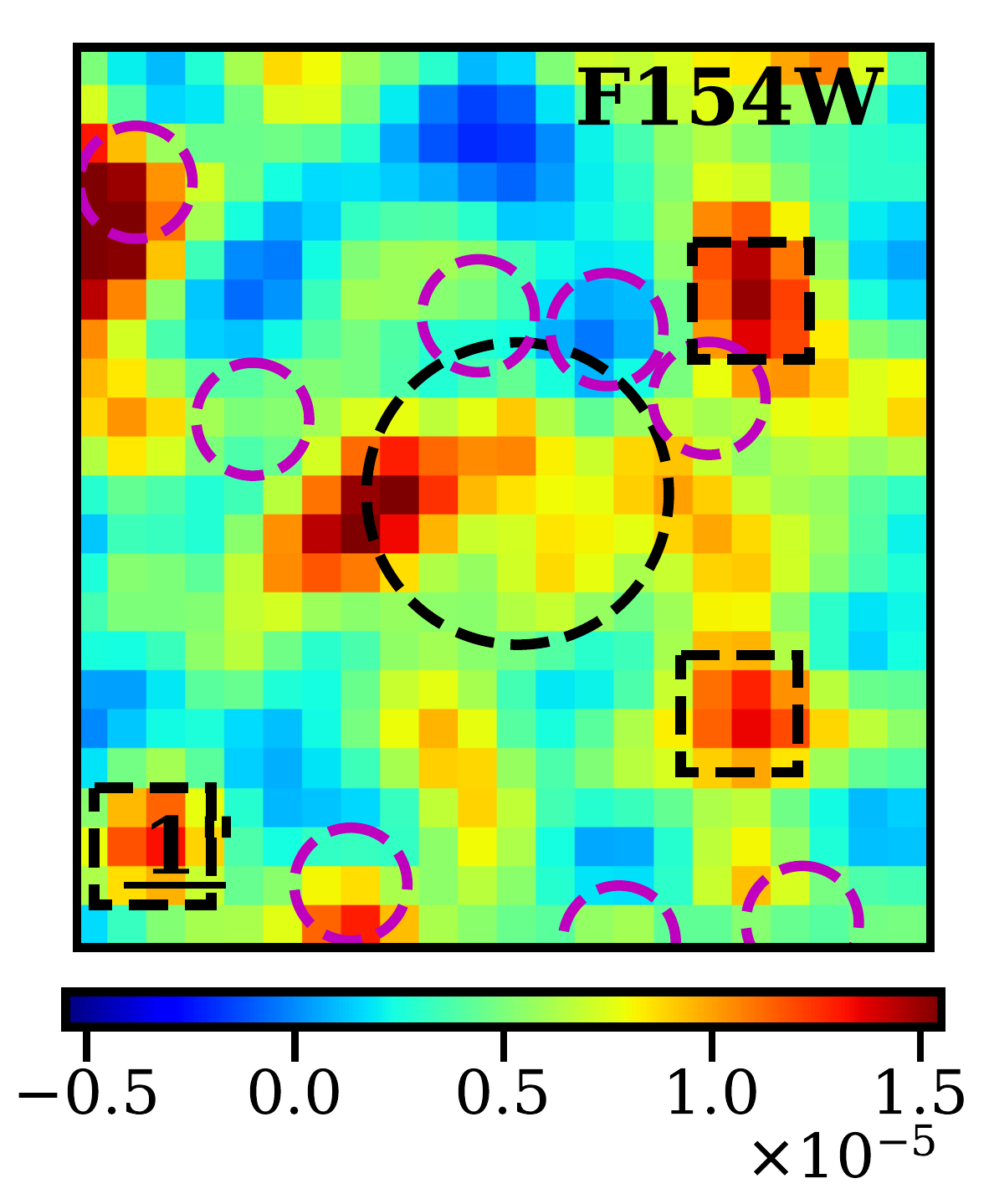}
    \includegraphics[width=0.312\columnwidth]{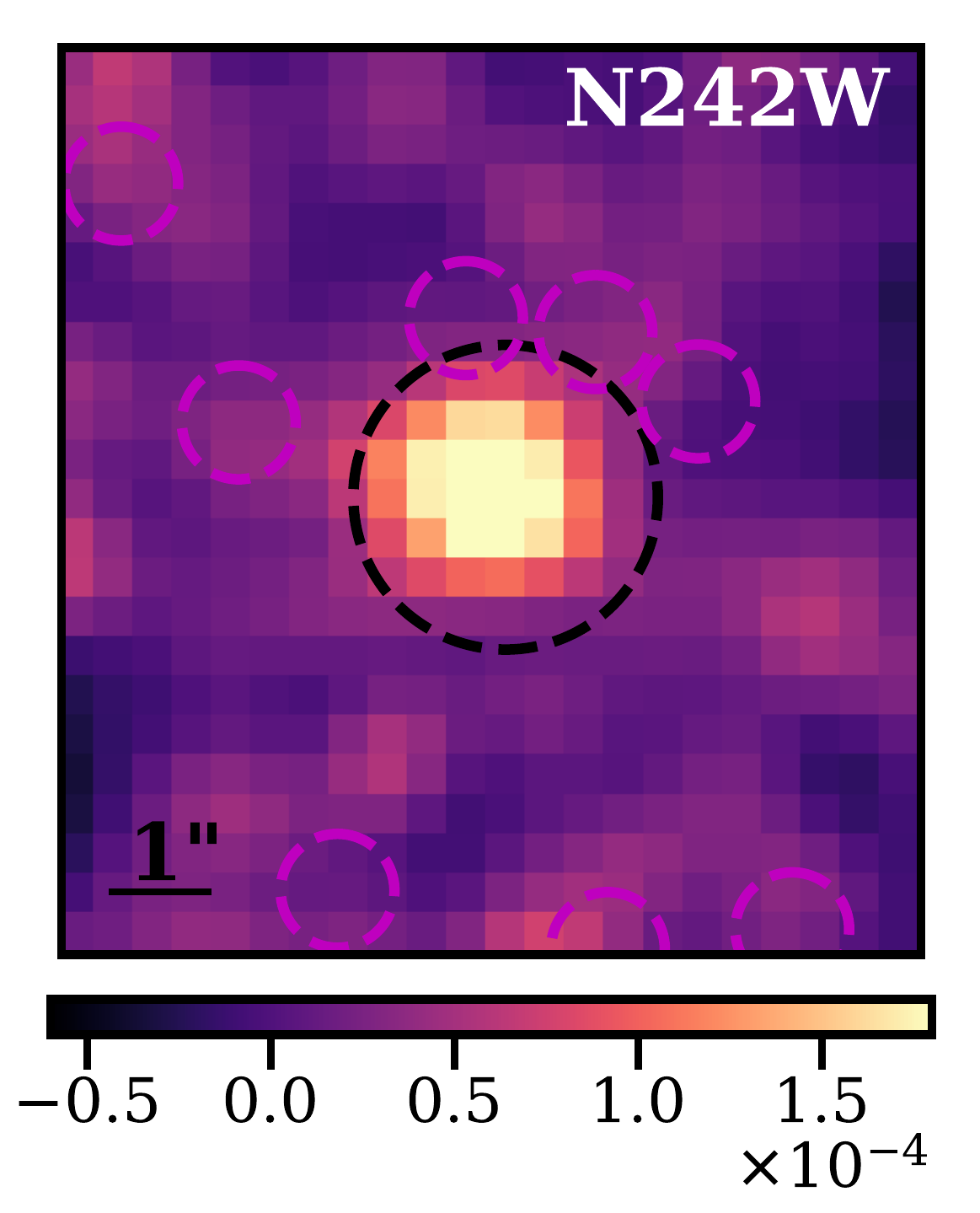}
    \includegraphics[width=0.32\columnwidth]{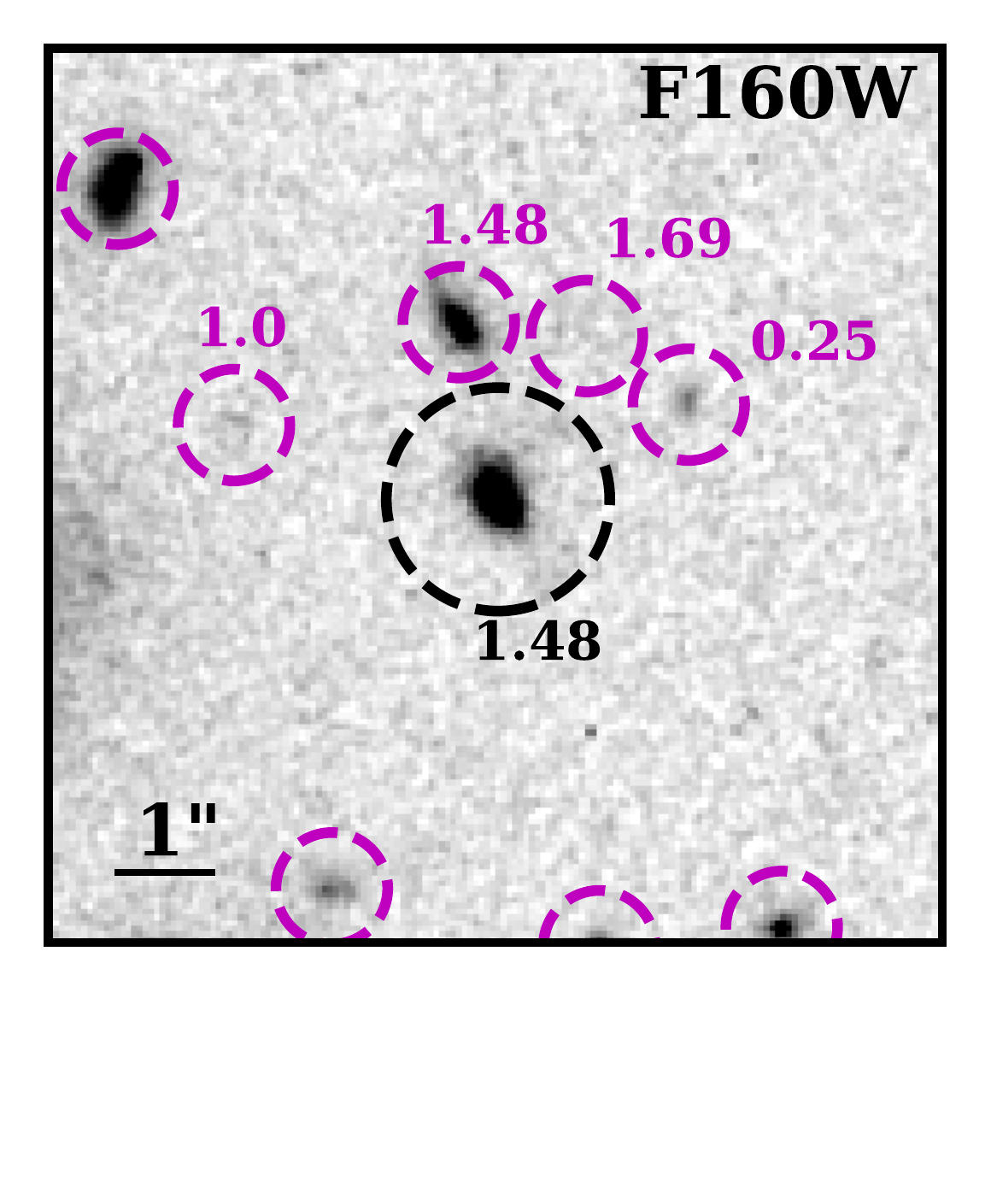}
    \includegraphics[width=0.32\columnwidth]{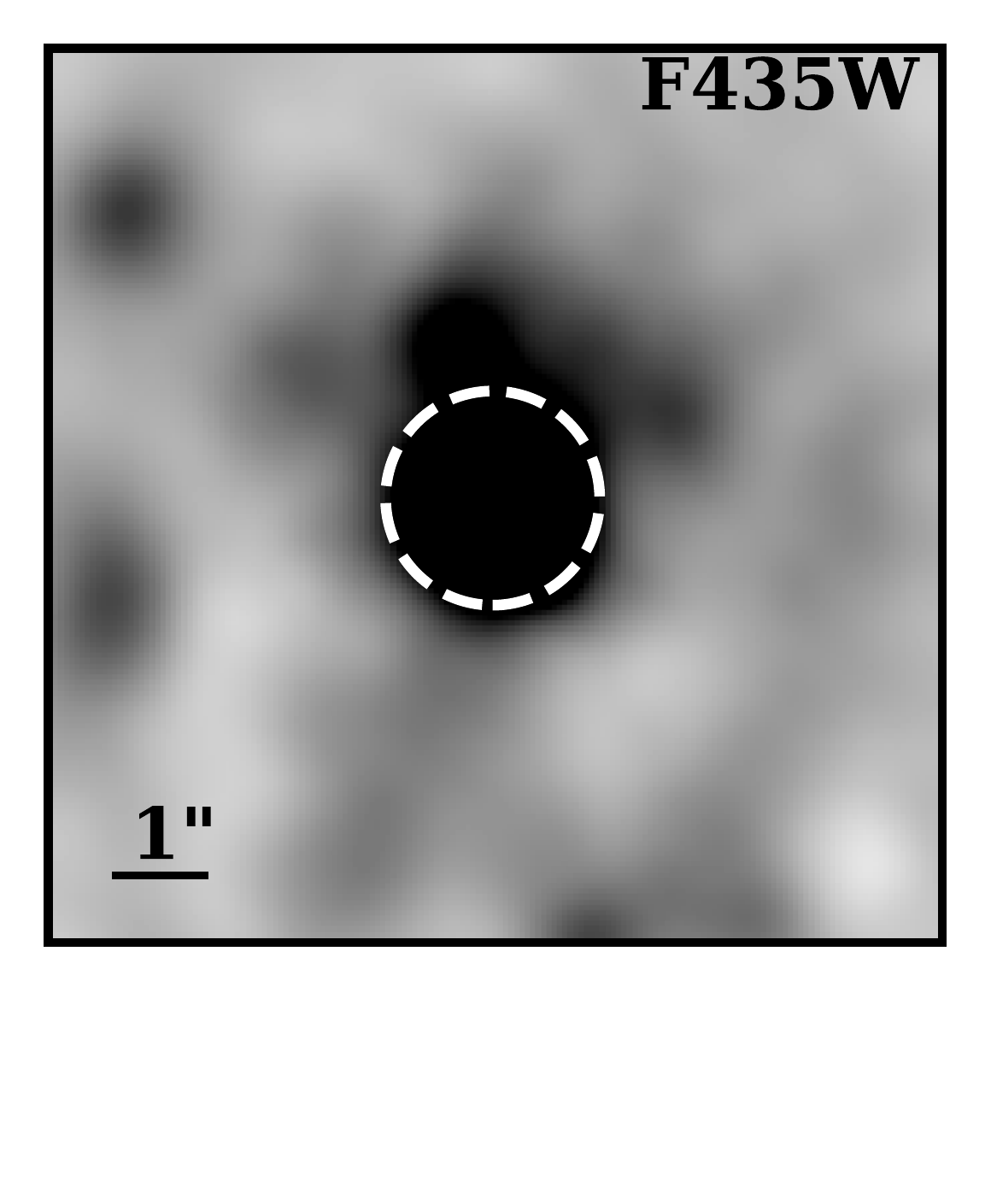}
    \includegraphics[width=0.80\columnwidth]{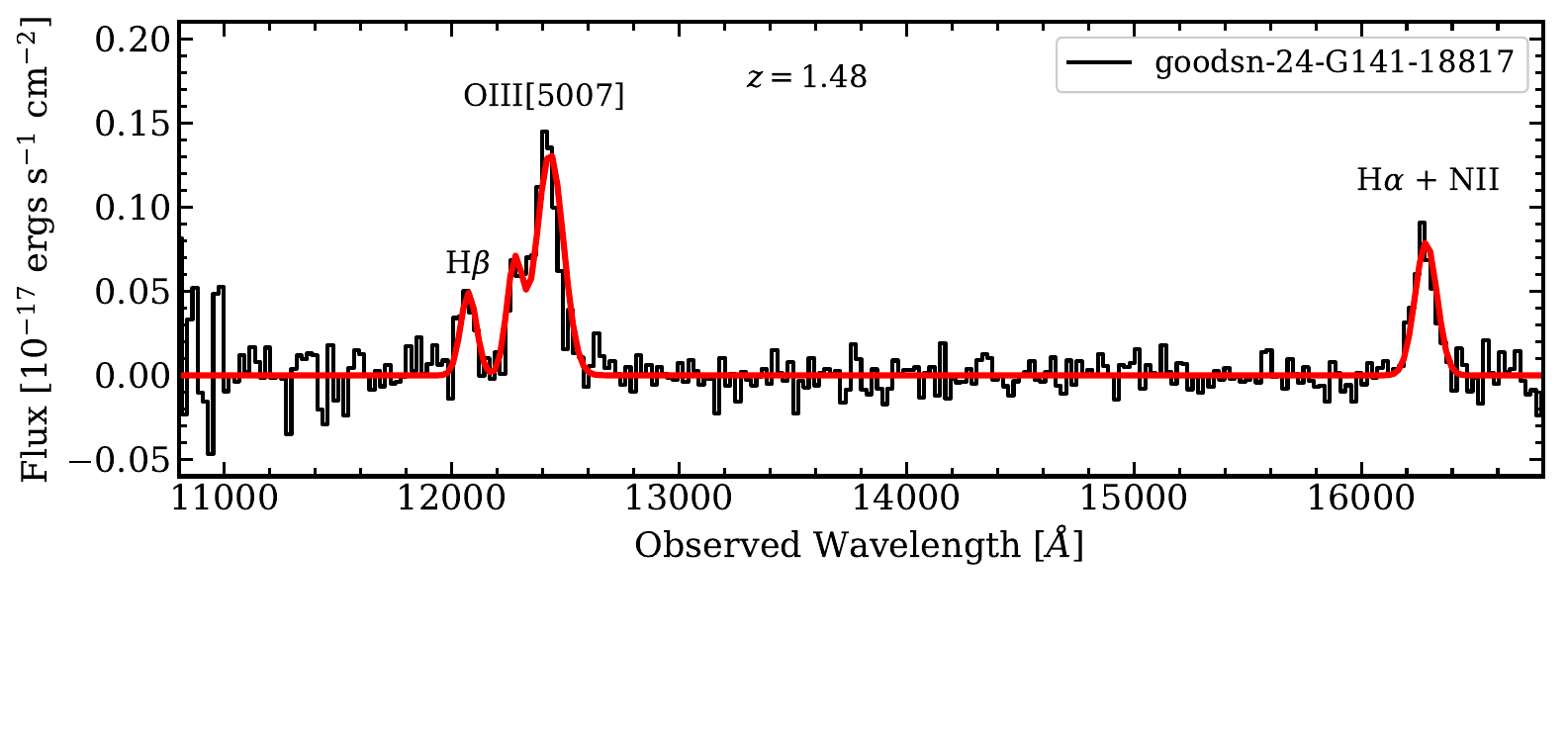}
        
    \caption{LyC leakers having off-centred or extended emission in FUV 154W band. The sequence and symbols are the same as in Figure~\ref{fig:spec}. Bright peaks in the F154W image top panel at north and north-east are due to sources observed in HST F435W(not seen in the 10$^{\prime\prime}\times$10$^{\prime\prime}$) cutouts shown, while in the third panel the spurious sources at north-west, south-west and south-east have magnitudes (SNR): 26.61 (2.56), 26.42 (2.88) and 27.0 (2.01) respectively.}
    \label{fig:spec1}
\end{figure*}

\subsection{IGM Attenuation Models} 
\label{sec:igm}

Since the average IGM absorption of LyC by hydrogen at a given redshift is often stronger than individual sightlines for LyC-detected galaxies, we explore the two IGM attenuation codes of \citet{Inoueetal2014} (hereafter \citetalias{Inoueetal2014}) and \citet{Bassettetal2021} (hereafter \citetalias{Bassettetal2021}) to model the distributions of IGM attenuation integrated through various sightlines at the redshifts of our LyC leakers. Both models use a Monte-carlo (MC) method to simulate the various sightlines, integrating the cumulative effects of damped Ly$\alpha$ absorbers (DLAs), Lyman limit systems (LLSs), and the Ly$\alpha$ forest (LAF) through the IGM at some redshift, using the spatial and density distribution statistics of these absorbers. 

The most notable differences between these two codes are the absorber distribution parameters. The \citetalias{Bassettetal2021} model adopts the distributions of \citet{Steideletal2018}, which in turn uses \ion{H}{I} distribution data from the Keck Baryonic Structure Survey \citep{Steidel2004, Steidel2010, Steidel2014, Rudie2012, Strom2017}, and also includes an optional CGM correction based on data from \citet{Rudie2013}, while \citetalias{Inoueetal2014} uses LyC photon mean-free path data from \citet{Prochaska2009}, \citet{OMeara2013}, \citet{Fumagalli2013}, and \citet{Worseck2014}. The parameters are summarized in table~1 of \citetalias{Inoueetal2014} and table~11 of \citet{Steideletal2018} for the \citetalias{Bassettetal2021} model. In \citetalias{Inoueetal2014}, the distributions for DLAs and LAF are computed separately and summed, while \citetalias{Bassettetal2021} use a piece-wise function for the absorber distributions based on absorber density and adds the optional CGM absorption, which we do not include. These differences can lead to a higher IGM transmission in the \citetalias{Bassettetal2021} model at $z$\,$\simeq$\,3--4 vs. \citetalias{Inoueetal2014} due to the bimodal peaks in the transmission distribution of the \citetalias{Bassettetal2021} model. Furthermore, as discussed in \citetalias{Bassettetal2021}, LyC detections from galaxies at $z$\,$\simeq$\,3--4 are more likely unimodal with transmission values peaking near 100\%. 

At the redshifts of our galaxies, we see the opposite effect, again due to the absorber distribution statistics. \citetalias{Inoueetal2014} uses an equivalent DLA redshift-distribution power law as \citetalias{Bassettetal2021} at our redshift range of interest, however \citetalias{Inoueetal2014} uses as a softer density distribution power law for DLAs ($g(N_{{H}{I}})\sim N_{{H}{I}}^{-0.9}exp(N_{{H}{I}}/10^{21})$ vs $g(N_{{H}{I}}) \sim N_{{H}{I}}^{-1.463}$ where $g(N_{{H}{I}})$ represents the $N_{{H}{I}}$ density distribution; see \citetalias{Inoueetal2014}). This results in fewer dense absorbers ($15 \lesssim log(N_{{H}{I}}/cm^{-2}) \lesssim 21$) in the line of sight for the \citetalias{Inoueetal2014} model. We show the Astrosat UVIT F154W filter-weighted transmission for all unique redshifts and $f_{esc}$ distributions of our LyC-leakers for both codes in Figure~\ref{fig:igmTran}. The transmission (T=$e^{-\tau}$) and $f_{esc}$ columns in Table \ref{tab:phot} correspond to the mean T and $f_{esc}$ which is averaged over all lines of sights for the respective IGM models. The mean transmission for the \citetalias{Inoueetal2014} model used in this work is $\sim 0.57$, while the same for the \citetalias{Bassettetal2021} model is $\sim 0.33$. These transmission values are comparable to the transmissions of LyC at higher redshift, eg: at $z\sim3-4$ \citetalias{Bassettetal2021}. However, the LyC transmission in \citetalias{Bassettetal2021} is with respect to rest wavelengths of $880-910 \AA$, whereas the LyC wavelength probed in this work is $\sim$650~\AA, a region of the LyC spectrum in which the transmission is lower.



\section{Extended LyC emission}


Previous works have indicated that the ionizing LyC and non-ionizing UV emissions can originate from different regions in star-forming galaxies \citep{Iwataetal2009, Vanzellaetal2010, Vanzellaetal2012}.  
Some of the LyC leakers in our sample show extended LyC emission beyond the UVIT PSF FWHM or off-centred emission ($\gtrsim 1^{\prime\prime}$) with respect to the optical and the N242W band. The typical offset with respect to the optical centroids (estimated from HST F435W band) is $\sim 0^{\prime\prime}.8$, while it is $> 1^{\prime\prime}.6$ for two cases (see Figure~\ref{fig:spec1}). Note that the rms of the astrometry in UVIT images is $\sim 0^{\prime\prime}$.25 \citep{Mondaletal2023}.


\citet{Iwataetal2009} found similar offsets in the UV continuum (UVC) and LyC for a sample of Lyman Alpha Emitters (LAEs) and  Lyman Break Galaxies (LBGs) at $z\sim$3, with substructures in the LyC. We do not identify any substructures in the LyC because of the resolution constraints; however, the offsets and extended emission are prominently seen.  For size comparison, we show the HST F435W band images (which probes the rest frame $\sim$1500-2300~\AA~ for the redshift range of our sample) convolved with UVIT F154W PSF in Figure~\ref{fig:spec} and~\ref{fig:spec1}. The ratio of LyC (F154W) to the UVC (F435W) are provided in Table~\ref{tab:phot}. Our observed LyC to UVC ratio is on average higher than what has been reported in the literature \citep{Steideletal2018}. However, it is worth noting that previous LyC studies such as those of \citet{Steideletal2018} at $z\sim 3$ probe LyC wavelengths at $\sim$ 900~\AA~ and the UVC at 1500~\AA, whereas in this work the LyC wavelengths probed for the leakers is $\sim$ 600-700~\AA~ and the UVC corresponds to rest-frame wavelengths 1500-2300~\AA~. It would be interesting to compare our observed ratio with the intrinsic ratio obtained from the best-fit intrinsic spectral energy distribution (SED) of these objects. We defer this exercise for the future.

We further produced the stack of the LyC signal signal of the six LyC leakers (the first six galaxies in Table~\ref{tab:phot} for which the LyC emission is within 1$^{\prime\prime}$.6) from the HST centroids. To construct the stack, we consider 10$^{\prime\prime}$x10$^{\prime\prime}$ size sky subtracted F154W band images of the objects. We then mask neighbouring sources outside a 1$^{\prime\prime}$.6 aperture radius from the centroid of the source using segmentation maps produced by SExtractor with a 1$\sigma$ detection threshold. The stacked image is then produced by the average count rate of the six LyC leaking galaxies. The stacked LyC shown in Figure~\ref{fig:stack} has a total flux of 2.75$\times$10$^{-4}$ c/s within the 1$^{\prime\prime}$.6 aperture corresponding to S/N$\simeq$7.49 and $\sigma\simeq$3.0$\times$10$^{-6}$ c/s.

In previous studies by \citet{Smithetal2018, Smithetal2020}, flat LyC profiles have  been observed with average stacked LyC profiles of galaxies and weak AGNs  where they associate the non-central LyC emission to the porous IGM and scattering process LyC photons undergo prior to the escape. Because of the resolution constraints we do not attempt to produce the surface brightness profile of the LyC here. Three of the galaxies show even larger offsets in LyC emission from the F160W centroids and extended emission. In Figure \ref{fig:spec1}, we show these three galaxies and their spectra. Although the galaxies show very extended and off-centred emission and the source of emission could be debatable, there are no nearby sources which could be considered as possible sources of emission. Therefore, the extended emission is most probably from the host galaxies. The spatial distribution of the LyC emitting sources from that of non-ionizing UV-emitting stars can cause these offsets.

\begin{figure}[t!]
    \centering
    \includegraphics[width=\columnwidth]{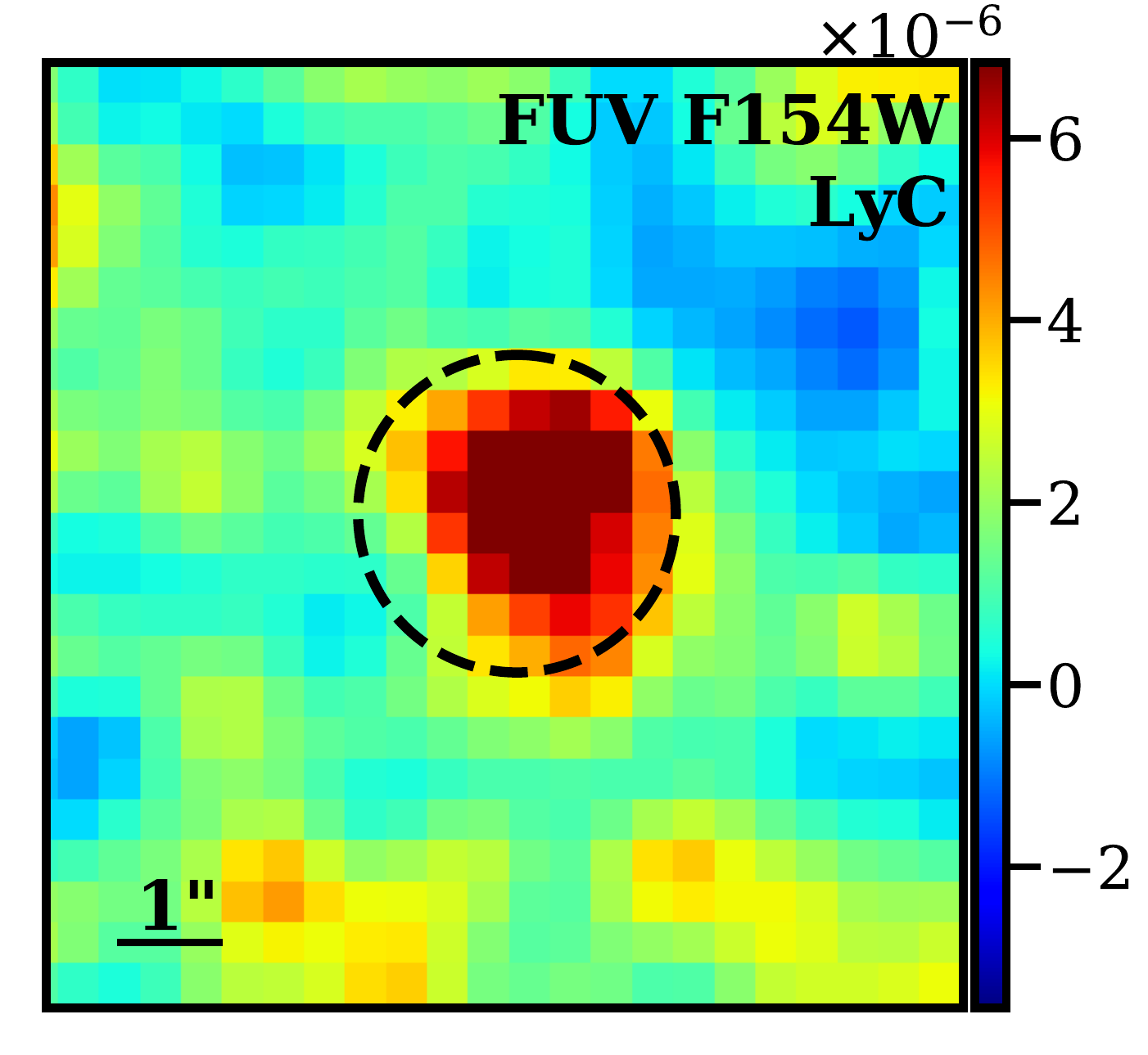}
    \caption{The stacked image of LyC (FUV F154W) of the six LyC leakers having emission within UVIT PSF FWHM $\sim$1.6$^{\prime\prime}$.}
    \label{fig:stack}
\end{figure}

\section{Foreground Interlopers and AGN diagnostics}
\label{sec:contamination}

As mentioned in section~\ref{sec:sample}, the final sample of ten LyC leakers is chosen such that there are no nearby sources within a radius of 1$^{\prime\prime}$.2 in the HST images (see third panel of Figures~\ref{fig:spec} $\&$ \ref{fig:spec1}). However, some fields are crowded, and surrounding objects, although not within 1$^{\prime\prime}$.2, are close to the LyC sources. In Figure~\ref{fig:spec} $\&$ \ref{fig:spec1}, the neighbouring sources are shown by dashed magenta circles on the HST F160W images. The photometric redshifts of the nearest sources are also given in magenta. The surrounding sources detected in HST detection images do not have emission in the FUV band as observed in all the middle image panels in Figure~\ref{fig:spec} $\&$ \ref{fig:spec1}. Also, the possibility of contamination from objects in the line of sight can be ruled out as the spectra do not show any other lines than the H$\alpha$ and O[III]5007 and the redshifts estimated from the grism spectra match with the photometric redshift from the \citet{Skeltonetal2014} catalog.  Hence, the estimated escape fraction is unlikely to be contaminated by the neighbouring or foreground sources. 

To check if any of the objects in our sample are possible AGNs (Active Galactic Nuclei), we explored the X-ray archival data. We find that one of the sources from the sample is identified as an AGN in the 2 Ms Chandra Deep Field-North Catalog \citep{Xueetal2016}. The other 9 sources are classified as normal galaxies.

\section{Possibility of more LyC emitter candidates in the AUDF-North}

There are 20 sources in the shortlisted sample of 39 galaxies having 2.5 $<$ SNR $<$ 3 (see SNR histogram in Figure~\ref{fig:mag-snr}). Considering the flux of the faintest source of the 20 galaxies and a mean local background, we estimate exposure time so that the sources have an SNR $>$ 3. With an observation with effective exposure time $>$ 46 ks in FUV, the 20 sources would be above SNR 3. Thus, we predict the possibility of tens of more LyC detections in the AUDF Fields in the upcoming observations and the capability of UVIT.

\section{Summary}
\label{sec:summary}

Using the deep imaging from UVIT and the grism spectra from the 3D-HST survey, we report the direct detection of LyC emission from 10 sources (9 galaxies and 1 AGN) in the GOODS-North field. The reported LyC leakers are at the redshift of 1.0 $< z < $ 1.6, near the epoch when the cosmic star formation peaked. Using two IGM attenuation models, we provide the absolute LyC escape fraction of the detected LyC emitters. The LyC emission is seen to be off-centred and extended as compared to the optical counterpart in most of the sources in the sample. The rest frame UV wavelengths  of the LyC leakers are found to be in the EUV regime (550-700\AA), far below the Lyman limit (912\AA) studied to date.  Further, we also predict the possibility of at least tens of more LyC detections in the Astosat UVIT deep fields, which would be possible with upcoming observations.

\begin{acknowledgments}
SD and KS acknowledges support from the Indian Space Research Organisation (ISRO) funding under project PAO/REF/CP167. This publication uses the UVIT data obtained from the Indian Space Science Data Centre (ISSDC) of ISRO, where the data for AstroSat mission is archived. The UVIT project is a collaboration between IIA, IUCAA, TIFR, ISRO from India and CSA from Canada.  This work is based on observations taken by the 3D-HST Treasury Program (GO 12177 and 12328) with the NASA/ESA HST, which is operated by the Association of Universities for Research in Astronomy, Inc., under NASA contract NAS5-26555.
\end{acknowledgments}

%

\facilities{3DHST(WFC3), ASTROSAT(UVIT)}


\software{astropy \citep{AstropyCollaboration},  
          Source Extractor \citep{BertinArnouts1996}
          }




\begin{thebibliography}{}
\expandafter\ifx\csname natexlab\endcsname\relax\def\natexlab#1{#1}\fi
\providecommand{\url}[1]{\href{#1}{#1}}
\providecommand{\dodoi}[1]{doi:~\href{http://doi.org/#1}{\nolinkurl{#1}}}
\providecommand{\doeprint}[1]{\href{http://ascl.net/#1}{\nolinkurl{http://ascl.net/#1}}}
\providecommand{\doarXiv}[1]{\href{https://arxiv.org/abs/#1}{\nolinkurl{https://arxiv.org/abs/#1}}}

\bibitem[{{Alavi} {et~al.}(2020){Alavi}, {Colbert}, {Teplitz}, {Siana}, {Scarlata}, {Rutkowski}, {Mehta}, {Henry}, {Dai}, {Haardt}, \& {Bagley}}]{Alavietal2020}
{Alavi}, A., {Colbert}, J., {Teplitz}, H.~I., {et~al.} 2020, \apj, 904, 59, \dodoi{10.3847/1538-4357/abbd43}

\bibitem[{{Astropy Collaboration} {et~al.}(2013){Astropy Collaboration}, {Robitaille}, {Tollerud}, {Greenfield}, {Droettboom}, {Bray}, {Aldcroft}, {Davis}, {Ginsburg}, {Price-Whelan}, {Kerzendorf}, {Conley}, {Crighton}, {Barbary}, {Muna}, {Ferguson}, {Grollier}, {Parikh}, {Nair}, {Unther}, {Deil}, {Woillez}, {Conseil}, {Kramer}, {Turner}, {Singer}, {Fox}, {Weaver}, {Zabalza}, {Edwards}, {Azalee Bostroem}, {Burke}, {Casey}, {Crawford}, {Dencheva}, {Ely}, {Jenness}, {Labrie}, {Lim}, {Pierfederici}, {Pontzen}, {Ptak}, {Refsdal}, {Servillat}, \& {Streicher}}]{AstropyCollaboration}
{Astropy Collaboration}, {Robitaille}, T.~P., {Tollerud}, E.~J., {et~al.} 2013, \aap, 558, A33, \dodoi{10.1051/0004-6361/201322068}

\bibitem[{{Bassett} {et~al.}(2021){Bassett}, {Ryan-Weber}, {Cooke}, {Me{\v{s}}tri{\'c}}, {Kakiichi}, {Prichard}, \& {Rafelski}}]{Bassettetal2021}
{Bassett}, R., {Ryan-Weber}, E.~V., {Cooke}, J., {et~al.} 2021, \mnras, 502, 108, \dodoi{10.1093/mnras/stab070}

\bibitem[{{Bertin} \& {Arnouts}(1996)}]{BertinArnouts1996}
{Bertin}, E., \& {Arnouts}, S. 1996, \aaps, 117, 393, \dodoi{10.1051/aas:1996164}

\bibitem[{{Bian} {et~al.}(2017){Bian}, {Fan}, {McGreer}, {Cai}, \& {Jiang}}]{Bianetal2017}
{Bian}, F., {Fan}, X., {McGreer}, I., {Cai}, Z., \& {Jiang}, L. 2017, \apjl, 837, L12, \dodoi{10.3847/2041-8213/aa5ff7}

\bibitem[{{Borgohain} {et~al.}(2022){Borgohain}, {Saha}, {Elmegreen}, {Gogoi}, {Combes}, \& {Tandon}}]{Borgohainetal2022}
{Borgohain}, A., {Saha}, K., {Elmegreen}, B., {et~al.} 2022, \nat, 607, 459, \dodoi{10.1038/s41586-022-04905-9}

\bibitem[{{Borthakur} {et~al.}(2014){Borthakur}, {Heckman}, {Leitherer}, \& {Overzier}}]{Borthakuretal2014}
{Borthakur}, S., {Heckman}, T.~M., {Leitherer}, C., \& {Overzier}, R.~A. 2014, Science, 346, 216, \dodoi{10.1126/science.1254214}

\bibitem[{{Brammer} {et~al.}(2012){Brammer}, {van Dokkum}, {Franx}, {Fumagalli}, {Patel}, {Rix}, {Skelton}, {Kriek}, {Nelson}, {Schmidt}, {Bezanson}, {da Cunha}, {Erb}, {Fan}, {F{\"o}rster Schreiber}, {Illingworth}, {Labb{\'e}}, {Leja}, {Lundgren}, {Magee}, {Marchesini}, {McCarthy}, {Momcheva}, {Muzzin}, {Quadri}, {Steidel}, {Tal}, {Wake}, {Whitaker}, \& {Williams}}]{Brammeretal2012}
{Brammer}, G.~B., {van Dokkum}, P.~G., {Franx}, M., {et~al.} 2012, \apjs, 200, 13, \dodoi{10.1088/0067-0049/200/2/13}

\bibitem[{{Capak} {et~al.}(2004){Capak}, {Cowie}, {Hu}, {Barger}, {Dickinson}, {Fernandez}, {Giavalisco}, {Komiyama}, {Kretchmer}, {McNally}, {Miyazaki}, {Okamura}, \& {Stern}}]{Capaketal2004}
{Capak}, P., {Cowie}, L.~L., {Hu}, E.~M., {et~al.} 2004, \aj, 127, 180, \dodoi{10.1086/380611}

\bibitem[{{Fletcher} {et~al.}(2019){Fletcher}, {Tang}, {Robertson}, {Nakajima}, {Ellis}, {Stark}, \& {Inoue}}]{Fletcheretal2019}
{Fletcher}, T.~J., {Tang}, M., {Robertson}, B.~E., {et~al.} 2019, \apj, 878, 87, \dodoi{10.3847/1538-4357/ab2045}

\bibitem[{{Flury} {et~al.}(2022){Flury}, {Jaskot}, {Ferguson}, {Worseck}, {Makan}, {Chisholm}, {Saldana-Lopez}, {Schaerer}, {McCandliss}, {Wang}, {Ford}, {Heckman}, {Ji}, {Giavalisco}, {Amorin}, {Atek}, {Blaizot}, {Borthakur}, {Carr}, {Castellano}, {Cristiani}, {De Barros}, {Dickinson}, {Finkelstein}, {Fleming}, {Fontanot}, {Garel}, {Grazian}, {Hayes}, {Henry}, {Mauerhofer}, {Micheva}, {Oey}, {Ostlin}, {Papovich}, {Pentericci}, {Ravindranath}, {Rosdahl}, {Rutkowski}, {Santini}, {Scarlata}, {Teplitz}, {Thuan}, {Trebitsch}, {Vanzella}, {Verhamme}, \& {Xu}}]{Flury22}
{Flury}, S.~R., {Jaskot}, A.~E., {Ferguson}, H.~C., {et~al.} 2022, \apjs, 260, 1, \dodoi{10.3847/1538-4365/ac5331}

\bibitem[{{Fumagalli} {et~al.}(2013){Fumagalli}, {O'Meara}, {Prochaska}, \& {Worseck}}]{Fumagalli2013}
{Fumagalli}, M., {O'Meara}, J.~M., {Prochaska}, J.~X., \& {Worseck}, G. 2013, \apj, 775, 78, \dodoi{10.1088/0004-637X/775/1/78}

\bibitem[{{Giavalisco} {et~al.}(2004){Giavalisco}, {Ferguson}, {Koekemoer}, {Dickinson}, {Alexander}, {Bauer}, {Bergeron}, {Biagetti}, {Brandt}, {Casertano}, {Cesarsky}, {Chatzichristou}, {Conselice}, {Cristiani}, {Da Costa}, {Dahlen}, {de Mello}, {Eisenhardt}, {Erben}, {Fall}, {Fassnacht}, {Fosbury}, {Fruchter}, {Gardner}, {Grogin}, {Hook}, {Hornschemeier}, {Idzi}, {Jogee}, {Kretchmer}, {Laidler}, {Lee}, {Livio}, {Lucas}, {Madau}, {Mobasher}, {Moustakas}, {Nonino}, {Padovani}, {Papovich}, {Park}, {Ravindranath}, {Renzini}, {Richardson}, {Riess}, {Rosati}, {Schirmer}, {Schreier}, {Somerville}, {Spinrad}, {Stern}, {Stiavelli}, {Strolger}, {Urry}, {Vandame}, {Williams}, \& {Wolf}}]{Giavaliscoetal2004}
{Giavalisco}, M., {Ferguson}, H.~C., {Koekemoer}, A.~M., {et~al.} 2004, \apjl, 600, L93, \dodoi{10.1086/379232}

\bibitem[{{Grogin} {et~al.}(2011){Grogin}, {Kocevski}, {Faber}, {Ferguson}, {Koekemoer}, {Riess}, {Acquaviva}, {Alexander}, {Almaini}, {Ashby}, {Barden}, {Bell}, {Bournaud}, {Brown}, {Caputi}, {Casertano}, {Cassata}, {Castellano}, {Challis}, {Chary}, {Cheung}, {Cirasuolo}, {Conselice}, {Roshan Cooray}, {Croton}, {Daddi}, {Dahlen}, {Dav{\'e}}, {de Mello}, {Dekel}, {Dickinson}, {Dolch}, {Donley}, {Dunlop}, {Dutton}, {Elbaz}, {Fazio}, {Filippenko}, {Finkelstein}, {Fontana}, {Gardner}, {Garnavich}, {Gawiser}, {Giavalisco}, {Grazian}, {Guo}, {Hathi}, {H{\"a}ussler}, {Hopkins}, {Huang}, {Huang}, {Jha}, {Kartaltepe}, {Kirshner}, {Koo}, {Lai}, {Lee}, {Li}, {Lotz}, {Lucas}, {Madau}, {McCarthy}, {McGrath}, {McIntosh}, {McLure}, {Mobasher}, {Moustakas}, {Mozena}, {Nandra}, {Newman}, {Niemi}, {Noeske}, {Papovich}, {Pentericci}, {Pope}, {Primack}, {Rajan}, {Ravindranath}, {Reddy}, {Renzini}, {Rix}, {Robaina}, {Rodney}, {Rosario}, {Rosati}, {Salimbeni}, {Scarlata}, {Siana}, {Simard}, {Smidt}, {Somerville}, {Spinrad},
  {Straughn}, {Strolger}, {Telford}, {Teplitz}, {Trump}, {van der Wel}, {Villforth}, {Wechsler}, {Weiner}, {Wiklind}, {Wild}, {Wilson}, {Wuyts}, {Yan}, \& {Yun}}]{Groginetal2011}
{Grogin}, N.~A., {Kocevski}, D.~D., {Faber}, S.~M., {et~al.} 2011, \apjs, 197, 35, \dodoi{10.1088/0067-0049/197/2/35}

\bibitem[{{Inoue} {et~al.}(2014){Inoue}, {Shimizu}, {Iwata}, \& {Tanaka}}]{Inoueetal2014}
{Inoue}, A.~K., {Shimizu}, I., {Iwata}, I., \& {Tanaka}, M. 2014, \mnras, 442, 1805, \dodoi{10.1093/mnras/stu936}

\bibitem[{{Iwata} {et~al.}(2009){Iwata}, {Inoue}, {Matsuda}, {Furusawa}, {Hayashino}, {Kousai}, {Akiyama}, {Yamada}, {Burgarella}, \& {Deharveng}}]{Iwataetal2009}
{Iwata}, I., {Inoue}, A.~K., {Matsuda}, Y., {et~al.} 2009, \apj, 692, 1287, \dodoi{10.1088/0004-637X/692/2/1287}

\bibitem[{{Izotov} {et~al.}(2016){Izotov}, {Orlitov{\'a}}, {Schaerer}, {Thuan}, {Verhamme}, {Guseva}, \& {Worseck}}]{Izotovetal2016}
{Izotov}, Y.~I., {Orlitov{\'a}}, I., {Schaerer}, D., {et~al.} 2016, \nat, 529, 178, \dodoi{10.1038/nature16456}

\bibitem[{{Izotov} {et~al.}(2018){Izotov}, {Schaerer}, {Worseck}, {Guseva}, {Thuan}, {Verhamme}, {Orlitov{\'a}}, \& {Fricke}}]{Izotovetal2018}
{Izotov}, Y.~I., {Schaerer}, D., {Worseck}, G., {et~al.} 2018, \mnras, 474, 4514, \dodoi{10.1093/mnras/stx3115}

\bibitem[{{Izotov} {et~al.}(2021){Izotov}, {Worseck}, {Schaerer}, {Guseva}, {Chisholm}, {Thuan}, {Fricke}, \& {Verhamme}}]{Izotov21}
{Izotov}, Y.~I., {Worseck}, G., {Schaerer}, D., {et~al.} 2021, \mnras, 503, 1734, \dodoi{10.1093/mnras/stab612}

\bibitem[{{Kennicutt}(1998)}]{Kennicutt98}
{Kennicutt}, Robert~C., J. 1998, \apj, 498, 541, \dodoi{10.1086/305588}

\bibitem[{{Koekemoer} {et~al.}(2011){Koekemoer}, {Faber}, {Ferguson}, {Grogin}, {Kocevski}, {Koo}, {Lai}, {Lotz}, {Lucas}, {McGrath}, {Ogaz}, {Rajan}, {Riess}, {Rodney}, {Strolger}, {Casertano}, {Castellano}, {Dahlen}, {Dickinson}, {Dolch}, {Fontana}, {Giavalisco}, {Grazian}, {Guo}, {Hathi}, {Huang}, {van der Wel}, {Yan}, {Acquaviva}, {Alexander}, {Almaini}, {Ashby}, {Barden}, {Bell}, {Bournaud}, {Brown}, {Caputi}, {Cassata}, {Challis}, {Chary}, {Cheung}, {Cirasuolo}, {Conselice}, {Roshan Cooray}, {Croton}, {Daddi}, {Dav{\'e}}, {de Mello}, {de Ravel}, {Dekel}, {Donley}, {Dunlop}, {Dutton}, {Elbaz}, {Fazio}, {Filippenko}, {Finkelstein}, {Frazer}, {Gardner}, {Garnavich}, {Gawiser}, {Gruetzbauch}, {Hartley}, {H{\"a}ussler}, {Herrington}, {Hopkins}, {Huang}, {Jha}, {Johnson}, {Kartaltepe}, {Khostovan}, {Kirshner}, {Lani}, {Lee}, {Li}, {Madau}, {McCarthy}, {McIntosh}, {McLure}, {McPartland}, {Mobasher}, {Moreira}, {Mortlock}, {Moustakas}, {Mozena}, {Nandra}, {Newman}, {Nielsen}, {Niemi}, {Noeske}, {Papovich},
  {Pentericci}, {Pope}, {Primack}, {Ravindranath}, {Reddy}, {Renzini}, {Rix}, {Robaina}, {Rosario}, {Rosati}, {Salimbeni}, {Scarlata}, {Siana}, {Simard}, {Smidt}, {Snyder}, {Somerville}, {Spinrad}, {Straughn}, {Telford}, {Teplitz}, {Trump}, {Vargas}, {Villforth}, {Wagner}, {Wandro}, {Wechsler}, {Weiner}, {Wiklind}, {Wild}, {Wilson}, {Wuyts}, \& {Yun}}]{Koekemoeretal2011}
{Koekemoer}, A.~M., {Faber}, S.~M., {Ferguson}, H.~C., {et~al.} 2011, \apjs, 197, 36, \dodoi{10.1088/0067-0049/197/2/36}

\bibitem[{{Koyama} {et~al.}(2019){Koyama}, {Shimakawa}, {Yamamura}, {Kodama}, \& {Hayashi}}]{Koyam19}
{Koyama}, Y., {Shimakawa}, R., {Yamamura}, I., {Kodama}, T., \& {Hayashi}, M. 2019, \pasj, 71, 8, \dodoi{10.1093/pasj/psy113}

\bibitem[{{Leitet} {et~al.}(2013){Leitet}, {Bergvall}, {Hayes}, {Linn{\'e}}, \& {Zackrisson}}]{Leitetetal2013}
{Leitet}, E., {Bergvall}, N., {Hayes}, M., {Linn{\'e}}, S., \& {Zackrisson}, E. 2013, \aap, 553, A106, \dodoi{10.1051/0004-6361/201118370}

\bibitem[{{Leitherer} {et~al.}(2016){Leitherer}, {Hernandez}, {Lee}, \& {Oey}}]{Leithereretal2016}
{Leitherer}, C., {Hernandez}, S., {Lee}, J.~C., \& {Oey}, M.~S. 2016, \apj, 823, 64, \dodoi{10.3847/0004-637X/823/1/64}

\bibitem[{{Liu} {et~al.}(2023){Liu}, {Jiang}, {Windhorst}, {Guo}, \& {Zheng}}]{Liuetal2023}
{Liu}, Y., {Jiang}, L., {Windhorst}, R.~A., {Guo}, Y., \& {Zheng}, Z. 2023, arXiv e-prints, arXiv:2310.07283, \dodoi{10.48550/arXiv.2310.07283}

\bibitem[{{Madau}(1995)}]{Madau95}
{Madau}, P. 1995, \apj, 441, 18, \dodoi{10.1086/175332}

\bibitem[{{Marques-Chaves} {et~al.}(2021){Marques-Chaves}, {Schaerer}, {{\'A}lvarez-M{\'a}rquez}, {Colina}, {Dessauges-Zavadsky}, {P{\'e}rez-Fournon}, {Saldana-Lopez}, \& {Verhamme}}]{MarquesChaves21}
{Marques-Chaves}, R., {Schaerer}, D., {{\'A}lvarez-M{\'a}rquez}, J., {et~al.} 2021, \mnras, 507, 524, \dodoi{10.1093/mnras/stab2187}

\bibitem[{{Marques-Chaves} {et~al.}(2022){Marques-Chaves}, {Schaerer}, {{\'A}lvarez-M{\'a}rquez}, {Verhamme}, {Ceverino}, {Chisholm}, {Colina}, {Dessauges-Zavadsky}, {P{\'e}rez-Fournon}, {Saldana-Lopez}, {Upadhyaya}, \& {Vanzella}}]{MarquesChaves22}
---. 2022, \mnras, 517, 2972, \dodoi{10.1093/mnras/stac2893}

\bibitem[{{Momcheva} {et~al.}(2016){Momcheva}, {Brammer}, {van Dokkum}, {Skelton}, {Whitaker}, {Nelson}, {Fumagalli}, {Maseda}, {Leja}, {Franx}, {Rix}, {Bezanson}, {Da Cunha}, {Dickey}, {F{\"o}rster Schreiber}, {Illingworth}, {Kriek}, {Labb{\'e}}, {Ulf Lange}, {Lundgren}, {Magee}, {Marchesini}, {Oesch}, {Pacifici}, {Patel}, {Price}, {Tal}, {Wake}, {van der Wel}, \& {Wuyts}}]{Momchevaetal2016}
{Momcheva}, I.~G., {Brammer}, G.~B., {van Dokkum}, P.~G., {et~al.} 2016, \apjs, 225, 27, \dodoi{10.3847/0067-0049/225/2/27}

\bibitem[{{Mondal} {et~al.}(2023){Mondal}, {Saha}, {Bhattacharya}, {Borgohain}, {Tandon}, {Rafelski}, {Jansen}, {Windhorst}, {Teplitz}, \& {Smith}}]{Mondaletal2023}
{Mondal}, C., {Saha}, K., {Bhattacharya}, S., {et~al.} 2023, \apjs, 264, 40, \dodoi{10.3847/1538-4365/aca7c4}

\bibitem[{{Nordon} {et~al.}(2013){Nordon}, {Lutz}, {Saintonge}, {Berta}, {Wuyts}, {F{\"o}rster Schreiber}, {Genzel}, {Magnelli}, {Poglitsch}, {Popesso}, {Rosario}, {Sturm}, \& {Tacconi}}]{Nordonetal2013}
{Nordon}, R., {Lutz}, D., {Saintonge}, A., {et~al.} 2013, \apj, 762, 125, \dodoi{10.1088/0004-637X/762/2/125}

\bibitem[{{O'Meara} {et~al.}(2013){O'Meara}, {Prochaska}, {Worseck}, {Chen}, \& {Madau}}]{OMeara2013}
{O'Meara}, J.~M., {Prochaska}, J.~X., {Worseck}, G., {Chen}, H.-W., \& {Madau}, P. 2013, \apj, 765, 137, \dodoi{10.1088/0004-637X/765/2/137}

\bibitem[{{Osterbrock} \& {Bochkarev}(1989)}]{OsterbrockBochkarev1989}
{Osterbrock}, D.~E., \& {Bochkarev}, N.~G. 1989, \sovast, 33, 694

\bibitem[{{Osterbrock} \& {Ferland}(2006)}]{OsterbrockFerland2006}
{Osterbrock}, D.~E., \& {Ferland}, G.~J. 2006, {Astrophysics of gaseous nebulae and active galactic nuclei}

\bibitem[{{Prichard} {et~al.}(2022){Prichard}, {Rafelski}, {Cooke}, {Me{\v{s}}tri{\'c}}, {Bassett}, {Ryan-Weber}, {Sunnquist}, {Alavi}, {Hathi}, {Wang}, {Revalski}, {Bajaj}, {O'Meara}, \& {Spitler}}]{Prichardetal2022}
{Prichard}, L.~J., {Rafelski}, M., {Cooke}, J., {et~al.} 2022, \apj, 924, 14, \dodoi{10.3847/1538-4357/ac3004}

\bibitem[{{Prochaska} {et~al.}(2009){Prochaska}, {Worseck}, \& {O'Meara}}]{Prochaska2009}
{Prochaska}, J.~X., {Worseck}, G., \& {O'Meara}, J.~M. 2009, \apjl, 705, L113, \dodoi{10.1088/0004-637X/705/2/L113}

\bibitem[{{Reddy} {et~al.}(2018){Reddy}, {Oesch}, {Bouwens}, {Montes}, {Illingworth}, {Steidel}, {van Dokkum}, {Atek}, {Carollo}, {Cibinel}, {Holden}, {Labb{\'e}}, {Magee}, {Morselli}, {Nelson}, \& {Wilkins}}]{Reddy18}
{Reddy}, N.~A., {Oesch}, P.~A., {Bouwens}, R.~J., {et~al.} 2018, \apj, 853, 56, \dodoi{10.3847/1538-4357/aaa3e7}

\bibitem[{{Reddy} {et~al.}(2020){Reddy}, {Shapley}, {Kriek}, {Steidel}, {Shivaei}, {Sanders}, {Mobasher}, {Coil}, {Siana}, {Freeman}, {Azadi}, {Fetherolf}, {Leung}, {Price}, \& {Zick}}]{Reddy20}
{Reddy}, N.~A., {Shapley}, A.~E., {Kriek}, M., {et~al.} 2020, \apj, 902, 123, \dodoi{10.3847/1538-4357/abb674}

\bibitem[{{Rivera-Thorsen} {et~al.}(2019){Rivera-Thorsen}, {Dahle}, {Chisholm}, {Florian}, {Gronke}, {Rigby}, {Gladders}, {Mahler}, {Sharon}, \& {Bayliss}}]{RiveraThorsen19}
{Rivera-Thorsen}, T.~E., {Dahle}, H., {Chisholm}, J., {et~al.} 2019, Science, 366, 738, \dodoi{10.1126/science.aaw0978}

\bibitem[{{Robertson} {et~al.}(2010){Robertson}, {Ellis}, {Dunlop}, {McLure}, \& {Stark}}]{Robertsonetal2010}
{Robertson}, B.~E., {Ellis}, R.~S., {Dunlop}, J.~S., {McLure}, R.~J., \& {Stark}, D.~P. 2010, \nat, 468, 49, \dodoi{10.1038/nature09527}

\bibitem[{{Rudie} {et~al.}(2013){Rudie}, {Steidel}, {Shapley}, \& {Pettini}}]{Rudie2013}
{Rudie}, G.~C., {Steidel}, C.~C., {Shapley}, A.~E., \& {Pettini}, M. 2013, \apj, 769, 146, \dodoi{10.1088/0004-637X/769/2/146}

\bibitem[{{Rudie} {et~al.}(2012){Rudie}, {Steidel}, {Trainor}, {Rakic}, {Bogosavljevi{\'c}}, {Pettini}, {Reddy}, {Shapley}, {Erb}, \& {Law}}]{Rudie2012}
{Rudie}, G.~C., {Steidel}, C.~C., {Trainor}, R.~F., {et~al.} 2012, \apj, 750, 67, \dodoi{10.1088/0004-637X/750/1/67}

\bibitem[{{Saha} {et~al.}(2020){Saha}, {Tandon}, {Simmonds}, {Verhamme}, {Paswan}, {Schaerer}, {Rutkowski}, {Borgohain}, {Elmegreen}, {Inoue}, {Combes}, {Elmegreen}, \& {Paalvast}}]{Sahaetal2020}
{Saha}, K., {Tandon}, S.~N., {Simmonds}, C., {et~al.} 2020, Nature Astronomy, 4, 1185, \dodoi{10.1038/s41550-020-1173-5}

\bibitem[{{Schlafly} \& {Finkbeiner}(2011)}]{SchlaflyFinkbeiner2011}
{Schlafly}, E.~F., \& {Finkbeiner}, D.~P. 2011, \apj, 737, 103, \dodoi{10.1088/0004-637X/737/2/103}

\bibitem[{{Shapley} {et~al.}(2016){Shapley}, {Steidel}, {Strom}, {Bogosavljevi{\'c}}, {Reddy}, {Siana}, {Mostardi}, \& {Rudie}}]{Shapleyetal2016}
{Shapley}, A.~E., {Steidel}, C.~C., {Strom}, A.~L., {et~al.} 2016, \apjl, 826, L24, \dodoi{10.3847/2041-8205/826/2/L24}

\bibitem[{{Siana} {et~al.}(2007){Siana}, {Teplitz}, {Colbert}, {Ferguson}, {Dickinson}, {Brown}, {Conselice}, {de Mello}, {Gardner}, {Giavalisco}, \& {Menanteau}}]{Sianaetal2007}
{Siana}, B., {Teplitz}, H.~I., {Colbert}, J., {et~al.} 2007, \apj, 668, 62, \dodoi{10.1086/521185}

\bibitem[{{Skelton} {et~al.}(2014){Skelton}, {Whitaker}, {Momcheva}, {Brammer}, {van Dokkum}, {Labb{\'e}}, {Franx}, {van der Wel}, {Bezanson}, {Da Cunha}, {Fumagalli}, {F{\"o}rster Schreiber}, {Kriek}, {Leja}, {Lundgren}, {Magee}, {Marchesini}, {Maseda}, {Nelson}, {Oesch}, {Pacifici}, {Patel}, {Price}, {Rix}, {Tal}, {Wake}, \& {Wuyts}}]{Skeltonetal2014}
{Skelton}, R.~E., {Whitaker}, K.~E., {Momcheva}, I.~G., {et~al.} 2014, \apjs, 214, 24, \dodoi{10.1088/0067-0049/214/2/24}

\bibitem[{{Smith} {et~al.}(2018){Smith}, {Windhorst}, {Jansen}, {Cohen}, {Jiang}, {Dijkstra}, {Koekemoer}, {Bielby}, {Inoue}, {MacKenty}, {O'Connell}, \& {Silk}}]{Smithetal2018}
{Smith}, B.~M., {Windhorst}, R.~A., {Jansen}, R.~A., {et~al.} 2018, \apj, 853, 191, \dodoi{10.3847/1538-4357/aaa3dc}

\bibitem[{{Smith} {et~al.}(2020){Smith}, {Windhorst}, {Cohen}, {Koekemoer}, {Jansen}, {White}, {Borthakur}, {Hathi}, {Jiang}, {Rutkowski}, {Ryan}, {Inoue}, {O'Connell}, {MacKenty}, {Conselice}, \& {Silk}}]{Smithetal2020}
{Smith}, B.~M., {Windhorst}, R.~A., {Cohen}, S.~H., {et~al.} 2020, \apj, 897, 41, \dodoi{10.3847/1538-4357/ab8811}

\bibitem[{{Sobral} {et~al.}(2012){Sobral}, {Best}, {Matsuda}, {Smail}, {Geach}, \& {Cirasuolo}}]{Sobraletal2012}
{Sobral}, D., {Best}, P.~N., {Matsuda}, Y., {et~al.} 2012, \mnras, 420, 1926, \dodoi{10.1111/j.1365-2966.2011.19977.x}

\bibitem[{{Steidel} {et~al.}(2018){Steidel}, {Bogosavljevi{\'c}}, {Shapley}, {Reddy}, {Rudie}, {Pettini}, {Trainor}, \& {Strom}}]{Steideletal2018}
{Steidel}, C.~C., {Bogosavljevi{\'c}}, M., {Shapley}, A.~E., {et~al.} 2018, \apj, 869, 123, \dodoi{10.3847/1538-4357/aaed28}

\bibitem[{{Steidel} {et~al.}(2010){Steidel}, {Erb}, {Shapley}, {Pettini}, {Reddy}, {Bogosavljevi{\'c}}, {Rudie}, \& {Rakic}}]{Steidel2010}
{Steidel}, C.~C., {Erb}, D.~K., {Shapley}, A.~E., {et~al.} 2010, \apj, 717, 289, \dodoi{10.1088/0004-637X/717/1/289}

\bibitem[{{Steidel} {et~al.}(2004){Steidel}, {Shapley}, {Pettini}, {Adelberger}, {Erb}, {Reddy}, \& {Hunt}}]{Steidel2004}
{Steidel}, C.~C., {Shapley}, A.~E., {Pettini}, M., {et~al.} 2004, \apj, 604, 534, \dodoi{10.1086/381960}

\bibitem[{{Steidel} {et~al.}(2014){Steidel}, {Rudie}, {Strom}, {Pettini}, {Reddy}, {Shapley}, {Trainor}, {Erb}, {Turner}, {Konidaris}, {Kulas}, {Mace}, {Matthews}, \& {McLean}}]{Steidel2014}
{Steidel}, C.~C., {Rudie}, G.~C., {Strom}, A.~L., {et~al.} 2014, \apj, 795, 165, \dodoi{10.1088/0004-637X/795/2/165}

\bibitem[{{Strom} {et~al.}(2017){Strom}, {Steidel}, {Rudie}, {Trainor}, {Pettini}, \& {Reddy}}]{Strom2017}
{Strom}, A.~L., {Steidel}, C.~C., {Rudie}, G.~C., {et~al.} 2017, \apj, 836, 164, \dodoi{10.3847/1538-4357/836/2/164}

\bibitem[{{Vanzella} {et~al.}(2010){Vanzella}, {Giavalisco}, {Inoue}, {Nonino}, {Fontanot}, {Cristiani}, {Grazian}, {Dickinson}, {Stern}, {Tozzi}, {Giallongo}, {Ferguson}, {Spinrad}, {Boutsia}, {Fontana}, {Rosati}, \& {Pentericci}}]{Vanzellaetal2010}
{Vanzella}, E., {Giavalisco}, M., {Inoue}, A.~K., {et~al.} 2010, \apj, 725, 1011, \dodoi{10.1088/0004-637X/725/1/1011}

\bibitem[{{Vanzella} {et~al.}(2012){Vanzella}, {Guo}, {Giavalisco}, {Grazian}, {Castellano}, {Cristiani}, {Dickinson}, {Fontana}, {Nonino}, {Giallongo}, {Pentericci}, {Galametz}, {Faber}, {Ferguson}, {Grogin}, {Koekemoer}, {Newman}, \& {Siana}}]{Vanzellaetal2012}
{Vanzella}, E., {Guo}, Y., {Giavalisco}, M., {et~al.} 2012, \apj, 751, 70, \dodoi{10.1088/0004-637X/751/1/70}

\bibitem[{{Vanzella} {et~al.}(2016){Vanzella}, {de Barros}, {Vasei}, {Alavi}, {Giavalisco}, {Siana}, {Grazian}, {Hasinger}, {Suh}, {Cappelluti}, {Vito}, {Amorin}, {Balestra}, {Brusa}, {Calura}, {Castellano}, {Comastri}, {Fontana}, {Gilli}, {Mignoli}, {Pentericci}, {Vignali}, \& {Zamorani}}]{Vanzellaetal2016}
{Vanzella}, E., {de Barros}, S., {Vasei}, K., {et~al.} 2016, \apj, 825, 41, \dodoi{10.3847/0004-637X/825/1/41}

\bibitem[{{Vanzella} {et~al.}(2018){Vanzella}, {Nonino}, {Cupani}, {Castellano}, {Sani}, {Mignoli}, {Calura}, {Meneghetti}, {Gilli}, {Comastri}, {Mercurio}, {Caminha}, {Caputi}, {Rosati}, {Grillo}, {Cristiani}, {Balestra}, {Fontana}, \& {Giavalisco}}]{Vanzellaetal2018}
{Vanzella}, E., {Nonino}, M., {Cupani}, G., {et~al.} 2018, \mnras, 476, L15, \dodoi{10.1093/mnrasl/sly023}

\bibitem[{{Worseck} {et~al.}(2014){Worseck}, {Prochaska}, {O'Meara}, {Becker}, {Ellison}, {Lopez}, {Meiksin}, {M{\'e}nard}, {Murphy}, \& {Fumagalli}}]{Worseck2014}
{Worseck}, G., {Prochaska}, J.~X., {O'Meara}, J.~M., {et~al.} 2014, \mnras, 445, 1745, \dodoi{10.1093/mnras/stu1827}

\bibitem[{{Xue} {et~al.}(2016){Xue}, {Luo}, {Brandt}, {Alexander}, {Bauer}, {Lehmer}, \& {Yang}}]{Xueetal2016}
{Xue}, Y.~Q., {Luo}, B., {Brandt}, W.~N., {et~al.} 2016, \apjs, 224, 15, \dodoi{10.3847/0067-0049/224/2/15}

\end{thebibliography}

\appendix
\counterwithin{table}{section}
\counterwithin{figure}{section}

\section{Supplementary plots and Table} 
\label{sec:supp_plots}

\begin{table}
\centering
\caption{SExtractor Parameters Used for Detection and Photometry in F154W. Note that the DETECT$\_$THRESH (1.2) is the relative detection threshold in terms of the background rms value above the mean local sky background.}
\begin{tabular}{ll}
\hline\hline
Parameter Name & Value \\
\hline
DETECT$\_$MINAREA & 4 \\
DETECT$\_$THRESH & 1.2 \\
FILTER$\_$NAME & default.conv \\
DEBLEND$\_$NTHRESH & 32 \\
DEBLEND$\_$MINCONT & 0.01 \\
CLEAN$\_$PARAM & 1.0 \\
PHOT$\_$FLUXFRAC & 0.5 \\
PHOT$\_$APERTURES & 6.71 \\
PHOT$\_$AUTOPARAMS & 2.5, 3.5 \\
PIXEL$\_$SCALE & 0.417 \\
BACK$\_$SIZE & 9 \\
BACK$\_$FILTERSIZE & 3 \\

 \hline
\end{tabular}
\label{tab:SExtractor}
\end{table}

\begin{table*}
\centering
\caption{Spectral Properties of the LyC emitters- Column(1): 3D-HST grism ID, Column(2): RA in hms, Column(3): Dec in dms, Column(4): spectroscopic redshift, Column(5),(6) \& (7): H$\alpha$, H$\beta$ and [OIII]5007 line fluxes, respectively in units of 10$^{-17}$ ergs~s$^{-1}$~cm$^{-2}$, corrected for foreground and internal dust extinction, column(8): SFR in units of M$_{\odot}$ yr$^{-1}$ derived from corrected H$\alpha$ line flux and column(9): specific star formation rate (sSFR) in units of Gyr$^{-1}$.}

\begin{tabular}{ccccccccc}
\hline\hline
grism ID & RA & Dec & {z} & {H$\alpha$ flux} & H$\beta$ flux & {[OIII] flux} & {H$\alpha$ SFR} & {sSFR}\\\cline{5-7}
 & hms & dms & & \multicolumn{3}{c}{10$^{-17}ergs~s^{-1}~cm^{-2}$} & M$_{\odot}$ yr$^{-1}$ & Gyr$^{-1}$\\
\hline
goodsn-33-G141-11332 & 12:36:43.42 & +62:11:51.56 & 1.24 & 44.69$\pm$2.46 & 8.93$\pm$7.86 & 52.93$\pm$6.10 & 31.46$\pm$1.73 & 7.37$\pm$0.41 \\ 
goodsn-17-G141-36246 & 12:36:52.46 & +62:20:12.51 & 1.26 & 6.13$\pm$1.55 & - & 12.78$\pm$3.11 & 4.49$\pm$1.14 & 3.65$\pm$0.92 \\ 
goodsn-36-G141-25099 & 12:37:21.18 & +62:15:53.51 & 1.52 & 2.05$\pm$1.45 & - & 15.83$\pm$4.08 & 2.41$\pm$1.70 & 1.18$\pm$0.83 \\ 
goodsn-18-G141-37037 & 12:37:08.89 & +62:20:45.11 & 1.43 & 10.83$\pm$1.46 & - & 18.28$\pm$2.12 & 10.94$\pm$1.47 & 6.90$\pm$0.93 \\ 
goodsn-35-G141-18763 & 12:37:02.91 & +62:14:04.60 & 1.24 & 31.96$\pm$3.01 & - & 23.51$\pm$5.83 & 22.74$\pm$2.14 & 3.52$\pm$0.33 \\ 
goodsn-11-G141-09345 & 12:35:54.96 & +62:11:18.36 & 1.21 & 12.07$\pm$2.67 & - & 8.58$\pm$4.70 & 8.09$\pm$1.79 & 3.37$\pm$0.75 \\ 
goodsn-18-G141-37738 & 12:37:07.48 & +62:21:47.84 & 1.45 & 119.35$\pm$2.67 & 24.96$\pm$3.11 & 251.82$\pm$3.81 & 124.45$\pm$2.78 & 1.06$\pm$0.02 \\ \hline
goodsn-17-G141-35906 & 12:37:11.62 & +62:19:58.62 & 1.30 & 15.52$\pm$2.07 &  13.29$\pm$4.41 &  41.06$\pm$5.79 & 12.41$\pm$1.66 & 6.51$\pm$0.87 \\ 
goodsn-46-G141-14637 & 12:37:34.75 & +62:12:52.38 & 1.49 & 27.28$\pm$2.47 & 14.29$\pm$4.94 & 32.66$\pm$4.62 & 30.34$\pm$2.74 & 4.09$\pm$0.37 \\ 
goodsn-24-G141-18817 & 12:36:40.52 & +62:14:03.57 & 1.49 & 23.44$\pm$1.96 & 15.61$\pm$3.04 & 78.74$\pm$3.91 & 25.85$\pm$2.16 & 19.16$\pm$1.60 \\  

 \hline
\end{tabular}
\label{tab:specs}
\end{table*}

Here, we provide some of the statistics and properties of the sample. Table~\ref{tab:specs} represents the quantities derived from the spectral analysis of the grism G141 spectra of the LyC leakers. The line fluxes mentioned in Table~\ref{tab:specs} are corrected for foreground and internal extinction using \citet{SchlaflyFinkbeiner2011} and UV $\beta$ slope \citet{Reddy18}, respectively. The SFR is obtained using the corrected H$\alpha$ line flux following \citet{Kennicutt98} that employs the Salpeter initial mass function. Whereas, to estimate the sSFR, we use the SED stellar mass from \citep{Momchevaetal2016} catalogs.

Figure~\ref{fig:audf-lyc} shows the 3D-HST grism footprint (blue polygon) over the UVIT F145W band image. The red circles represent the positions of the LyC leakers reported in the paper. Figure~\ref{fig:redshift} left panel shows the redshift distribution of the 178 shortlisted candidates having H$\alpha$ and [OIII] line flux SNR greater than 3 (blue histogram) and the 39 sources identified in UVIT F154W band (black histogram). In the right panel of Figure~\ref{fig:redshift}, we show the main sequence relation (SFR vs stellar mass) for the ten LyC leakers along with the non-leaker galaxies at the same redshift (1$<z<1.5$). For SFR estimates of non-leakers we use the H$\alpha$ line fluxes from \citet{Momchevaetal2016} catalogs.

Figure~\ref{fig:mag-snr} left panel shows the FUV 154W and N242W band magnitude distributions of the 39 sources identified in FUV 154W band of the 178 sources, while on the right is shown the F154W SNR histogram of the 39 sources.

\begin{figure}
    \centering
    \includegraphics[width=\columnwidth]{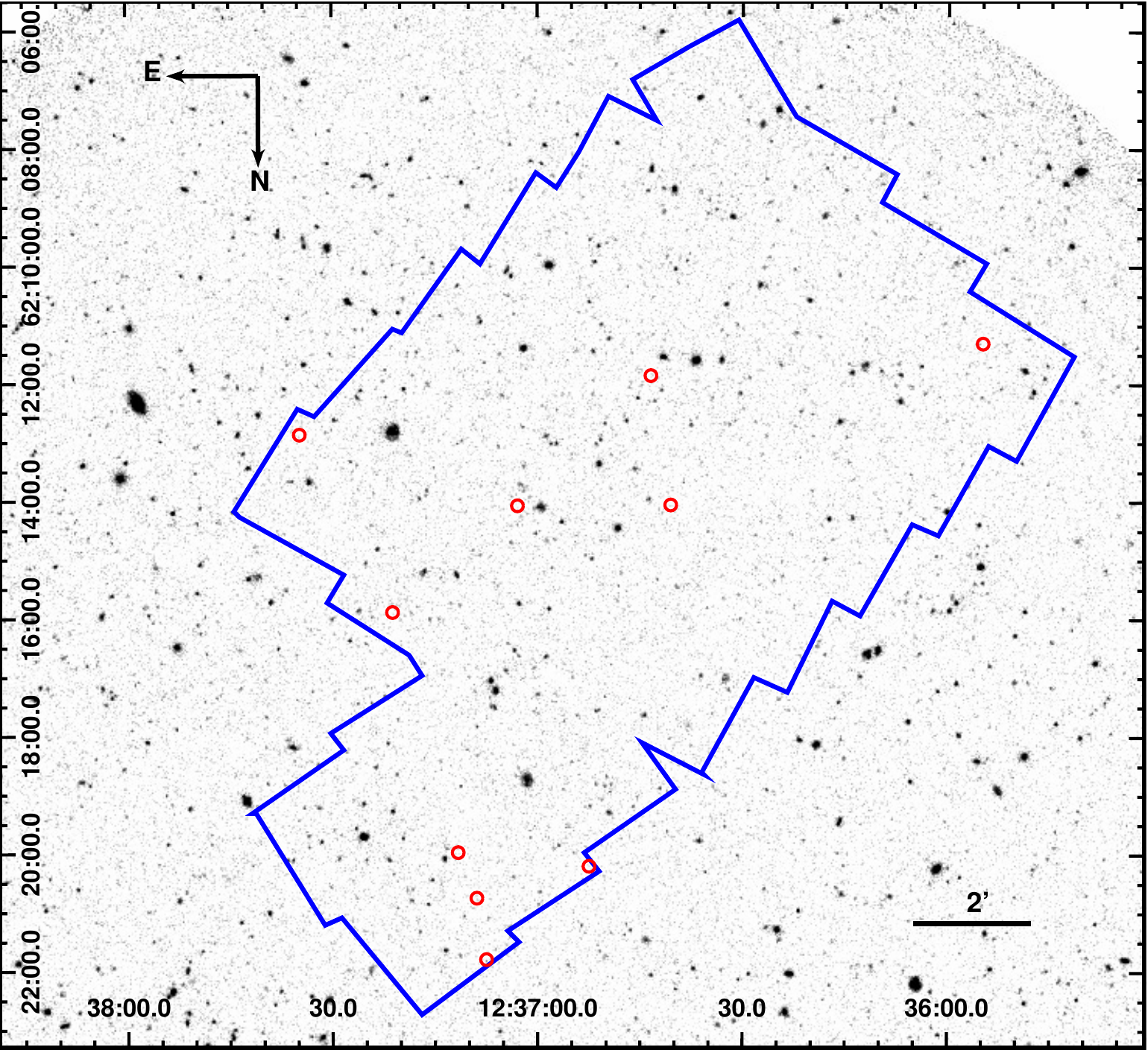}
    \caption{LyC leakers indicated by red circles in the UVIT AUDF-North field. Blue polygon is the 3D-HST GOODS North spectral coverage.}
    \label{fig:audf-lyc}
\end{figure}

\begin{figure*}
    \centering
    \includegraphics[width=\columnwidth]{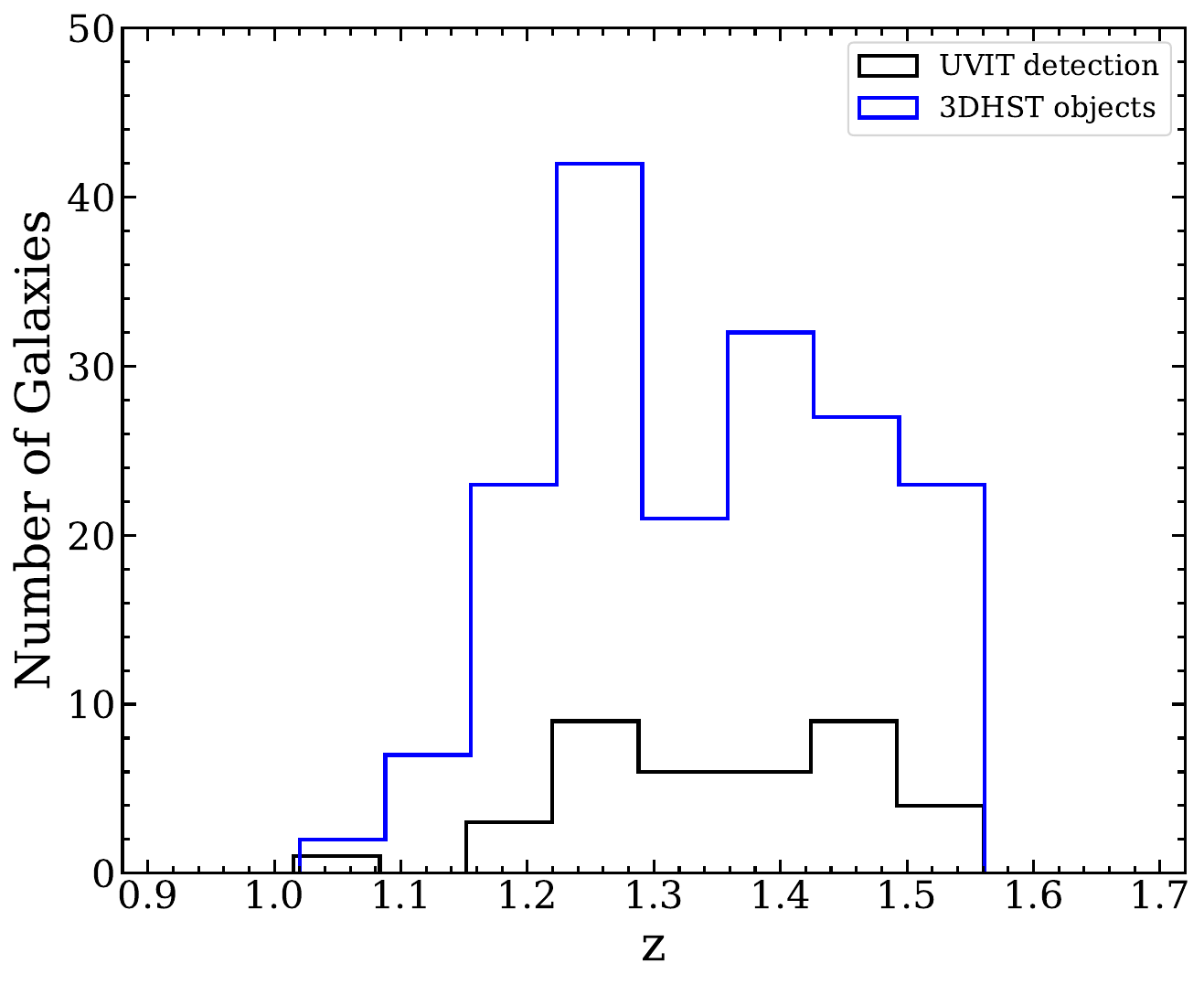}
    \includegraphics[width=\columnwidth]{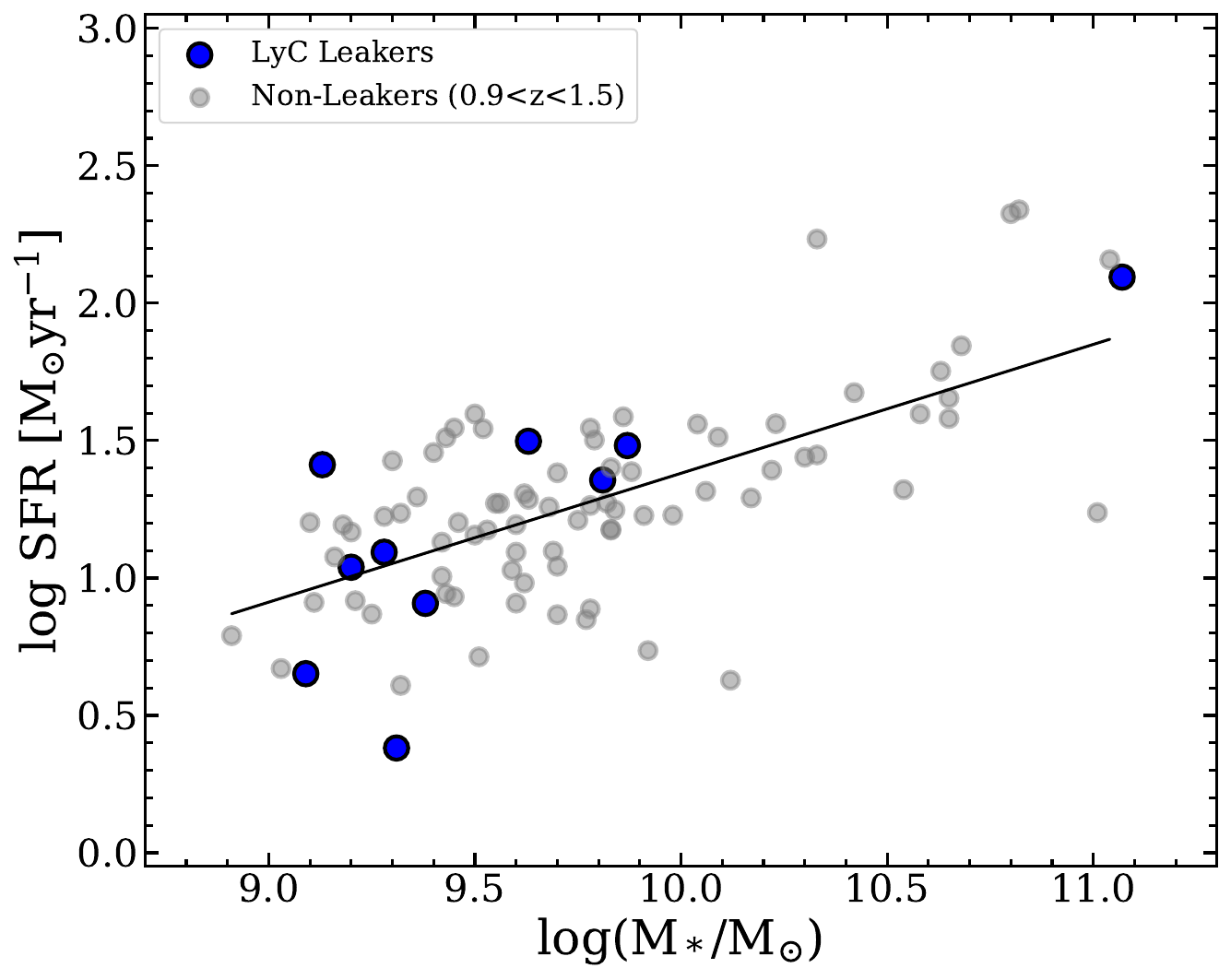}
    \caption{Left- Redshift histogram of sources having H$\alpha$ and O[III] flux SNR $>$ 3 and redshift $<$ 2 (Blue) and those having detection in UVIT FUV (39 sources) out of 178 sources selected from 3D-HST. Right- SFR-stellar mass plot of the LyC leakers (blue filled circles) along with other non-leaker galaxies (gray filled circles) at 1$<z<1.5$. The solid black line is a linear fit to the sample with a slope of 0.46.}
    \label{fig:redshift}
\end{figure*}


\begin{figure*}[t!]
    \centering
    \includegraphics[width=\columnwidth]{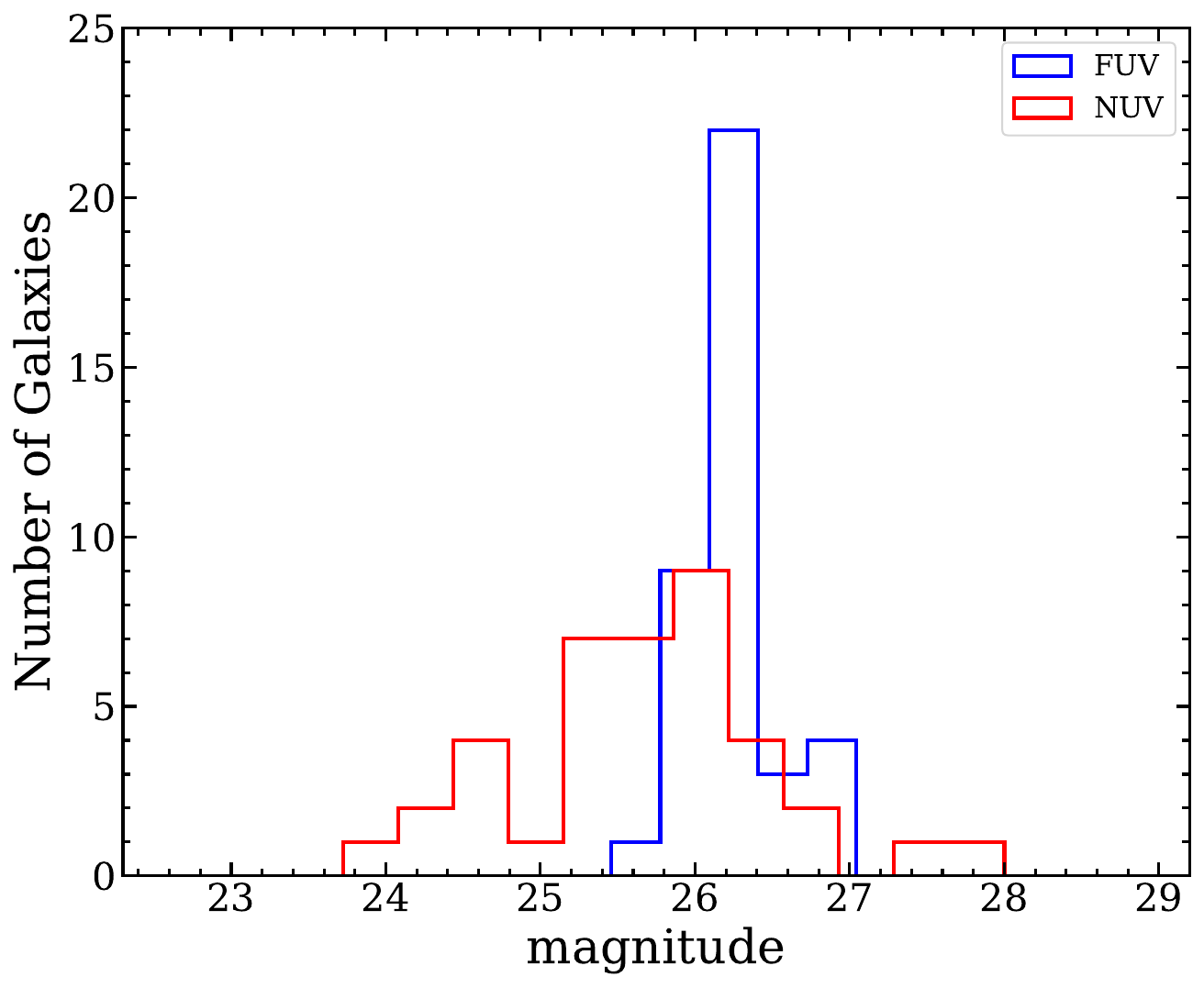}
    \includegraphics[width=\columnwidth]{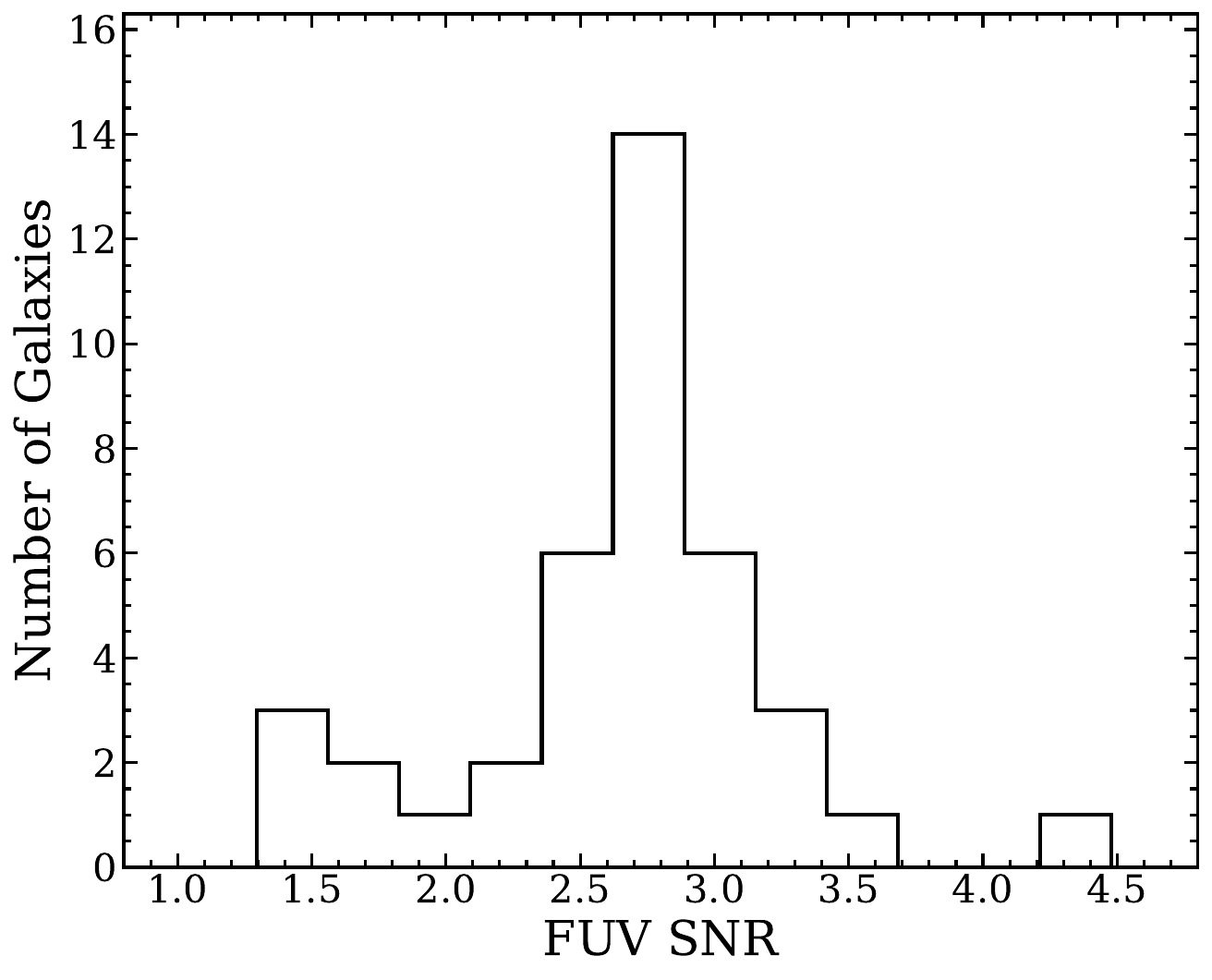}
    \caption{Left-FUV magnitude distribution of clean sources having detection in UVIT FUV (39 sources) out of 178 sources selected from 3D-HST. Right-FUV SNR distribution of 39 clean sources.}
    \label{fig:mag-snr}
\end{figure*}

\end{document}